\documentclass[a4paper,10pt]{amsart}
\usepackage[margin=1in]{geometry}
\usepackage{soul}
\usepackage{lineno,hyperref}
\modulolinenumbers[1]
\usepackage{url}
\usepackage{amsmath,amssymb,amsthm,amsfonts,stmaryrd,mathtools}
\usepackage{bm}
\usepackage{xfrac}
\usepackage{array}
\usepackage{blkarray}
\usepackage{arydshln}
\usepackage{multirow}
\usepackage{empheq}
\usepackage{enumitem}
\usepackage{comment}
  \setlist{nosep} % or \setlist{noitemsep} to leave space around whole list
\usepackage{color}
\usepackage{adjustbox}
\usepackage{hyperref}
\definecolor{darkgreen}{rgb}{0.0, 0.5, 0}
\usepackage[numbers,sort]{natbib}
\usepackage{cleveref}
\usepackage[makeroom]{cancel}
\usepackage{enumitem}
\usepackage{natbib}

\usepackage{xcolor}
\usepackage[foot]{amsaddr}

\makeatletter \@addtoreset{equation}{section}

\makeatother

\newtheorem{corollary}{Corollary}[section]
\newtheorem{definition}{Definition}[section]
\newtheorem{remark}{Remark}[section]
\newtheorem{example}{Example}[section]
\newtheorem{prop}{Proposition}[section]

\usepackage{tikz}
\hypersetup{
	colorlinks=true,
    linkcolor=red,
    citecolor=blue,
}

\newcommand{\bx}{\mathbf{x}}
\newcommand{\bp}{\mathbf{p}}
\newcommand{\bz}{\mathbf{z}}
\newcommand{\bs}{\mathbf{s}}
\newcommand{\be}{\mathbf{e}}
\newcommand{\db}[2]{\left\{\left\{#1,#2\right\}\right\}}
\newcommand{\m}[1]{\mathcal{#1}}
\newcommand{\sym}[2]{\left(#1\right) \odot \left(#2\right)}
\newcommand{\sympl}[2]{#1 \odot #2}
\newcommand{\symL}[2]{\left(#1\right) \odot #2}

\title{A Geometric Perspective on Kinetic Matter-Radiation Interaction and Moment Systems}
\author{Brian K. Tran$^\dagger$ and Joshua W. Burby$^\ddagger$ and Ben S. Southworth$^\dagger$}
\address[$\dagger$]{Los Alamos National Laboratory, Theoretical Division. Los Alamos, New Mexico 87545.}
\address[$\ddagger$]{University of Texas at Austin.
Austin, Texas, USA 78712.}
\email{btran@lanl.gov, joshua.burby@austin.utexas.edu, southworth@lanl.gov}

\allowdisplaybreaks

\begin{document}

\begin{abstract}
We provide a geometric perspective on the kinetic interaction of matter and radiation, based on a pair bracket approach. We discuss the interaction of kinetic theories via dissipative brackets, with our fundamental example being the coupling of matter, described by the Boltzmann equation, and radiation, described by the radiation transport equation. We explore the transition from kinetic systems to their corresponding moment systems, provide a Hamiltonian description of such moment systems, and give a geometric interpretation of the moment closure problem for kinetic theories. As an application, we discuss in detail diffusion radiation hydrodynamics as an example of a pair bracket formulation on a space of moments corresponding to kinetic matter-radiation interaction. Additionally, using the variable moment closure framework of \citet{Burby2023}, we show how to construct Hamiltonian moment closures for kinetic transport equations with arbitrary Hamiltonian. Using this general construction, we derive novel Hamiltonian moment closures for pure radiation transport.
\end{abstract}
\maketitle

{\hypersetup{linkcolor=black}\tableofcontents}

\flushbottom
\maketitle

\thispagestyle{empty}

\section{Introduction}\label{sec:intro}
We are interested in kinetic systems describing the interaction of matter and radiation. We approach the study of such systems from a geometric perspective, through Hamiltonian Lie--Poisson and dissipative bracket formulations. Despite matter and radiation being of fundamentally different nature, we will see that their associated kinetic descriptions have many common features when viewed from a geometric perspective. This will particularly manifest when we study the moments associated to kinetic systems, where we will see that moment spaces, moment evolution equations and moment closures can be described in a unified way through the Hamiltonian. 

There has been a significant amount of work on moment closures for kinetic systems (see, e.g., \cite{Grad1949, GaHa2013, CaiFan2013, CaiFan2014, All2019, Abdel2023, Ted2023, Lev1996, Tronci_geom_2008, Holm_Tronci_2009, Tassi_2014, BT_2012}) and recently data-driven moment closures \cite{sadr2021,huang2022,Por2023, Dona2023, huang2023, Burby2023, Li_2023}. An important feature of recent work on moment closures is constructing or learning moment closures which ensure energy conservation, entropy dissipation and hyperbolicity (see, e.g., \cite{All2019, CaiFan2014, huang2023}). We approach the moment closure problem from a geometric, particularly pair bracket, perspective; as a consequence, all such geometric moment closures are automatically energy conservative and entropy dissipative. Furthermore, hyperbolicity of such moment closures could be studied by looking at the dissipative-free Hamiltonian limit, utilizing techniques from Hamiltonian PDE \cite{hyperbolic1, hyperbolic2, hyperbolic3, hyperbolic4}. This geometric interpretation of the moment closure problem opens new avenues for deriving moment closures analytically; we provide two examples where we derive moment closures from thermodynamic principles and geometric principles. It also opens the possibility of approaching data-driven structure-preserving moment closures from the learning of metriplectic brackets \cite{gruber24}.

This paper is organized as follows. In \Cref{sec:kinetic-theory}, we give an overview of kinetic transport from the Hamiltonian perspective and introduce interaction through dissipative brackets. As examples, we discuss the kinetic description of matter in \Cref{sec:hamiltonian-fluid} and the kinetic description of radiation in \Cref{sec:hamiltonian-radiation}. In \Cref{sec:rad-hydro}, we consider the coupling of these two kinetic theories, and the coupling via interaction brackets in \Cref{sec:interaction-brackets}. In \Cref{sec:moment-equations}, we turn our attention to the moments associated to a kinetic system. Particularly, we characterize the moment kernels for a kinetic system in terms of its underlying Hamiltonian and discuss natural weighted variants of these moment kernels. In \Cref{sec:interaction-moment-spaces}, we consider pair bracket formulations on spaces of moments and consider diffusion radiation hydrodynamics as an example. Subsequently, in \Cref{sec:geometric-moment-closures}, we consider the moment closure problem and provide a geometric notion of a moment closure. As an example of such a geometric moment closure, in \Cref{sec:arbitrary-transport-closure}, we show how to construct Hamiltonian moment closures for transport equations with arbitrary Hamiltonian, using the variable moment closure framework of \cite{Burby2023}. Using this general construction, we present novel degree one and two closures for pure radiation transport in \Cref{example:pure-rad-closures}. Interestingly, these radiation closures take a distinctly different form than classical moment closures used in transport problems. 

\subsection{Kinetic theory}\label{sec:kinetic-theory}
Throughout, we let $Q$ be an open, bounded, and connected subset of $\mathbb{R}^n$. We use the natural identifications $TQ = Q \times \mathbb{R}^n$ and $T^*Q= Q \times \mathbb{R}^n$ for the tangent and cotangent bundles of $Q$, respectively. For $\mathbf{v} \in T_\bx Q = \mathbb{R}^n$ and $\bp \in T^*_\bx Q = \mathbb{R}^n$, the fiberwise duality pairing $T^*_\bx Q \times T_\bx Q\rightarrow \mathbb{R}$ is then simply the standard product on $\mathbb{R}^n$, $\bp \cdot \mathbf{v} = \bp^T \mathbf{v}.$

We briefly recall the transport of a phase space distribution $g(t,\bx,\bp)$ \cite{Kre2010} advected along the dynamics of a Hamiltonian system with coordinates $(\bx,\bp)$. Consider a particle at time $t$, with position $\bx(t) \in \Omega$ and momenta $\bp(t) \in T_{\bx(t)}^*Q$. Let $H: T^*Q \rightarrow \mathbb{R}$ be the particle Hamiltonian describing the energy of such a particle. Then, the dynamics of the particle are given by Hamilton's equations
\begin{subequations}\label{eq:Hamilton}
    \begin{align}
        \dot{\bx}(t) &= \nabla_{\bp}H(\bx,\bp), \\
        \dot{\bp}(t) &= -\nabla_{\bx}H(\bx,\bp),
    \end{align}
\end{subequations}
where $\dot{\bx} := d\bx/dt$ and $\dot{\bp} := d\bp/dt$. In the absence of interaction terms, the statement that $g$ is advected (or transported) along the dynamics of the Hamiltonian system is $\frac{d}{dt}g(t,\bx(t),\bp(t)) = 0$. Expanding this equation by the chain rule and using \Cref{eq:Hamilton} yields the expression
\begin{align*}
    0 &= \frac{d}{dt}g(t,\bx(t),\bp(t)) \\
      &= \frac{\partial}{\partial t} g +  \nabla_{\bx}g \cdot \dot{\bx} + \dot{\bp}\cdot\nabla_{\bp}g \\
      &= \frac{\partial}{\partial t} g + \nabla_{\bx}g\cdot\nabla_{\bp}H - \nabla_{\bx}H\cdot\nabla_{\bp}g,
\end{align*}
Thus, we see that the transport of $g$ can be expressed via the canonical Poisson structure $[\cdot,\cdot]$ on $T^*Q$, 
\begin{equation}\label{eq:transport-poisson}
    \frac{\partial}{\partial t}g + [g,H] = 0,
\end{equation}
where the canonical Poisson bracket \cite{RaMa1987,MaRa1999} is given by $[F,G] := \nabla_\bx F\cdot \nabla_\bp G - \nabla_\bx G\cdot\nabla_\bp F$. We refer to \eqref{eq:transport-poisson} as transport of $g$, since it can be expressed as a transport equation on $T^*Q$; namely, \eqref{eq:transport-poisson} can equivalently be expressed
\begin{equation}\label{eq:transport-poisson-canonical}
    \frac{\partial}{\partial t}g + \nabla_\bx \cdot (g\nabla_\bp H) - \nabla_\bp\cdot(g \nabla_\bx H) = 0.
\end{equation}

\textbf{Evolution of Functionals.}
Along with the evolution of the phase space distribution $g(t,\bx,\bp)$ itself, it is useful to describe the evolution of functionals of $g$; i.e., maps from the space of phase space distributions into $\mathbb{R}$ (see also \cite{manifoldstensor}). To be more precise, let $C(T^*Q) := C^\infty(T^*Q)$ be the space of smooth functions on $T^*Q$ and let $C^*(T^*Q) := C^\infty(T^*Q)'$ be its topological dual consisting of compactly supported distributions on $T^*Q$. The duality pairing for $g \in C^*(T^*Q)$ and $h \in C(T^*Q)$ can be \emph{formally} expressed as integration
$$\langle g, h \rangle = \int_{T^*Q} h(\bx,\bp) g(\bx,\bp) d\bp d\bx,$$
where $d\bp d\bx$ denotes the Louiville volume form on $T^*Q$. Functionals are then maps $C^*(T^*Q) \rightarrow \mathbb{R}$. For example, the Hamiltonian (or total energy) functional $\m{H}: C^*(T^*Q) \rightarrow \mathbb{R}$ given by
\begin{equation}\label{eq:hamiltonian-functional}
    \mathcal{H}[g] := \int_{T^*Q} H g d\bp d\bx,
\end{equation}
which is well-defined assuming the Hamiltonian $H \in C(T^*Q)$, and the total momentum vector-valued functional $\mathcal{P}: C^*(T^*Q) \rightarrow \mathbb{R}^n$ given by
\begin{equation}\label{eq:momentum-functional}
    \mathcal{P}[g] := \int_{T^*Q} \bp g d\bp d\bx,
\end{equation}
The evolution of a functional $\m{F}[g]$ can be defined in terms of a Lie--Poisson bracket \cite{MaRa1999}, $\db{\cdot}{\cdot}$, defined by
\begin{equation}\label{eq:lie-poisson-bracket}
    \db{\mathcal{F}}{\mathcal{G}}[g] := \int_{T^*Q} g \left[ \frac{\delta\mathcal{F}}{\delta g}, \frac{\delta \mathcal{G}}{\delta g} \right] d\bp d\bx,
\end{equation}
where $\delta\m{F}/\delta g \in C(T^*Q)$ denotes the functional derivative of $\m{F}$ with respect to $g$, if it exists, defined by
\begin{equation}\label{eq:functional-diff}
    \frac{d}{d\epsilon}\Big|_{\epsilon = 0} \m{F}[g + \epsilon v] = \int_{T^*Q} \frac{\delta \m{F}[g]}{\delta g} v\, d\bp d\bx = \left\langle v, \frac{\delta \m{F}}{\delta g}\right\rangle,
\end{equation}
for all $v \in C^*(T^*Q)$ (see, e.g., \cite{Dreizler2011}). We will assume that all functionals are differentiable in this sense. For example, this is true for all linear functionals of the form
$$ \m{F}[g] = \int_{T^*Q} k(\bx,\bp) g(\bx,\bp) d\bp d\bx,$$
provided $k \in C(T^*Q)$. In this case, $\delta\m{F}/\delta g = k$.

The evolution of a functional $\m{F}[g]$, where $g$ is now a time-dependent phase space distribution, i.e., $g(t) \in C^*(T^*Q)$, such that $g$ evolves under \Cref{eq:transport-poisson}, can be computed by the chain rule for functional derivatives,
\begin{align*} \frac{\partial}{\partial t} \m{F}[g] &= \int_{T^*Q} \frac{\delta \m{F}}{\delta g} \frac{\partial g}{\partial t} d\bp d\bx =  -\int_{T^*Q} \frac{\delta \m{F}}{\delta g} [g, H] d\bp d\bx 
\\ &= -\int_{T^*Q} \frac{\delta \m{F}}{\delta g} \left[g, \frac{\delta \m{H}}{\delta g} \right] d\bp d\bx, 
\end{align*}
which, after integration by parts, is equivalently expressed in terms of the Lie--Poisson bracket of $\m{F}$ with the energy functional $\m{H}$,
\begin{equation}\label{eq:lie-poisson-evolution}
    \frac{\partial}{\partial t} \mathcal{F}[g] = \db{\mathcal{F}}{\mathcal{H}}[g].
\end{equation}
For a more in-depth discussion of the Lie--Poisson bracket in kinetic theory, see \cite{pavelka2016, manifoldstensor}. 

As an example, since we will perform similar calculations throughout, let us compute the evolution equation for a linear functional of $g$. To do so, consider the linear functional
$$ \m{F}_\varphi [g] := \int_{T^*Q} \varphi(\bx,\bp) g(\bx,\bp) d\bp d\bx,$$
where $\varphi$ is a fixed but arbitrary element of $C(T^*Q)$. A direct calculation by \eqref{eq:functional-diff} yields
$$ \frac{\delta \m{F}_\varphi}{\delta g} = \varphi. $$

Now, we can compute the evolution of $\m{F}_\varphi[g]$, 
\begin{align*}
    \frac{\partial}{\partial t} \m{F}_\varphi [g] = -\int_{T^*Q} \frac{\delta \m{F}_\varphi}{\delta g} \left[g, \frac{\delta \m{H}}{\delta g} \right] d\bp d\bx = -\int_{T^*Q} \varphi [g, H] d\bp d\bx.
\end{align*}
Noting that the left hand side is $\frac{\partial}{\partial t}\m{F}_\varphi [g] = \int_{T^*Q} \varphi(\bx,\bp) \frac{\partial}{\partial t} g(t,\bx,\bp) d\bp d\bx$, we have
$$ \int_{T^*Q} \varphi \frac{\partial}{\partial t} g d\bp d\bx = -\int_{T^*Q} \varphi [g, H] d\bp d\bx. $$
Since $\varphi \in C(T^*Q)$ is arbitrary, this yields $\partial g/\partial t = -[g,H]$ which is of course \eqref{eq:transport-poisson}. 

\textbf{Dissipative Bracket.}
One method to introduce interaction into the transport equation is by a dissipative bracket \cite{Kaufman1984, Morrison1984, Dukek1997}. The evolution of a functional including interactions is expressed
\begin{align}\label{eq:dissipative-bracket}
    \frac{\partial}{\partial t} \mathcal{F} = \db{\mathcal{F}}{\mathcal{H}} + (\mathcal{F},\mathcal{S}),
\end{align}
where $\db{\cdot}{\cdot}$ is the Lie--Poisson bracket and we have introduced the entropy functional $\mathcal{S}[g]$ and the \textit{dissipative bracket} $(\cdot,\cdot)$ which is a symmetric positive semi-definitive bilinear form. If the entropy functional is invariant under the Lie--Poisson evolution by $\m{H}$, i.e., $\db{\mathcal{S}}{\mathcal{H}}=0$, then the interactions through the dissipative bracket generate entropy:
$$ \frac{\partial}{\partial t} \mathcal{S} =  (\mathcal{S},\mathcal{S}) \geq 0 $$
(note that in the literature, this inequality is usually referred to as entropy ``dissipation", where, depending on convention, the inequality $\leq 0$ may be used instead). Furthermore, if $(\mathcal{H},\mathcal{S}) = 0$, then the interactions are energy conservative, 
$$ \frac{\partial}{\partial t}\mathcal{H} = \db{\mathcal{H}}{\mathcal{H}} + (\mathcal{H},\mathcal{S}) = 0, $$
where we used that $\db{\mathcal{H}}{\mathcal{H}}=0$ by antisymmetry of the Lie--Poisson bracket. Note that one could relax this condition and not require that the dissipative bracket is energy-conservative, i.e., $(\mathcal{H},\mathcal{S})$ may be non-zero, as this allows one to consider the system interacting with an energetic environment to which energy may be transferred. We will not consider this case here. 

To summarize, we say that a system admits a \textit{pair bracket formulation} if the equations of motion are given by \eqref{eq:dissipative-bracket}, where $\db{\cdot}{\cdot}$ is an antisymmetric bracket, $(\cdot,\cdot)$ is a symmetric positive semi-definite bracket, with an energy functional $\mathcal{H}$ and an entropy functional $\mathcal{S}$ such that $\{\m{S},\m{H}\}=0$ and $(\m{H},\m{S})=0$.

A stronger notion of a pair bracket formulation is a \textit{metriplectic formulation} \cite{Morrison_met_1986}, where the entropy and energy functionals satisfy the stronger requirement that they are Casimirs of the antisymmetric bracket and dissipative bracket, respectively, i.e., $\db{\m{F}}{\m{S}}=0$ and  $(\m{F},\m{H})=0$ for all functionals $\m{F}$. For our purposes, the weaker pair bracket formulation suffices.

In the context of kinetic theory, the dissipative term $(\mathcal{F},\mathcal{S})$ in \Cref{eq:dissipative-bracket} can be interpreted as a collision operator \cite{Kre2010, Lev1996}. There is freedom in choosing such collision operators, or dissipative terms, to describe different interactions in kinetic theories, but they are usually constrained to maintain mass, momentum and energy conservation, entropy dissipation \cite{Lev1996} and possibly other physical symmetries.

In the next sections \ref{sec:hamiltonian-fluid} and \ref{sec:hamiltonian-radiation}, we will recall facts about the Hamiltonian formulations of kinetic matter and radiation. Note that the kinetic equations for matter and radiation are governed by the same equation $\dot{g} + [g,H] = 0$; however, the differences between the two systems lie in the Hamiltonian $H$ as well as the interpretation of the phase space distribution $g$. As we will see, the matter Hamiltonian is homogeneous of degree two in $\bp$ whereas the radiation Hamiltonian is homogeneous of degree one in $\bp$ (for a discussion of properties of radiation transport that arise from this homogeneity of degree one, see \cite{phasespacelifts}). Subsequently, in \Cref{sec:rad-hydro}, we will discuss a geometric approach to the coupling between the kinetic matter and radiation and their corresponding moments.

\subsubsection{Hamiltonian formulation of kinetic matter}\label{sec:hamiltonian-fluid}
We consider a distribution of massive particles, say electrons, in phase space $T^*Q$, described by a distribution function $f(t,\bx,\bp)$, where the dynamics of the particles are governed by the particle Hamiltonian
    $$ H_e(\bx,\bp) = \frac{\|\bp\|^2}{2m} + V(\bx), $$
where $V: Q \rightarrow \mathbb{R}$ is a potential function. By \Cref{eq:transport-poisson}, the transport equation for a distribution $f(t,\bx,\bp)$ is
    \begin{equation}\label{eq:transport-fluid}
        \frac{\partial}{\partial t}f(t,\bx,\bp) + \frac{\bp}{m} \cdot \nabla_{\bx} f(t,\bx,\bp) - \nabla_{\bx} V(\bx) \cdot \nabla_{\bp} f(t,\bx,\bp) = 0, 
    \end{equation}
 which is the collisionless Boltzmann equation \cite{Kre2010}. More generally, the right hand side of \Cref{eq:transport-fluid} may be non-zero due to collision or interaction terms; we will introduce such terms when coupling this system to the radiation transport equation.

 The Hamiltonian functional associated with the system \eqref{eq:transport-fluid} is
 \begin{equation}\label{eq:electron-hamiltonian-functional}
     \m{H}_e[f] = \int_{T^*Q} f(t,\bx,\bp) H_e(\bx,\bp) d\bp d\bx = \int_{T^*Q} f(t,\bx,\bp) \left(\frac{\|\bp\|^2}{2m} + V(\bx)\right) d\bp d\bx.
 \end{equation}
Denoting the Lie--Poisson bracket as
\begin{equation}\label{eq:electron-lie-poisson-bracket}
    \db{\m{F}}{\m{G}}_e = \int_{T^*Q} f \left[ \frac{\delta\mathcal{F}}{\delta f}, \frac{\delta \mathcal{G}}{\delta f} \right] d\bp d\bx,
\end{equation}
the evolution of a functional $\m{F}[f]$ is given by
\begin{equation}\label{eq:electron-evolution}
    \frac{\partial}{\partial t} \m{F}[f] = \db{\m{F}}{\m{H}_e}_e . 
\end{equation}

As an example, we compute the evolution of the spatial mass, momentum and energy densities, given by the following fiber integrals.

\begin{example}[Fluid Moments]\label{ex:fluid-moments}
Consider the spatial mass, momentum and energy densities, defined as
\begin{align*}
    \rho_e(t,\bx) &:= \int_{T^*_\bx Q} m f(t,\bx,\bp) d\bp, \\
    \mathbf{F}_e(t,\bx) &:= \int_{T^*_\bx Q} \bp f(t,\bx,\bp) d\bp,\\
    \mathcal{E}_e(t,\bx) &:= \int_{T^*_\bx Q} H_e(\bx,\bp) f(t,\bx,\bp) d\bp.
\end{align*}
Note that these fluid moments depend on $f$ and we have suppressed this notation as is standard. 

Analogous to the calculation performed in \Cref{sec:kinetic-theory}, let us compute the equations of motion for these fluid moments. Consider a functional of $f$ of the form
$$ \m{F}[f] = \widetilde{\m{F}}[\rho_e, \mathbf{F}_e, \mathbf{\m{E}_e}], $$
i.e., the functional $\m{F}$ only depends on $f$ through the moments $(\rho_e, \mathbf{F}_e, \mathbf{\m{E}_e})$. Then,
$$ \frac{\delta \m{F}}{\delta f} = m \frac{\delta \widetilde{\m{F}}}{\delta \rho_e} + \bp \cdot \frac{\delta \widetilde{\m{F}}}{\delta \mathbf{F}_e} + H_e \frac{\delta \widetilde{\m{F}}}{\delta \m{E}_e}. $$
In particular, consider the linear functional of the moments given by
\begin{align*}
    \mathcal{F}_{a\mathbf{b}c}[f] &= \widetilde{\m{F}}_{a\mathbf{b}c}[\rho_e, \mathbf{F}_e, \mathbf{\m{E}_e}] \\
    &:= \int_Q a(\bx) \rho_e(t,\bx) d\bx + \int_Q \mathbf{b}(\bx) \cdot \mathbf{F}_e(t,\bx) d\bx + \int_Q c(\bx) \m{E}_e(t,\bx) d\bx,
\end{align*}
where $a$, the components of $\mathbf{b}$, and $c$ are fixed but arbitrary elements of $C(Q)$, where $C(Q)$ denotes the space of smooth functions on $Q$ with mild growth, analogous to the definition of $C(T^*Q)$ in \Cref{sec:kinetic-theory}. The functional derivative of this linear functional is
$$ \frac{\delta \m{F}_{a\mathbf{b}c}}{\delta f} = m a + \mathbf{p}\cdot \mathbf{b} + H_e c. $$
The evolution of this functional is then
\begin{align*}
    \frac{\partial \m{F}_{a\mathbf{b}c}}{\partial t} &= \db{\m{F}_{a\mathbf{b}c}}{\m{H}_e}_e = - \int_{T^*Q} \frac{\delta \m{F}_{a\mathbf{b}c}}{\delta f} [f, H_e] d\bp d\bx \\
    &= - \int_{T^*Q} m a [f, H_e]  d\bp d\bx - \int_{T^*Q} \bp\cdot\mathbf{b} [f, H_e]  d\bp d\bx - \int_{T^*Q} H_e c [f,H_e]  d\bp d\bx \\
    &= - m \int_{T^*Q} a \left( \nabla_{\bx} f \cdot \frac{\bp}{m} \right) d\bp d\bx  \\
        & \qquad - \int_{T^*Q} \mathbf{b} \cdot \left[\left( \nabla_{\bx} ( f \bp ) \right) \cdot \frac{\bp}{m} -  f \nabla_\bx V \right] d\bp d\bx \\
        & \qquad - \int_{T^*Q} c \nabla_\bx \left(\frac{\bp}{m} f H_e \right) d\bp d\bx \\
    &= - \int_{Q} a \nabla_{\bx} \cdot \mathbf{F}_e d\bx  \\
        & \qquad - \int_{Q} \mathbf{b} \cdot \left[ -m^{-1} \nabla_\bx \mathbb{S}_e - m^{-1} \rho \nabla_\bx V \right] d\bx \\
        & \qquad - \int_{Q} c \nabla_\bx \cdot \mathbf{G}_e d\bx,
\end{align*}
where we introduced the stress tensor
$$ \mathbb{S}_e(t,\bx) := \int_{T^*_\bx Q} \bp \otimes \bp f(t,\bx,\bp) d\bp d\bx $$
and the energy flux
$$ \mathbf{G}_e(t,\bx) := \int_{T^*_\bx Q} \frac{\bp}{m} f(t,\bx,\bp) H_e(\bx,\bp) d\bp. $$
Noting that the left hand side of the evolution equation is equivalently
$$ \frac{\partial \m{F}_{a\mathbf{b}c}}{\partial t} = \int_Q a \frac{\partial}{\partial t}\rho_e d\bx + \int_Q \mathbf{b} \cdot \frac{\partial}{\partial t}\mathbf{F}_e d\bx + \int_Q c \frac{\partial}{\partial t}\m{E}_e d\bx,$$
by comparing both sides and noting that $a$, $\mathbf{b}$, and $c$ are arbitrary, we arrive at the evolution equations for the moments
\begin{align*}
    \frac{\partial}{\partial t} \rho_e &= - \nabla_\bx \cdot \mathbf{F}_e, \\
    \frac{\partial}{\partial t} \mathbf{F}_e &= -m^{-1} \nabla_\bx \mathbb{S}_e - m^{-1} \rho \nabla_\bx V, \\
    \frac{\partial}{\partial t} \m{E}_e &= - \nabla_\bx \cdot \mathbf{G}_e.
\end{align*}
All three moment evolution equations are of course local conservation laws for mass, momentum and energy, respectively (note that for the local momentum conservation law, there is an additional term correpsonding to the potential $V$). Furthermore, each evolution equation involves the divergence of a tensor of one higher rank and one higher degree in $\bp$; we will discuss this ``moment closure problem" in more detail in \Cref{sec:moment-equations}.

\end{example}

\subsubsection{Hamiltonian formulation of radiation transport}\label{sec:hamiltonian-radiation}
Radiation transport describes the asymptotic short-wavelength limit of Maxwell's equations (see, e.g., \cite{phasespacelifts, wigner}); the lift of Maxwell's equations to phase space $T^*Q$ in this asymptotic limit is necessary to properly describe quadratic observables \cite{wigner}. The spatial domain is again $Q$. The fundamental quantity in radiative transfer is the specific intensity $I$ which depends on time $t$, position $\bx$, frequency $\nu \in \mathbb{R}_+$, and direction $\mathbf{\Omega} \in \mathbb{S}^{n-1}$ \cite{castor_2004, mihalas_1984, Chan1960}.To get a precise definition of the quantities in radiative transfer, we first consider the radiation transport equation in vacuum, which takes the form
\begin{equation}\label{eq:rad-transport-vacuum}
    \frac{1}{c} \frac{\partial}{\partial t} I(t,\bx,\nu,\mathbf{\Omega}) + \mathbf{\Omega} \cdot \nabla_\bx I(t,\bx,\nu,\mathbf{\Omega}) = 0,
\end{equation}
where $c>0$ is the speed of light.

To provide a geometric description, it will be useful to think of the frequency $\nu$ and direction $\mathbf{\Omega}$ as constituting a momentum vector. Thus, for radiation, we define the momentum as 
$$ \bp := \frac{\nu}{c}\Omega. $$
Assuming the momentum is non-zero, we have the coordinate transformations
\begin{subequations}\label{eq:coordinate-transform}
\begin{align}
    \mathbf{\Omega} &= \frac{\bp}{\|\bp\|},\\
    \nu &= \|\bp\|c.
\end{align}
\end{subequations}
The collection of all position and momenta form the cotangent bundle $T^*Q$; coordinatized by canonical coordinates $(\bx,\mathbf{p}) \in T^*Q$. We additionally coordinatize the cotangent bundle via $(\bx,\nu,\mathbf{\Omega}) \in Q \times \mathbb{R}_+ \times \mathbb{S}^{n-1}$ which we refer to as frequency-angular coordinates on $T^*Q$. We will mostly utilize the former set of coordinates as they are the canonical coordinates on the cotangent bundle, but include the frequency-angular coordinates here as they are more common coordinates in the radiation transport literature \cite{castor_2004,mihalas_1984,Chan1960}.

In radiation transport, we are interested in quantities integrated over the frequency and angular variables, i.e., over the momentum variable. Recalling that radiation transport is the asymptotic short-wavelength limit of Maxwell's equations, such integrals over the momentum variable relate the phase space description of radiation transport back to the observables of classical electromagnetism. Furthermore, when coupling radiation transport to, for example, the evolution of a fluid, the coupling occurs through such integrals. Such integration can be considered as fiber integrals with respect to the bundle $T^*Q \rightarrow Q$. Since we have two sets of coordinates, we can perform such fiber integrals with respect to either coordinate system. Fiber integrals with respect to momentum coordinates are more natural from a geometric perspective; however, fiber integrals with respect to frequency-angular coordinates are more common in the radiation transport literature, so we include it here as well for completeness. The fiber integral of an integrable function $S:T^*Q \rightarrow \mathbb{R}$ over $T_x^*Q$ is expressed
\begin{align*}
    \int_{T_\bx^*Q} S(\bx,\bp) d\bp.
\end{align*}
Note that, although the integration domain $T_\bx^*Q = \mathbb{R}^n$, we use the notation $T_\bx^*Q$ throughout to emphasize the fact that it is a fiber integral over momenta. Via the coordinate transformation \eqref{eq:coordinate-transform}, the fiber integral is equivalently
$$ \int_{T_\bx^*Q} S(\bx,\bp) d\bp  = \int_{\mathbb{S}^{n-1}}\int_{\mathbb{R}_+} S(\bx,\nu,\mathbf{\Omega}) J(\nu) d\nu d\mathbf{\Omega}, $$
where $S(\bx,\nu,\mathbf{\Omega})$ is the expression for $S(\bx,\bp)$ under the coordinate transformation \eqref{eq:coordinate-transform} and $J(\nu)$ denotes the Jacobian associated to the coordinate transformation \eqref{eq:coordinate-transform}, e.g., for $n=3$, $J(\nu) = \nu^2/c^3$. In the radiation transport literature, scalar densities of the form $\widetilde{S}(\bx,\nu,\mathbf{\Omega}) := S(\bx,\nu,\mathbf{\Omega}) J(\nu)$ with respect to the measure $d\nu d\mathbf{\Omega}$ are often used due to their physical interpretation with respect to frequency and angle; however, in this work, we prefer to use scalar densities $S(\bx,\bp)$ with respect to the measure $d\bp$ as $(\bx,\bp)$ are canonical coordinates on $T^*Q$.

Now, we discuss the Lie--Poisson formulation of radiation transport.

\textbf{Lie--Poisson structure of radiation transport.} The Hamiltonian of corresponding to radiation transport is \cite{phasespacelifts}
$$ H_r(\bx,\bp) = \|\bp\|c, $$
or, equivalently, in frequency-angular coordinates,
$$ H_r(\bx,\nu,\mathbf{\Omega}) = \nu. $$
Note that the above Hamiltonian is independent of $\bx$.

As we will see, for a geometric description of radiation transport, it is useful to consider a phase space distribution describing the density of states, as opposed to the specific intensity; to this end, we introduce the electromagnetic phase space density $\Psi(t,\bx,\bp)$. From \Cref{sec:kinetic-theory}, the transport of the photon number density (assuming $\|\bp\| \neq 0$) can be expressed in the Poisson formulation as
\begin{align*}
    0 &= \frac{\partial}{\partial t}\Psi + [\Psi, H_r ] = \frac{\partial}{\partial t}\Psi + \nabla_\bx \Psi \cdot \nabla_\bp H_r - \nabla_\bx H_r \cdot \nabla_\bp \Psi \\
      &= \frac{\partial \Psi}{\partial t} + \frac{\bp c}{\|\bp\|} \cdot \nabla_\bx \Psi.
\end{align*}
In frequency-angular coordinates, this can be equivalently expressed
\begin{align*}
    \frac{1}{c}\frac{\partial}{\partial t} \Psi(t,\bx,\nu,\mathbf{\Omega}) + \mathbf{\Omega} \cdot \nabla_\bx \Psi(t,\bx,\nu,\mathbf{\Omega}) = 0.
\end{align*}
As mentioned previously, the typical quantity in the radiation transport literature is the specific intensity $I(t,\bx,\nu,\mathbf{\Omega})$. This is related to $\Psi$ via \cite{mihalas_1984}
$$I(t,\bx,\nu,\mathbf{\Omega}) = ch\nu \Psi(t,\bx,\nu,\mathbf{\Omega}) = c H_r(\bx,\nu,\mathbf{\Omega}) \Psi(t,\bx,\nu,\mathbf{\Omega}),$$
or, equivalently,
$$I(t,\bx,\bp) = c H_r(\bx,\bp) \Psi(t,\bx,\bp).$$
Substituting this into the previous equation, or, alternatively, computing $\partial I/\partial t + [I,H_r] = 0$, reproduces the radiation transport equation, \Cref{eq:rad-transport-vacuum}. We have thus expressed the transport of both $I$ and $\Psi$ in terms of a Poisson structure. Although using the specific intensity $I$ or the electromagnetic phase space density $\Psi$ are equivalent, we state both since $I$ is more commonly used in the radiation transport literature, whereas we use $\Psi$ since it describes the density of states in phase space and hence, will give a natural expression for the Hamiltonian functional in terms of the Hamiltonian (see \Cref{eq:radiation-hamiltonian-functional} below).

The radiation transport equation can also be expressed via a Lie--Poisson structure. To do so, we will need to introduce the Hamiltonian functional for radiation transport, which is the total energy of the system. Intuitively, the total energy of the system is given by the electromagnetic phase space density $\Psi$ times the energy $H_r$, summed over all states, i.e.,
\begin{align}\label{eq:radiation-hamiltonian-functional}
    \mathcal{H}_r[\Psi] := \int_{T^*Q} \Psi(t,\bx,\bp) H_r(\bx,\bp) d\bp d\bx,
\end{align}
Denoting the corresponding Lie--Poisson bracket as
\begin{equation}\label{eq:radiation-lie-poisson-bracket}
    \db{\mathcal{F}}{\mathcal{G}}_r[\Psi] := \int_{T^*Q} \Psi \left[ \frac{\delta \mathcal{F}}{\delta \Psi}, \frac{\delta \mathcal{G}}{\delta \Psi} \right] d\bp d\bx, 
\end{equation}
the evolution of a functional $\m{F}[\Psi]$ is then given by
\begin{equation}\label{eq:radiation-evolution}
    \frac{\partial}{\partial t} \m{F} = \db{\m{F}}{\m{H}_r}_r. 
\end{equation}
As an example, we compute evolution equations for the electromagnetic energy density and the Poynting vector.
\begin{example}[Radiation Moments]\label{ex:radiation-moments}
    Consider the electromagnetic energy density and Poynting vector for radiation, respectively,
    \begin{align*}
         E_r(t,\bx) &:= \int_{T^*_\bx Q} H_r(\bx,\bp)\Psi(t,\bx,\bp) d\bp, \\
         \mathbf{F}_r(t,\bx) &:= \frac{1}{c} \int_{T^*_\bx Q} \mathbf{\Omega} H_r \Psi d\bp = \int_{T^*_\bx Q} \bp \Psi d\bp.
    \end{align*}
    The evolution equations for these spatial densities can be computed analogously to \Cref{ex:fluid-moments}. Introduce the linear functional
    \begin{align*}
        \m{F}_{a\mathbf{b}}[\Psi] &= \widetilde{\m{F}}_{a\mathbf{b}}[E_r, \mathbf{F}_r] \\
        &:= \int_Q a(\bx) E_r(t,\bx) d\bx + \int_Q \mathbf{b}(\bx) \cdot \mathbf{F}_r(t,\bx) d\bx,
    \end{align*}
    where again $a$ and the components of $\mathbf{b}$ are fixed but arbitrary elements of $C(Q)$. The functional derivative of this linear functional is
    \begin{align*}
        \frac{\delta \m{F}_{a\mathbf{b}}}{\delta \Psi} &= H_r \frac{\delta \widetilde{\m{F}}_{a\mathbf{b}}}{\delta E_r} + \frac{1}{c} H_r\mathbf{\Omega} \cdot \frac{\delta \widetilde{\m{F}}_{a\mathbf{b}}}{\delta \mathbf{F}_r} = H_r a + \frac{1}{c} H_r \mathbf{\Omega} \cdot b.
    \end{align*}
    The evolution of this functional is then
    \begin{align*}
        \frac{\partial \m{F}_{a\mathbf{b}}}{\partial t} &= \db{\m{F}_{a\mathbf{b}}}{\m{H}_r}_r = - \int_{T^*Q} \frac{\delta \m{F}_{a\mathbf{b}}}{\delta \Psi} [\Psi, H_r] d\bp d\bx \\
        &= - \int_{T^*Q} H_r a [\Psi, H_r] d\bp d\bx - \int_{T^*Q} \frac{1}{c} H_r \mathbf{\Omega} \cdot b [\Psi, H_r] d\bp d\bx \\
        &= - \int_{T^*Q} a \left( \nabla_\bx (H_r \Psi ) \cdot \frac{\bp}{\|\bp\|} \right) d\bp d\bx - \int_{T^*Q} \mathbf{b} \cdot \left( \nabla_\bx (H_r \mathbf{\Omega} \Psi) \cdot \mathbf{\Omega} \right) d\bp d\bx \\
        &= - \int_{Q} a (\nabla_\bx \cdot c^2 \mathbf{F}_r) d\bx - \int_Q \mathbf{b} \cdot (\nabla_\bx \cdot \mathbb{P}) d\bx,
    \end{align*}
    where we introduced the radiation momentum flux (pressure) tensor
    $$ \mathbb{P}_r := \int_{T^*_\bx Q} \mathbf{\Omega} \otimes \mathbf{\Omega} H_r \Psi d\bp. $$
    By an analogous argument to \Cref{ex:fluid-moments}, this yields the evolution equations for the radiation moments,
    \begin{align*}
        \frac{\partial}{\partial t} E_r &= - \nabla_\bx \cdot c^2 \mathbf{F}_r, \\
        \frac{\partial}{\partial t} \mathbf{F}_r &= - \nabla_\bx \cdot \mathbb{P}.
    \end{align*}
    Note that each moment evolution equation involves the divergence of a tensor of one higher rank and one higher degree in $\mathbf{\Omega} = \bp/\|\bp\|$. 
\end{example}

\begin{remark}\label{rmk:moment-comparison}
    It is interesting to compare this example to \Cref{ex:fluid-moments}, where each evolution equation similarly involved the divergence of a tensor of one higher rank but one higher degree in $\mathbf{p}$ instead of $\mathbf{\Omega} = \bp/\|\bp\|$. This can be understood from the fact the Hamiltonian for an electron $H_e(\bx,\bp) = \|\bp\|^2/2m + V(\bx)$ and the radiation Hamiltonian $H_r(\bx,\bp) = \|\bp\|c$ involve different powers of $\|\bp\|$; i.e., the kinetic term in $H_e$ is homogeneous of degree $2$ in $\bp$ whereas $H_r$ is homogeneous of degree $1$ in $\bp$. This manifested in the calculations of this example as well as \Cref{ex:fluid-moments} through the Poisson bracket which involves derivatives of the respective Hamiltonian with respect to $\bp$, 
    \begin{align*}
        \nabla_\bp H_r &= c\, \mathbf{\Omega}, \\
        \nabla_\bp H_e &= \frac{1}{m} \bp.
    \end{align*}
    We discuss this in more detail in \Cref{sec:moment-equations}.
\end{remark}

\section{Geometric Formulation of Kinetic Matter-Radiation Interactions}\label{sec:rad-hydro}

In this section, we will describe the interaction of the kinetic matter and radiation through a pair bracket formulation. Subsequently, in \Cref{sec:moment-equations}, we turn our attention to the moment systems that arise from kinetic systems and give a geometric notion of a moment closure. 

To begin, we first provide a Lie--Poisson formulation of the kinetic matter and radiation equations in the non-interacting limit where the evolution of $f$ and $\Psi$ are decoupled. 

We define the bracket $\db{\cdot}{\cdot}_0$ on the space of functionals of both the electron distribution function $f$ and the radiation distribution function $\Psi$ to be the sum of $\db{\cdot}{\cdot}_e$ and $\db{\cdot}{\cdot}_r$, i.e., for two functionals $\m{F}[f;\Psi]$ and $\m{G}[f;\Psi]$,
\begin{equation}\label{eq:non-interacting-lie-poisson-bracket}
    \db{\m{F}}{\m{G}}_0 [f; \Psi] := \int_{T^*Q} f \left[ \frac{\delta\mathcal{F}}{\delta f}, \frac{\delta \mathcal{G}}{\delta f} \right] d\bp d\bx + \int_{T^*Q} \Psi \left[ \frac{\delta\mathcal{F}}{\delta \Psi}, \frac{\delta \mathcal{G}}{\delta \Psi} \right] d\bp d\bx.
\end{equation}
The total Hamiltonian functional, corresponding to the total energy of the system, is given by the sum of the electron and radiation Hamiltonian functionals, i.e.,
\begin{equation}\label{eq:total-hamiltonian-functional}
    \m{H}_{\text{tot}} [f; \Psi] := \m{H}_e[f] + \m{H}_r[\Psi].
\end{equation}
The evolution of a functional $\m{F}[f;\Psi]$ is then given by 
\begin{equation}\label{eq:non-interacting-evolution}
    \frac{\partial}{\partial t} \m{F}[f;\psi] = \db{\m{F}}{\m{H}_{\text{tot}}}_0.
\end{equation}
This bracket describes the non-interacting limit, since for a functional $\m{F}[f]$ of $f$ only and similarly for a functional $\m{G}[\Psi]$ of $\Psi$ only, \eqref{eq:non-interacting-evolution} yields
\begin{align*}
    0 &= \frac{\partial}{\partial t} \m{F}[f] - \db{\m{F}[f]}{\m{H}_{\text{tot}}[f;\Psi]}_0 = \frac{\partial}{\partial t} \m{F}[f] - \db{\m{F}[f]}{\m{H}_e[f]}_e, \\
    0 &= \frac{\partial}{\partial t} \m{G}[\Psi] - \db{\m{G}[\Psi]}{\m{H}_{\text{tot}}[f;\Psi]}_0 = \frac{\partial}{\partial t} \m{G}[\Psi] - \db{\m{G}[\Psi]}{\m{H}_r[\Psi]}_r,
\end{align*}
using the fact that the total Hamiltonian functional decomposes into the sum of the electron and radiation Hamiltonian functionals, \Cref{eq:total-hamiltonian-functional} and the formula \eqref{eq:non-interacting-lie-poisson-bracket}.

\subsection{Interaction brackets}\label{sec:interaction-brackets}
A general procedure to define an interaction between two kinetic systems is by introducing a dissipative bracket $(\cdot,\cdot)_{\text{tot}}$ with an entropy functional $\m{S}_{\text{tot}}[f;\Psi]$ and additionally, one may introduce a perturbation to the Lie--Poisson bracket, $\db{\cdot}{\cdot}_{\text{tot}} = \db{\cdot}{\cdot}_0 + \db{\cdot}{\cdot}_{\text{int}}$, where $\db{\cdot}{\cdot}_{\text{int}}$ is antisymmetric and its corresponding Hamiltonian flow preserves the entropy functional $\db{\m{S}_{\text{tot}}}{\m{H}_{\text{tot}}}_{\text{int}}=0$.

Given such data, the pair bracket structure of the dynamics is given by
\begin{equation}\label{eq:double-bracket-dynamics}
    \frac{\partial}{\partial t} \m{F} = \db{\m{F}}{\m{H}_{\text{tot}}}_{\text{tot}} + (\m{F},\m{S}_{\text{tot}})_{\text{tot}},
\end{equation}
where $\mathcal{F}$ is an arbitrary functional of $f$ and $\Psi$. We refer to \Cref{eq:double-bracket-dynamics} as the kinetic matter-radiation system, which depends on the choice of interaction brackets.

We will focus on the dissipative bracket $(\cdot,\cdot)_{\text{tot}}$ to account for interactions. A general procedure for constructing dissipative brackets is presented in \cite{Kaufman1984}. Following the ideas in \cite{Kaufman1984}, a fairly general form for a dissipative bracket can be expressed
\begin{align}\label{eq:general-dissipative-bracket}
    (\m{F},\m{G})_{\text{tot}} = \int_{T^*Q} \cdots \int_{T^*Q} \int_{T^*Q} \gamma \Delta_{1\dots k}(\m{F}) \Delta_{1 \dots k}(\m{G}) d\bp d\bx d\bx^{\{1\}} d\bp^{\{1\}} \cdots d\bx^{\{k\}} d\bp^{\{k\}};
\end{align}
here, $\Delta_{1\dots k}$ is a differential operator containing functional derivatives with respect to the distributions (in our case, $f$ and $\Psi$) at $(t,\bx^{\{j\}},\bp^{\{j\}})$, $j=1,\dots,k$ and $\gamma$ is a \textit{scattering kernel}, which is a non-negative real-valued function, depending generally on $(\bx,\bp)$, $(\bx^{\{j\}},\bp^{\{j\}}), j=1,\dots,k$, as well as the distributions. Note that the bracket defined by \Cref{eq:general-dissipative-bracket} is clearly symmetric; furthermore, the requirement that $\gamma \geq 0$ ensures that the bracket is indeed dissipative $(\m{S}_{\text{tot}}, \m{S}_{\text{tot}})_{\text{tot}} \geq 0$. Intuitively, adding multiple phase space integrals in \Cref{eq:general-dissipative-bracket} allows one to account for non-local or scattering terms between the distribution at $(t,\bx,\bp)$ and the distribution at $(t,\bx^{\{j\}},\bp^{\{j\}})$ as $(\bx^{\{j\}},\bp^{\{j\}})$ varies over phase space. The number of such additional phase space integrals will depend on the desired form for the interaction, as we will see in the example below. The nature of the interaction can be determined by defining the differential operator $\Delta_{1\dots k}$ as well as the scattering kernel $\gamma$. For example, if one wants to consider only spatially local interactions, the scattering kernel should contain a factor $\delta(\bx-\bx^{\{1\}})\cdots \delta (\bx-\bx^{\{2\}})$. One may wish to enforce further properties on the dissipative bracket, such as energy conservation and momentum conservation. For example, this can be done by choosing the differential operator such that it is momentum conservative, $\Delta_{1\dots k}(\m{P}_{\text{tot}}) = 0$, where $\m{P}_{\text{tot}}$ is the sum of the electron and radiation momentum functionals, and including a factor of $\delta(\Delta_{1\dots k}(\m{H}_{\text{tot}}))$ in the scattering kernel to ensure energy conservation \cite{Kaufman1984}.

\begin{example}\label{ex:dissipative-bracket}
As an example with just radiation, consider the radiation transport equation for the specific intensity $I$ with scattering and absorption terms of the form 
\begin{equation}\label{eq:rte-scattering-absoprtion}
    \frac{\partial}{\partial t} I(t,\bx,\bp) + c\bold{\Omega} \cdot \nabla I = \int_{\|\bp^{\{1\}}\|=\|\bp\|} \alpha(\bx,\bp^{\{1\}},\bp) I(t,\bx,\bp^{\{1\}}) d\bp^{\{1\}} - a(\bx, \bp) I(t,\bx,\bp),
\end{equation}
\sloppy which is the standard form for radiative transfer in a medium \cite{mihalas_1984, bosboom2023}. Here, $\alpha(\bx,\bp^{\{1\}},\bp)$ is the scattering coefficient which is symmetric in $\bp^{\{1\}},\bp$, which ensures momentum conservation \cite{mihalas_1984}, and $a$ is the absorption coefficient defined by
$a(\bx,\bp) := \int_{\|\bp_1\|=\|\bp\|}\alpha(\bx,\bp_1,\bp) d\bp_1,$
which ensures energy conservation \cite{mihalas_1984}. The terms on the left hand side of \eqref{eq:rte-scattering-absoprtion} are given by the Lie--Poisson structure, $\partial I/\partial t - \db{I}{\m{H}_r}_r$, so we have the task of constructing a dissipative bracket for the terms on the right hand side. 

We begin with the general form of the dissipative bracket discussed above, \Cref{eq:general-dissipative-bracket}. Since there are two terms on the right hand side describing scattering and absorption processes, we take $k=2$.
\begin{align}\label{eq:rte-2-bracket}
        (\m{F},\m{G})_{\textup{tot}} &= \int_{T^*Q} \int_{T^*Q} \int_{T^*Q} \gamma \Delta_{12}(\m{F}) \Delta_{12}(\m{G})  d\bp d\bx d\bx^{\{1\}} d\bp^{\{1\}} d\bx^{\{2\}} d\bp^{\{2\}}.
\end{align}
Since these processes are local in space, we include in the scattering kernel $\gamma$ a factor of $\delta(\bx-\bx^{\{1\}}) \delta(\bx - \bx^{\{2\}})$. Furthermore, since the absorption term is local in momentum, we additionally include a factor of $\delta(\bp - \bp^{\{2\}})$. To conserve energy, we include a factor of $\Delta_{12}(\m{H}_r)$
This yields
\begin{align*}
    \gamma :&= \beta(\bx,\bp^{\{1\}},\bp^{\{2\}}, \Psi_1, \Psi_2) \delta(\bp-\bp^{\{2\}}) \delta(\bx-\bx^{\{1\}}) \delta(\bx - \bx^{\{2\}}) \delta\left( \Delta_{12} (\m{H}_r) \right),
\end{align*}
where we denote $\Psi_j := \Psi(t,\bx^{\{j\}},\bp^{\{j\}})$ for brevity, and the function $\beta$ is to be determined. For the entropy functional, we choose the wave entropy \cite{Kaufman1984}, $\m{S}_r[\Psi] = \int_{T^*Q} \log(\Psi) d\bp d\bx$; note that $\m{S}_r$ satisfies $\db{\m{S}_r}{\m{H}_r}_r = 0$ so no additional terms are generated from the Lie--Poisson bracket. Finally, for the differential operator $\Delta_{12}$, we take the following linear operator which is antisymmetric in $\Psi_1, \Psi_2$,
$$\Delta_{12    }(\m{F}) := \frac{\delta \m{F}}{\delta \Psi_1} - \frac{\delta \m{F}}{\delta \Psi_2},$$
which will conserve the momentum functional provided that $\beta$ is symmetric in $\bp^{\{1\}}$ and $\bp^{\{2\}}$. Since the right hand side of \eqref{eq:rte-scattering-absoprtion} is linear in $I$ (or, equivalently, $\Psi$) whereas the differential operator applied to the entropy functional has terms inversely proportional to $\Psi$, $\Delta_{12}(\m{S}_r) = 1/\Psi_1 - 1/\Psi_2$, we include in $\beta$ a quadratic factor of $\Psi_1\Psi_2$, i.e., we define $\beta(\bx,\bp^{\{1\}},\bp^{\{2\}}, \Psi_1, \Psi_2) = \widetilde{\beta}(\bx,\bp^{\{1\}},\bp^{\{2\}})\Psi_1\Psi_2.$

Substituting these choices into \eqref{eq:rte-2-bracket} gives the expression for the dissipative bracket
\begin{align*}
    (\m{F},\m{G})_{\textup{tot}} &= 2c^{-1} \int_{T^*Q} \left(\int_{\|\bp^{\{1\}}\| = \|\bp\| }  \widetilde{\beta}(\bx,\bp^{\{1\}},\bp) d\bp^{\{1\}} \right) \Psi \Psi \frac{\delta \m{F}}{\delta \Psi} \frac{\delta \m{G}} {\delta \Psi} d\bp d\bx \\
    &\quad - 2c^{-1} \int_{T^*Q} \int_{\|\bp^{\{1\}}\| = \|\bp\| } \widetilde{\beta}(\bx,\bp^{\{1\}},\bp) \Psi_1 \Psi \frac{\delta \m{F}}{\delta \Psi_1} \frac{\delta \m{G}}{\delta \Psi} d\bp^{\{1\}} d\bp d\bx.
\end{align*}
To compute the dissipation terms for the specific intensity, take a linear functional $\m{F}_\varphi[\Psi] = \int_{T^*Q} \varphi I d\bp d\bx = c \int_{T^*Q} \varphi H_r \Psi d\bp d\bx $ where $\varphi$ is a fixed but arbitrary element of $C(T^*Q)$ and $\m{G} = \m{S}_r$ in the above expression which, using a similar argument to \Cref{ex:fluid-moments}, yields the right hand side of \eqref{eq:rte-scattering-absoprtion} with the choice $ \widetilde{\beta}(\bx,\bp^{\{1\}},\bp) = -(c/2) \alpha(\bx,\bp^{\{1\}},\bp)$. Note that this produces the same dissipative term derived in \cite{bosboom2023}, although there, a different dissipative bracket and entropy functional, given by the $L^2(T^*Q)$ norm squared of the specific intensity, are used.

More generally, we can consider an interaction bracket for radiation and matter. For example, we consider a resonant two-wave interaction with matter. In \Cref{eq:double-bracket-dynamics}, we take $k=3$ and define the differential operator
$$ \Delta_{123}(\m{F}) := \frac{\delta \m{F}}{\delta \Psi_1} - \frac{\delta \m{F}}{\delta \Psi_2} - (\bp^{\{1\}} - \bp^{\{2\}}) \cdot \frac{\partial}{\partial \bp^{\{3\}}} \frac{\delta \m{F}}{\delta f_3} $$
where $\Psi_j := \Psi(t,\bx^{\{j\}},\bp^{\{j\}})$ and $f_3 := f(t,\bx^{\{3\}},\bp^{\{3\}})$. This differential operator is chosen so that the total momentum functional,
$$ \m{P}_{\text{tot}}[f;\Psi] = \int_{T^*Q} \bp \Psi d\bp d\bx + \int_{T^*Q} \bp f d\bp d\bx, $$
is conserved. Analogous choices to the case of pure radiation can be made to define the scattering kernel $\gamma$, e.g., including spatially localizing factors $\delta(\bx-\bx^{\{1\}})\delta(\bx-\bx^{\{2\}})\delta(\bx-\bx^{\{3\}})$ as well as an energy conservative factor
$$ \delta(\Delta_{123}(\m{H}_{\text{tot}})) = \delta\left(c\|\bp^{\{1\}}\| - c\|\bp^{\{2\}}\| - (\bp^{\{1\}}-\bp^{\{2\}})\cdot \frac{\bp^{\{3\}}}{m} \right). $$
The total entropy functional can be chosen to be, for example, the sum of the wave entropy for the radiation distribution and the Boltzmann entropy for the electron distribution $S_e[f] := \int_{T^*Q} f \log(f) d\bp d\bx$.
\end{example}
\begin{remark}
    In the above example, the derivation of the interaction bracket proceeded by pattern matching, i.e., finding a bracket that matches known physical laws. It is more interesting to derive brackets from physical or mathematical principles. This is done subsequently in \Cref{sec:diffusion} where we derive interaction brackets for radiation diffusion hydrodynamics using thermodynamical principles and in \Cref{sec:arbitrary-transport-closure} where the brackets arise from geometric principles.
\end{remark}

There are many choices that can be made to define an interaction bracket, resulting in different dissipative dynamics, i.e., different collision operators. Note that, although we chose linear differential operators in the above examples for simplicity, one can generally consider nonlinear differential operators, e.g., dissipative brackets involving nonlinear differential operators are discussed in \cite{Dukek1997} in the context of stoichiometric reactions of kinetic systems. Rather than continuing to discuss possible choices of interaction brackets, we will move on to discuss moment systems associated to kinetic systems from a geometric perspective, which will be our main focus for the remainder. 

\subsection{Geometric perspective on moment systems}\label{sec:moment-equations}
We now introduce the notions of moment spaces and the kinetic moments associated to a kinetic system. Analogous to the definition of $C(T^*Q)$ and $C^*(T^*Q)$ in \Cref{sec:kinetic-theory}, let $C(Q) := C^\infty(Q)$ denote the space of smooth functions on $Q$ and $C^*(Q)$ denote its topological dual consisting of compactly supported distributions on $Q$. Throughout, we use the summation convention where repeated lower and upper indices are summed from $1$ to $n = \dim(Q)$. We define the space of degree $k$ moments on $Q$, denoted $\mathfrak{M}_k$, as the space of symmetric degree $k$ covariant tensor fields on $Q$ whose coefficient functions are in $C^*(Q)$, i.e., the space of tensors of the form
$$ M(\bx) = M_{i_1 \dots i_k}(\bx) dx^{i_1} \otimes \cdots \otimes dx^{i_k}, $$
where $\{dx^i\}_{i=1}^n$ is the standard basis for $T^*_\bx Q = \mathbb{R}^n$, $M_{i_1 \dots i_k}(\bx)$ is symmetric in $(i_1 \dots i_k)$, and $ M_{i_1 \dots i_k} \in C^*(Q)$. Similarly, we define the space of degree $k$ comoments on $Q$, denoted $\mathfrak{C}_k$, as the space of symmetric degree $k$ contravariant tensor fields on $Q$ whose coefficient functions are in $C(Q)$, i.e., the space of tensors of the form
$$ D(\bx) = D^{i_1 \dots i_k}(\bx) \be_{i_1} \otimes \cdots \otimes \be_{i_k}, $$
where $\{\be_i\}_{i=1}^n$ is the standard basis for $T_\bx Q = \mathbb{R}^n$, $D^{i_1 \dots i_k}(\bx)$ is symmetric in $(i_1 \dots i_k)$, and $D^{i_1 \dots i_k} \in C(Q)$. The duality pairing between $M \in \mathfrak{M}_k$ and $D \in \mathfrak{C}_k$ is given by
$$ \langle M, D\rangle = \int_Q M_{i_1 \dots i_k}(\bx) D^{i_1 \dots i_k}(\bx) d\bx. $$

We define the space of truncated moments at degree $k$ as
\begin{equation}\label{eq:space-of-truncated-moments}
    \mathfrak{M}^k_0 := \oplus_{i=0}^k \mathfrak{M}_i,
\end{equation}
i.e., the direct sum of the space of moments of degree between $0$ and $k$.

As discussed in \Cref{ex:fluid-moments} and \Cref{ex:radiation-moments}, one can utilize the Lie--Poisson formulation to derive the evolution equation of moments of the phase space distribution when no dissipative terms are present. As we saw, the evolution of such moments involved divergences of tensors of one higher degree in $\bz := \nabla_\bp H(\bx,\bp)$. Interestingly, different Hamiltonians thus produce different tensors with which to multiply the distribution and integrate; for example, for the particle kinetic system, we used (see \Cref{ex:fluid-moments})
$$ \bp \otimes \cdots \otimes \bp $$
(note that $\bp$ equals $\nabla_\bp H_e$ up to a constant factor) whereas for the radiation kinetic system, we used (see \Cref{ex:radiation-moments})
$$ \mathbf{\Omega} \otimes \cdots \otimes \mathbf{\Omega}$$
(similarly, $\mathbf{\Omega}$ equals $\nabla_\bp H_r$ up to a constant factor). The difference between the kinetic moments for these two Hamiltonians can be understood from the homogeneity of the Hamiltonian in $\bp$; the electron Hamiltonian is homogeneous of degree $2$ in $\bp$ whereas the radiation Hamiltonian is homogeneous of degree $1$ in $\bp$. This observation motivates the following definition of kinetic moments.

\begin{definition}[Kinetic Moments]\label{def:kinetic-moments}
    For a phase space distribution $g \in C^*(T^*Q)$ and a separable Hamiltonian $H \in C(T^*Q)$, $H(\bx,\bp) = K(\bp) + U(\bx)$, the $k^{th}$ kinetic moment of $g$ is
    \begin{align}\label{eq:def-kinetic-moments}
        M^k[g] &\in \mathfrak{M}_k, \nonumber \\ 
        M^k[g](\bx) & := \int_{T^*_\bx Q} \underbrace{ \mathbf{z} \otimes \cdots \otimes \mathbf{z}}_{k \textup{ times}} g(\bx,\bp) d\bp, 
    \end{align}
    where throughout we denote
    $$ \mathbf{z} := \nabla_\bp H(\bx,\bp) = \nabla_\bp K(\bp). $$
    We will often denote the $k^{th}$ kinetic moment as simply $M^k(\bx)$ when the distribution $g$ is clear in context. Additionally, we refer to 
    $$ \underbrace{\bz \otimes \cdots \otimes \bz}_{k \textup{ times}} = \nabla_\bp H \otimes \cdots \otimes \nabla_\bp H$$
    as the $k^{th}$ moment kernel corresponding to $H$.
\end{definition}

To add additional flexibility in the definition of kinetic moments, one can in principle consider weighting the distribution $g(\bx,\bp)$ by some function $G(\bx,\bp)$, prior to multiplying by a moment kernel and integrating over $\bp$. We have already seen an example of this in \Cref{ex:radiation-moments}, where the distribution $\Psi$ was weighted by the radiation Hamiltonian $H_r(\bx,\bp)$. In essence, although one could use the moments associated to $\Psi$, the moments associated to $H_r \Psi$ (which is the same as the specific intensity $I$ up to a multiplicative constant) could also be used and are more commonly used in the radiation transport literature due to their physical interpretation (e.g., the zeroth moment of $H_r \Psi$ is the electromagnetic energy density). As such, we introduce the notion of a weighted kinetic moment.

\begin{definition}[Weighted Kinetic Moments]\label{def:weighted-kinetic-moments}
    Let $G \in C(T^*Q)$. For a phase space distribution $g \in C^*(T^*Q)$ and a separable Hamiltonian $H \in C(T^*Q)$, $H(\bx,\bp) = K(\bp) + U(\bx)$, the $k^{th}$ $G$-weighted kinetic moment associated to $g$ is
    \begin{align}\label{eq:def-weighted-kinetic-moments}
        G^k[g] &\in \mathfrak{M}_k, \nonumber \\ 
        G^k[g](\bx) & := \int_{T^*_\bx Q} \underbrace{\mathbf{z} \otimes \cdots \otimes \mathbf{z}}_{k \textup{ times}} g(\bx,\bp) G(\bx,\bp) d\bp.
    \end{align}
    We will often denote the $k^{th}$ $G$-weighted kinetic moment as simply $G^k(\bx)$ when the distribution $g$ is clear in context.
    
    In particular, when the weight is chosen to be the Hamiltonian, $G = H$, we refer to the $H$-weighted kinetic moments as energy-weighted kinetic moments and denote them as $H^k$.
\end{definition}

Now, we derive the evolution equations for the weighted kinetic moments and as an immediate corollary, the evolution equations for the unweighted kinetic moments. Although one can derive the equations of motion for a general weight $G$, we will focus on the particular case where $G$ Poisson commutes with $H$, i.e., $[G,H] = 0$. As we will see, such a choice of $G$ is natural in the sense that the weight does not generate additional terms in the evolution equations, when compared with the unweighted evolution equations.

To derive the equations of motion, we introduce the following symmetric tensor product operation.

\begin{definition}[Symmetric Tensor Product]\label{def:symm-tensor-product}
    For a contravariant tensor $A$ of degree $k$ and a vector (i.e., tensor of degree 1) $\mathbf{v}$ on $Q$, the symmetric tensor product of $A$ and $\mathbf{v}$, denoted $\sympl{A}{\mathbf{v}}$, is defined on a tensor-product basis as
    \begin{equation}\label{eq:symm-tensor-product}
        \symL{\mathbf{e}_{i_1} \otimes \cdots \otimes \mathbf{e}_{i_k}}{v} :=\mathbf{v} \otimes \mathbf{e}_{i_1} \otimes \cdots \otimes \mathbf{e}_{i_k} + \mathbf{e}_{i_1} \otimes \mathbf{v} \otimes \cdots \otimes \mathbf{e}_{i_k} + \ldots + \mathbf{e}_{i_1} \otimes \cdots \otimes \mathbf{e}_{i_k} \otimes \mathbf{v},
    \end{equation}
    and extended by linearity to $A$. In particular, note that if $A$ is symmetric, then $\sympl{A}{v}$ is symmetric.
\end{definition}

Then, we have the following evolution equations.

\begin{prop}\label{prop:g-weighted-moments}
    Assume that $G \in C(T^*Q)$ Poisson commutes with $H$, i.e., $[G,H] = 0$.
        Then, for a kinetic system with phase space distribution $g(t,\bx,\bp)$ and separable Hamiltonian $H(\bx,\bp) = K(\bp) + U(\bx)$, the Lie--Poisson evolution of the $G$-weighted kinetic moments (see \Cref{sec:kinetic-theory}) is equivalently given by
    \begin{align}\label{eq:g-weighted-moment-evolution}
        \frac{\partial}{\partial t} G^k &+ \nabla_\bx \cdot  G^{k+1} \nonumber \\
        &= - \int_{T^*_\bx Q} g(t,\bx,\bp) G(\bx,\bp)\, \sym{\underbrace{\bz \otimes \cdots \otimes \bz}_{k-1 \textup{ times}}}{\nabla_{\bp}\nabla_{\bp}H \cdot \nabla_\bx U} d\bp,
    \end{align}
    where $\nabla_\bp \nabla_\bp H$ denotes the Hessian of $H$ with respect to $\bp$ and for $k=0$, we use the convention that the right hand side is zero.
    \begin{proof}
    We prove this by a similar calculation to \Cref{example:fluid-closures} and \Cref{example:pure-rad-closures}; consider the linear functional of $g$ given by
    \begin{align*}
        \m{F}^k_D [g](t,\bx) &:= \int_{T^*Q} z_{i_1}\cdots z_{i_k} D^{i_1 \dots i_k}(\bx) G(\bx,\bp) g(t,\bp,\bx)\, d\bp d\bx \\
        &= \int_Q G^k_{i_1 \dots i_k}(t,\bx) D^{i_1 \dots i_k}(\bx) d\bx,
    \end{align*}
    where $D \in \mathfrak{C}_k$ is a fixed but arbitrary degree $k$ comoment, $\{z_i\}_{i=1}^n$ denotes the components of $\bz$, and in the equality we used the definition of $G$-weighted kinetic moments (\Cref{def:weighted-kinetic-moments}). We now compute the Lie--Poisson evolution equation for this functional. Note that its functional derivative with respect to $g$ is $\delta \m{F}^k_D/\delta g = z_{i_1} \cdots z_{i_k} D^{i_1 \dots i_k}(\bx) G(\bx,\bp).$ The evolution equation is then
    \begin{align*}
        \frac{\partial}{\partial t} \m{F}^k_D &= \{ \m{F}^k_D, \m{H} \} = \int_{T^*Q} g(t,\bx,\bp) \left[ \frac{\delta \m{F}^k_D}{\delta g}, \frac{\delta \m{H}}{\delta g} \right] d\bp d\bx \\
        &= \int_{T^*Q}  g(t,\bx,\bp) \left[ z_{i_1} \cdots z_{i_k} D^{i_1 \dots i_k}(\bx) G(\bx,\bp) , H(\bx,\bp) \right] d\bp d\bx \\
        &= \int_{T^*Q}  g(t,\bx,\bp) G(\bx,\bp) \left[ z_{i_1} \cdots z_{i_k} D^{i_1 \dots i_k}(\bx) , H(\bx,\bp) \right] d\bp d\bx 
    \end{align*}
    where, in the last equality, we used the Leibniz rule satisfied by the Poisson bracket \cite{RaMa1987}; one of the resulting terms by applying the Leibniz rule is proportional to $[G,H]$ and thus vanishes since, by assumption, $G$ Poisson commutes with $H$. Proceeding further, we have
    \begin{align*}
        \frac{\partial}{\partial t}& \m{F}^k_D = \int_{T^*Q}  g(t,\bx,\bp) G(\bx,\bp) \left[ z_{i_1} \cdots z_{i_k} D^{i_1 \dots i_k}(\bx) , H(\bx,\bp) \right] d\bp d\bx \\
        &= \int_{T^*Q}  g(t,\bx,\bp) G(\bx,\bp) \nabla_\bx ( z_{i_1} \cdots z_{i_k} D^{i_1 \dots i_k}(\bx)) \cdot \underbrace{\nabla_\bp H(\bx,\bp)}_{= \bz}  d\bp d\bx \\
            & \qquad - \int_{T^*Q}  g(t,\bx,\bp) G(\bx,\bp) \nabla_\bp ( z_{i_1} \cdots z_{i_k} D^{i_1 \dots i_k}(\bx)) \cdot \underbrace{\nabla_\bx H(\bx,\bp)}_{=\nabla_\bx U(\bx)} d\bp d\bx \\
        &= - \int_Q  \left(\nabla_\bx^{i_{k+1}} \int_{T^*_\bx Q} z_{i_1} \cdots z_{i_k}z_{i_{k+1}} g(t,\bx,\bp) G(\bx,\bp) d\bp \right) D^{i_1\dots i_k}(\bx) d\bx \\
            & \quad - \int_Q  \left( \int_{T^*_\bx Q} g G \sym{\underbrace{(\bz \otimes \cdots \otimes \bz)}_{k-1 \textup{ times}}}{\nabla_{\bp}\nabla_{\bp}H \cdot \nabla_\bx U} d\bp \right)_{i_1\dots i_k}D^{i_1 \dots i_k}(\bx) d\bx  \\
        &=- \int_Q \Big(\nabla \cdot G^{k+1}(t,\bx)\Big)_{i_1 \dots i_k} D^{i_1 \dots i_k}(\bx) d\bx \\
        & \quad - \int_Q  \left( \int_{T^*_\bx Q} g G \sym{\underbrace{(\bz \otimes \cdots \otimes \bz)}_{k-1 \textup{ times}}}{\nabla_{\bp}\nabla_{\bp}H \cdot \nabla_\bx U} d\bp \right)_{i_1\dots i_k}D^{i_1 \dots i_k}(\bx) d\bx 
    \end{align*}
    where we used integration by parts for the first integral and the definition of the symmetric tensor product (\Cref{def:symm-tensor-product}) for the second integral (noting that, for the case $k=0$, the second integral vanishes since $\nabla_\bp D(\bx) = 0$). Finally, note that the left hand side of this equation is equivalently
    $$ \frac{\partial}{\partial t} \m{F}^k_D = \int_Q \left(\frac{\partial}{\partial t}G^k_{i_1 \dots i_k}(t,\bx)\right) D^{i_1 \dots i_k}(\bx) d\bx. $$
    Since $D \in \mathfrak{C}_k$ is arbitrary, comparing terms yields \eqref{eq:g-weighted-moment-evolution}. \qed
    \end{proof} 
\end{prop}

As an immediate corollary, we obtain the evolution equations for the unweighted kinetic moments and the energy-weighted kinetic moments, since these are both examples of $G$-weighted kinetic moments that Poisson commute with $H$, with $G \equiv 1$ and $G = H$, respectively. 

\begin{corollary} \label{corr:moment-system-evolution}
    The Lie--Poisson evolution of the kinetic moments (\Cref{def:kinetic-moments}) is given by
    \begin{equation}\label{eq:moment-system-evolution}
        \frac{\partial}{\partial t} M^k + \nabla_\bx \cdot M^{k+1} = - \int_{T^*_\bx Q} g(t,\bx,\bp)\, \sym{\underbrace{\bz \otimes \cdots \otimes \bz}_{k-1 \textup{ times}}}{\nabla_{\bp}\nabla_{\bp}H \cdot \nabla_\bx U} d\bp.
    \end{equation}
    Similarly, the Lie--Poisson evolution of the energy-weighted kinetic moments (\Cref{def:weighted-kinetic-moments}) is given by
        \begin{equation}\label{eq:weighted-moment-system-evolution}
        \frac{\partial}{\partial t} H^k + \nabla_\bx \cdot H^{k+1} = - \int_{T^*_\bx Q} g(t,\bx,\bp) H(\bx,\bp)\, \sym{\underbrace{\bz \otimes \cdots \otimes \bz}_{k-1 \textup{ times}}}{\nabla_{\bp}\nabla_{\bp}H \cdot \nabla_\bx U} d\bp,
    \end{equation}
    again with the convention that the right hand side is zero for $k=0$.
    \begin{proof}
        This immediately follows from \Cref{prop:g-weighted-moments}, with the observation that the constant weight $1$ Poisson commutes with $H$, $[1,H] = 0$, as well as the energy weight $H$, $[H,H]=0$.
    \qed \end{proof}
\end{corollary}

From \Cref{prop:g-weighted-moments} and \Cref{corr:moment-system-evolution}, we can make two interesting observations. First, note that in \Cref{def:kinetic-moments}, there is no freedom in the choice of moment kernels in the sense that, once the zeroth moment is specified, \Cref{corr:moment-system-evolution} shows that the evolution equation for the zeroth moment depends on the divergence of the first moment formed using the first moment kernel $\bz = \nabla_\bp H$; subsequently, the evolution equation of the first moment depends on the divergence of the second moment formed using the second moment kernel $\bz \otimes \bz$, and similarly for higher moments. A similar calculation shows that this is more generally true for all Hamiltonians, not just separable Hamiltonians, where now the moment kernels generally depend on $\bx$ and $\bp$ from $\bz = \nabla_\bp H(\bx,\bp)$, with the only modification being more potential terms on the right hand side. Thus, the moment kernels to all degrees are determined by the Hamiltonian $H$; this arose from the Poisson bracket appearing inside the Lie--Poisson bracket providing a factor of $\bz$ at each degree of the moment evolution equations. 

Second, although there is no flexibility in the choice of moment kernel once the Hamiltonian is specified, there is still flexibility in the choice of weighting the distribution by a function of $(\bx,\bp)$, prior to forming the moments. In particular, for the $G$-weighted kinetic moments where $G$ Poisson commutes with $H$, the evolution equation \eqref{eq:g-weighted-moment-evolution} is formally the same as the evolution equation for the unweighted kinetic moments \eqref{eq:moment-system-evolution}, with $g$ replaced by $gG$. This arose from the fact the extra term that appears in the $G$-weighted moment evolution equations (see the proof of \Cref{prop:g-weighted-moments}) is proportional to $[G,H]$ and thus vanishes. Such choices of weights $G$ that Poisson commute with $H$, i.e., invariants of the Hamiltonian system, are natural in the sense that the $G-$weighted moment evolution equations do not have any additional terms relative to the unweighted moment evolution equations.

As an example, let us compute the unweighted and energy-weighted moment evolution equations for the matter kinetic theory (\Cref{sec:hamiltonian-fluid}) and radiation kinetic theory (\Cref{sec:hamiltonian-radiation}).

\begin{corollary}\label{corr:moment-system-evolution-specific}
    The evolution of the kinetic moments $M^k_e$ and the energy-weighted kinetic moments $H^k_e$ associated to the matter kinetic system with distribution $f$ and Hamiltonian $H_e(\bx,\bp) = \|\bp\|^2/2m + V(\bx)$ is given by 
    \begin{subequations}\label{eq:particle-kinetic-moment-evolution}
    \begin{align}
        \frac{\partial}{\partial t} M^k_e + \nabla_\bx \cdot M^{k+1}_e &= - m^{-1} \sympl{M^{k-1}_e}{\nabla_\bx U},\label{eq:particle-kinetic-moment-evolution-a} \\
        \frac{\partial}{\partial t} H^k_e + \nabla_\bx \cdot H^{k+1}_e &= - m^{-1} \sympl{H^{k-1}_e}{\nabla_\bx U}.\label{eq:particle-kinetic-moment-evolution-b}
    \end{align}
    \end{subequations}

    Similarly, the evolution of the kinetic moments $M^k_r$ and the energy-weighted kinetic moments $H^k_r$ associated to the radiation kinetic system with distribution $\Psi$ and Hamiltonian $H_r(\bx,\bp) = c \|\bp\|$ is given by 
    \begin{subequations}\label{eq:radiation-kinetic-moment-evolution}
    \begin{align}
        \frac{\partial}{\partial t} M^k_r + \nabla_\bx \cdot M^{k+1}_r &= 0,\label{eq:radiation-kinetic-moment-evolution-a} \\
        \frac{\partial}{\partial t} H^k_r + \nabla_\bx \cdot H^{k+1}_r &= 0. \label{eq:radiation-kinetic-moment-evolution-b}
    \end{align}
    \end{subequations}
    
    \begin{proof}
        For \Cref{eq:particle-kinetic-moment-evolution-a}, the right hand side of \Cref{eq:moment-system-evolution} can be expressed, using $\nabla_\bp \nabla_\bp H_e = m^{-1} I$ (where $I$ denotes the identity),
        \begin{align*}
            - \int_{T^*_\bx Q} g(t,\bx,\bp)\, &\sym{\underbrace{\bz \otimes \cdots \otimes \bz}_{k-1 \textup{ times}}}{\nabla_{\bp}\nabla_{\bp}H_e \cdot \nabla_\bx V} d\bp \\ &= - m^{-1} \int_{T^*_\bx Q} g(t,\bx,\bp)\, \symL{\underbrace{\bz \otimes \cdots \otimes \bz}_{k-1 \textup{ times}}}{\nabla_\bx V(\bx)} d\bp \\
            &= - m^{-1} \symL{ \int_{T^*_\bx Q} g(t,\bx,\bp) \underbrace{\bz \otimes \cdots \otimes \bz}_{k-1 \textup{ times}} d\bp}{\nabla_\bx V(\bx)} \\
            &= - m^{-1} \sympl{M^{k-1}_e}{\nabla_\bx V},
        \end{align*}
        where here $\bz = \nabla_\bp H_e = \bp/m$. In the second equality, we used that the integration with respect to $\bp$ can be distributed across the symmetric tensor product (\Cref{def:symm-tensor-product}) by linearity, noting that $\nabla_\bx V(\bx)$ does not depend on $\bp$. By \Cref{prop:g-weighted-moments}, the evolution equations for the energy-weighted moment are then just given by replacing the unweighted moments with the energy-weighted moments in the evolution equation, which yields \Cref{eq:particle-kinetic-moment-evolution-b}.

        For the radiation kinetic moments, there is no potential in the Hamiltonian $H_r(\bx,\bp) = c\|\bp\|$ and thus, \Cref{eq:radiation-kinetic-moment-evolution-a} and \Cref{eq:radiation-kinetic-moment-evolution-b} immediately follow from \Cref{eq:moment-system-evolution} and \Cref{eq:weighted-moment-system-evolution}, respectively (where here $\bz = \nabla_\bp H_r = c \mathbf{\Omega}$). \qed 
        \end{proof}
\end{corollary}

\subsection{Pair bracket formulation on moments}\label{sec:interaction-moment-spaces}
From \Cref{corr:moment-system-evolution-specific}, we see from \Cref{eq:particle-kinetic-moment-evolution} and \Cref{eq:radiation-kinetic-moment-evolution}, the evolution of the $k^{th}$ moment depends on the $(k+1)^{st}$ moment (as well as the $(k-1)^{st}$ moment for the electron kinetic moments). Instead of solving for the full phase space distribution, a common approach for approximately solving a kinetic system is to solve the moment equations after truncating at some finite number of moments \cite{Lev1996}, say $m$. This leads to the problem that the truncated evolution equation are not closed. To resolve this, we will first consider pair bracket formulations on a space of truncated moments and thus, the evolution equations on the space of truncated moments are automatically closed; we consider, in detail, an example in \Cref{sec:diffusion}. However, generally, a pair bracket formulation on a space of truncated moments need not be related to any underlying kinetic system. To address this, in \Cref{sec:geometric-moment-closures}, we will introduce the notion of a \emph{geometric moment closure} for which the dynamics on the truncated moment space arise as the pullback of the dynamics from the kinetic system.

Instead of imposing a pair bracket structure at the level of the phase space distributions, as discussed in \Cref{sec:interaction-brackets}, we can impose a pair bracket structure at the level of the moments. Consider a collection of truncation degrees $k(1),\dots,k(s) \geq 0$, $s \geq 1$. Then, we define the corresponding \emph{total truncated moment space}
\begin{equation}\label{eq:total-truncated-moment-space}
    \mathfrak{M} = \mathfrak{M}^{k(1)}_0 \times \cdots \times \mathfrak{M}^{k(q)}_0.
\end{equation}
We can then impose the evolution of functionals on $\mathfrak{M}$ through a pair bracket formulation,
\begin{equation}\label{eq:pair-bracket-form-moments}
    \frac{\partial}{\partial t} \m{G} = \{\m{G}, \m{H}_{\mathfrak{M}}\}_{\mathfrak{M}} + (\m{G}, \m{S}_{\mathfrak{M}})_{\mathfrak{M}},
\end{equation}
Similarly to the kinetic case, the dynamics at the level of the moments is specified by a Hamiltonian functional $\m{H}_{\mathfrak{M}}: \mathfrak{M} \rightarrow \mathbb{R}$, an entropy functional $\m{S}_{\mathfrak{M}}: \mathfrak{M} \rightarrow \mathbb{R}$, an antisymmetric bracket $\{\cdot, \cdot\}_{\mathfrak{M}}$ defined on functionals on $\mathfrak{M}$, and a dissipative bracket $(\cdot,\cdot)_{\mathfrak{M}}$ defined on functionals on $\mathfrak{M}$ such that $\{\m{S}_{\mathfrak{M}},\m{H}_{\mathfrak{M}}\}_{\mathfrak{M}}=0$ and $(\m{H}_{\mathfrak{M}},\m{S}_{\mathfrak{M}})_{\mathfrak{M}}=0$.

\begin{example}[Two-temperature diffusion radiation hydrodynamics]\label{sec:diffusion}
As an example of a pair bracket formulation on a space of moments, we consider the two-temperature gray diffusion radiation hydrodynamics system, which takes the form \cite{imex-radhydro}
\begin{subequations}\label{eq:2T-equations}
\begin{align}
\frac{\partial \rho}{\partial t} + \nabla_\bx \cdot \bold{P} &= 0, \label{2T a} \\
\frac{\partial}{\partial t} \bold{P} + \nabla_\bx \cdot [\rho^{-1} \bold{P} \otimes \bold{P} ] &= - \nabla_\bx (p_e + p_r)\bold{I}, \label{2T b}\\
\frac{\partial}{\partial t} E_e + \nabla_\bx \cdot [ E_e \rho^{-1}\bold{P} ] + \nabla_\bx \cdot \bm{F}_e &= -p_e \nabla_\bx \cdot \rho^{-1}\bold{P} - G_{er}(T_e,T_r), \label{2T c} \\
\frac{\partial}{\partial t} E_r + \nabla_\bx \cdot [ E_r \rho^{-1}\bold{P} ] + \nabla_\bx \cdot \bm{F}_r &= -p_r \nabla_\bx \cdot \rho^{-1}\bold{P} + G_{er}(T_e,T_r), \label{2T d}
\end{align}
\end{subequations}
where $p_e$ is the electron fluid pressure, $p_r$ is the radiation pressure, $T_e$ is the electron temperature and $T_r$ is the radiation temperature which will be defined in terms of the moment variables through equations of state. Additionally, $G_{er}$ denotes a thermal interaction term, which we take to be of the form
$$ G_{er} = \sigma_P a c (T_e^4 - T^4_r) $$
corresponding to a black body equation of state $E_r = aT_r^4$, where $a$ is a positive constant and the opacity $\sigma_P$ is a positive real-valued function of $\bx, E_e, E_r$, and finally, $\bm{F}_e$ and $\bm{F}_r$ are the electron and radiation thermal fluxes, respectively, which have the form
\begin{gather}
\bm{F}_e = -K_e(T_e)\,\nabla_\bx T_e, \quad \bm{F}_r = -\mathcal{D}(T_e)\,\nabla_\bx E_r = - a\m{D}(T_e) T_r^3 \nabla_\bx T_r,\label{flux_formulas_basic-0}
\end{gather}
where $K_e$ and $\m{D}$ are temperature-dependent diffusion coefficients. The space of moments is
$$ \mathfrak{M} := \mathfrak{M}^1_0 \times \mathfrak{M}^0_0 \times \mathfrak{M}^0_0 \ni ( \mathbf{P}, \rho, E_e, E_r), $$ 
noting that, for the rest of this section, we will write the degree one moment $\mathbf{P}$ before the degree zero moment $\rho$; we do this so that the Lie group and Lie algebra structures we define below are simpler to write. We take the following Hamiltonian functional on $\mathfrak{M}$, which is the sum of hydrodynamical kinetic energy, hydrodynamical internal energy, and radiation internal energy,
\begin{align}\label{eq:ham-functional-moments-matter-rad}
    \m{H}_{\mathfrak{M}} &: \mathfrak{M} \rightarrow \mathbb{R}, \nonumber \\
    \m{H}_{\mathfrak{M}}[\bold{P},\rho,E_e,E_r] &:= \frac{1}{2} \int_Q \left\| \frac{\bold{P}}{\rho} \right\|^2 \rho d\bx + \int_Q E_e d\bx + \int_{Q} E_r d\bx. 
\end{align}

We will now introduce a Lie--Poisson bracket for the advection terms, followed by dissipative brackets for the thermal flux and thermal interaction terms. 

\textbf{Lie--Poisson bracket for the non-dissipative terms.} When dissipative effects and species interactions are neglected, the 2T system reduces to a Lie--Poisson Hamiltonian system on the dual to the Lie algebra $\mathfrak{g} = \mathfrak{C}_1(Q)\rtimes (C(Q)\times C(Q)\times C(Q))\ni (\mathbf{u},g,g_e,g_r)$. The underlying Lie group $G=\text{Diff}(Q)\rtimes(C(Q)\times C(Q)\times C(Q))\ni (\varphi,G,G_e,G_r)$ is the semidirect product of the diffeomorphism group $\text{Diff}(Q)\ni \varphi$ with the abelian group $C(Q)\times C(Q)\times C(Q)\ni (G,G_e,G_r)$. The diffeomorphisms $\text{Diff}(Q)$ act from the left on $C(Q)\times C(Q)\times C(Q)$ by pushforward. The group product in $G$ is therefore
\begin{align*}
(\varphi^1,G^1,G_e^1,G_r^1)&(\varphi^2,G^2,G_e^2,G_r^2) \\ 
&= (\varphi^1\circ\varphi^2,G^1 + \varphi^1_*G^2,G_e^1 + \varphi^1_*G_e^2,G_r^1 + \varphi^1_*G_r^2).
\end{align*}
The corresponding Lie product $[\cdot,\cdot]_{\mathfrak{g}}$ on $\mathfrak{g}$ is
\begin{align}
\left[\begin{pmatrix} \mathbf{u}^1\\ g^1\\ g_e^1\\ g_r^1\end{pmatrix},\begin{pmatrix} \mathbf{u}^2\\ g^2\\ g_e^2\\ g_r^2\end{pmatrix}\right]_{\mathfrak{g}} = -\begin{pmatrix}[\mathbf{u}^1,\mathbf{u}^2]_{\mathfrak{X}}\\ \mathcal{L}_{\mathbf{u}^1}g^2-\mathcal{L}_{\mathbf{u}^2}g^1\\ \mathcal{L}_{\mathbf{u}^1}g_e^2-\mathcal{L}_{\mathbf{u}^2}g_e^1\\ \mathcal{L}_{\mathbf{u}^1}g_r^2-\mathcal{L}_{\mathbf{u}^2}g_r^1 \end{pmatrix},
\end{align}
where $\mathcal{L}$ denotes the Lie derivative and $[\cdot,\cdot]_{\mathfrak{X}}$ denotes the Lie bracket of vector fields. The Lie-Poisson bracket between functionals $\m{F},\m{G}:\mathfrak{g}^*\rightarrow\mathbb{R}$ on $\mathfrak{g}^*$, for $\xi^* = (\mathbf{P},\rho,\Sigma_e,\Sigma_r)\in \mathfrak{M}_1(Q)\times C^*(Q) \times C^*(Q) \times C^*(Q)$, is
\begin{equation}\label{eq:Lie-Poisson-bracket}
\begin{aligned}
\db{\m{F}}{\m{G}}_{\mathfrak{g}^*}(\xi^*) &= \left\langle\xi^*,\left[\frac{\delta \m{F}}{\delta \xi^*},\frac{\delta \m{G}}{\delta \xi^*}\right]_{\mathfrak{g}} \right\rangle \\
& = -\int_Q \mathbf{P}\cdot \left[\frac{\delta \m{F}}{\delta\mathbf{P}},\frac{\delta \m{G}}{\delta\mathbf{P}}\right]_{\mathfrak{X}} d\bx - \int_Q \rho\,\left(\mathcal{L}_{\delta \m{F}/\delta\mathbf{P}}\frac{\delta \m{G}}{\delta \rho} -\mathcal{L}_{\delta \m{G}/\delta\mathbf{P}}\frac{\delta \m{F}}{\delta \rho} \right) d\bx \\
& \qquad -\int_Q \Sigma_e\,\left(\mathcal{L}_{\delta \m{F}/\delta\mathbf{P}}\frac{\delta \m{G}}{\delta \Sigma_e} -\mathcal{L}_{\delta \m{G}/\delta\mathbf{P}}\frac{\delta \m{F}}{\delta \Sigma_e} \right) d\bx \\
& \qquad - \int_Q \Sigma_r\,\left(\mathcal{L}_{\delta \m{F}/\delta\mathbf{P}}\frac{\delta \m{G}}{\delta \Sigma_r}  -\mathcal{L}_{\delta \m{G}/\delta\mathbf{P}}\frac{\delta \m{F}}{\delta \Sigma_r} \right) d\bx.
\end{aligned}
\end{equation}
Here, $\Sigma_\nu$ is the species entropy density. For now, we use the fluid variables $(\mathbf{P},\rho)$ with the entropy variables $(\Sigma_e,\Sigma_r)$ since the entropy is advected and thus arises from the above Lie--Poisson bracket; subsequently, we will transform back to the energy density variables $E_e,E_r$. 

We define the species-$\nu$ specific entropy, $s_\nu := \Sigma_\nu/\rho$. The species-$\nu$ specific internal energy $\m{U}_\nu := E_\nu/\rho$ is specified by an equation of state $\m{U}_\nu = \mathcal{U}_\nu(\rho,s_\nu)$. The Hamiltonian functional for the 2T system \eqref{eq:ham-functional-moments-matter-rad} in terms of the entropy variables is
\begin{align}\label{eq:hamiltonian-matter-radiation-moments}
\mathcal{H}_{\mathfrak{M}}[\mathbf{P},\rho,\Sigma_e,\Sigma_r] & = \frac{1}{2}\int_Q \left\|\frac{\mathbf{P}}{\rho}\right\|^2 \rho \,d\bx +\int_Q \mathcal{U}_e\left(\rho,\frac{\Sigma_e}{\rho}\right)\rho\, d\bx + \int_Q \mathcal{U}_r\left(\rho,\frac{\Sigma_r}{\rho}\right)\rho\, d\bx.
\end{align}
Its functional derivatives are given by
\begin{align*}
\frac{\delta \mathcal{H}_{\mathfrak{M}}}{\delta \mathbf{P}} = \frac{\mathbf{P}}{\rho},\quad \frac{\delta \mathcal{H}_{\mathfrak{M}}}{\delta \rho} = -\frac{1}{2}\left\|\frac{\mathbf{P}}{\rho}\right\|^2 +  \m{F}_e  + \m{F}_r,\quad \frac{\delta\mathcal{H}_{\mathfrak{M}}}{\delta \Sigma_e} = T_e,\quad \frac{\delta\mathcal{H}_{\mathfrak{M}}}{\delta \Sigma_r} = T_r,
\end{align*}
where we have introduced specific Gibbs free energies and temperatures according to
\begin{align*}
\m{F}_\nu = \mathcal{U}_\nu + D\,\partial_D\mathcal{U}_\nu- s_{\nu}\,\partial_{s_\nu}\mathcal{U}_\nu,\quad T_\nu = \partial_{s_\nu}\mathcal{U}_\nu.
\end{align*}

Denoting the velocity field $\mathbf{u} := \mathbf{P}/\rho$, the equations of motion corresponding to the Lie--Poisson bracket $\db{\cdot}{\cdot}_{\mathfrak{g}^*}$ and Hamiltonian functional $\mathcal{H}_{\mathfrak{M}}$ are given by
\begin{align*}
(\partial_t + \mathcal{L}_{\mathbf{u}})\rho &= 0,\\
(\partial_t + \mathcal{L}_{\mathbf{u}})\mathbf{P} &= -\nabla_{\bx}(p_e + p_r) + \rho \nabla_{\bx}\left(\frac{1}{2}|\mathbf{u}|^2\right),\\
(\partial_t + \mathcal{L}_{\mathbf{u}})\Sigma_e &= 0, \\
(\partial_t + \mathcal{L}_{\mathbf{u}})\Sigma_r &= 0.
\end{align*}
where the species-$\nu$ pressure is defined according to $p_\nu = \rho^2\,\partial_\rho\mathcal{U}_\nu$. Transforming the above bracket $\db{\cdot}{\cdot}_{\mathfrak{g}^*}$ into the moment variables $(\mathbf{P}, \rho, E_e, E_r) \in \mathfrak{M}$ yields the 2T advection bracket; for functionals $\m{F}$ and $\m{G}$ on $\mathfrak{M}$,
\begin{align}\label{eq:2T-advection-bracket}
    \db{\m{F}}{\m{G}}_{\textup{2T}}[\mathbf{P}, \rho, E_e, E_r] &= -\int_Q \mathbf{P}\cdot \left[\frac{\delta \m{F}}{\delta\mathbf{P}},\frac{\delta \m{G}}{\delta\mathbf{P}}\right]_{\mathfrak{X}} d\bx \nonumber \\ 
    &\quad\ - \int_Q \rho\,\left(\mathcal{L}_{\delta \m{F}/\delta\mathbf{P}}\frac{\delta \m{G}}{\delta \rho} -\mathcal{L}_{\delta \m{G}/\delta\mathbf{P}}\frac{\delta \m{F}}{\delta \rho} \right) d\bx \nonumber \\
    & \quad\ -\int_Q \Sigma_e\,\left(\mathcal{L}_{\delta \m{F}/\delta\mathbf{P}}\left(T_e\frac{\delta \m{G}}{\delta E_e}\right) -\mathcal{L}_{\delta \m{G}/\delta\mathbf{P}}\left(T_e \frac{\delta \m{F}}{\delta E_e}\right) \right) d\bx \nonumber \\
    & \quad\ - \int_Q \Sigma_r\,\left(\mathcal{L}_{\delta \m{F}/\delta\mathbf{P}} \left(T_r\frac{\delta \m{G}}{\delta E_r}\right) -\mathcal{L}_{\delta \m{G}/\delta\mathbf{P}}\left(T_r\frac{\delta \m{F}}{\delta \Sigma_r}\right) \right) d\bx, 
\end{align}
where now the equation of state specifies the species entropy in terms of the density and internal energy $\Sigma_\nu = \Sigma_\nu(\rho, E_\nu)$, equivalently, $s_\nu = s_\nu(\rho, \m{U}_\nu)$, and the species temperature defined by $T_\nu^{-1} = \partial_{\m{U}_\nu} s_\nu$.   In the moment variable representation, the equations corresponding to this bracket with respect to the Hamiltonian functional on $\mathfrak{M}$, \eqref{eq:ham-functional-moments-matter-rad}, are instead
\begin{align*}
\frac{\partial \rho}{\partial t} + \nabla_\bx \cdot \bold{P} &= 0,  \\
\frac{\partial}{\partial t} \bold{P} + \nabla_\bx \cdot [\rho^{-1} \bold{P} \otimes \bold{P} ] &= - \nabla_\bx (p_e + p_r), \\
\frac{\partial}{\partial t} E_e + \nabla_\bx \cdot [ E_e \rho^{-1}\bold{P} ]  &= -p_e \nabla_\bx \cdot \rho^{-1}\bold{P},  \\
\frac{\partial}{\partial t} E_r + \nabla_\bx \cdot [ E_r \rho^{-1}\bold{P} ]  &= -p_r \nabla_\bx \cdot \rho^{-1}\bold{P},
\end{align*}
which are precisely the 2T equations \eqref{2T a}-\eqref{2T d}, ignoring the thermal flux and interaction terms.
Note that with the equation of state $s_\nu = s_\nu(\rho, E_\nu/\rho)$, the pressure is determined by the density and internal energy densities according to
\begin{align*}
p_\nu= -E_\nu -\rho\,\frac{\partial_\rho s_\nu}{\partial_{E_\nu} s_\nu}.
\end{align*}

\textbf{Dissipative bracket for the thermal flux terms.}
Now, we derive a dissipative bracket for the thermal flux terms appearing in the 2T system, \Cref{2T a}-\Cref{2T d}. Assuming a blackbody equation of state $E_r = a\,T_r^4$ and fluxes given by
\begin{gather}
\bm{F}_e = -K_e(T_e)\,\nabla_\bx T_e, \quad \bm{F}_r = -\mathcal{D}(T_e)\,\nabla_\bx E_r,\label{flux_formulas_basic}
\end{gather}
the flux terms arise from a dissipative bracket constructed as follows. 

The dissipative bracket will be defined as a symmetric positive semi-definite bilinear form on the space of functionals on $\mathfrak{M}$, where as before $\mathfrak{M}\ni (\mathbf{P},\rho,E_e,E_r)$ denotes the space of tuples comprising momentum density $\mathbf{P}$, fluid density $\rho$, and species-wise internal energy densities $E_\nu$. Assume species-wise equations of state for the entropy density, $\Sigma_\nu = \Sigma_\nu(\rho,E_\nu)$. The corresponding form of the First Law of Thermodynamics is
\begin{align}
d\Sigma_\nu = \frac{1}{T_\nu}\,dE_\nu - \frac{\mathcal{F}_\nu}{T_\nu}d\rho,\label{E_D_first_law}
\end{align}
where again $\m{F}_\nu$ denotes Gibbs free energy. \Cref{E_D_first_law} allows us to compute functional derivatives of the total entropy functional with respect to the moment variables $\bold{P}, \rho, E_e, E_r$. 

The entropy functional $\mathcal{S}_{\mathfrak{M}}: \mathfrak{M} \rightarrow\mathbb{R}$ is given by the sum of the electron and radiation entropy functionals,
\begin{align}\label{eq:entropy-radiation-matter-moments}
\mathcal{S}_{\mathfrak{M}}[\mathbf{P},\rho,E_e,E_r] & = \int_Q \Sigma_e\left(\rho,E_e\right)\,d\bx  + \int_Q \Sigma_r\left(\rho,E_r\right)\,d\bx.
\end{align}
By \Cref{E_D_first_law}, its functional derivatives are given by
\begin{equation}\label{eq:entropy-functional-derivatives}
\frac{\delta \mathcal{S}_{\mathfrak{M}}}{\delta \mathbf{P}} = 0,\quad \frac{\delta \mathcal{S}_{\mathfrak{M}}}{\delta \rho} = -\sum_\nu \frac{\mathcal{F}_\nu}{T_\nu},\quad \frac{\delta\mathcal{S}_{\mathfrak{M}}}{\delta E_e} = \frac{1}{T_e},\quad \frac{\delta\mathcal{S}_{\mathfrak{M}}}{\delta E_r} = \frac{1}{T_r}.
\end{equation}
These formulas are significant because they indicate that functional derivatives of the entropy functional coincide with the \textbf{thermodynamic forces},
\begin{gather*}
X_\rho = \partial_{\rho}\Sigma,\quad X_{E_e} = \partial_{E_e}\Sigma,\quad X_{E_r} = \partial_{E_r}\Sigma,\quad \text{ where } \Sigma := \Sigma_e + \Sigma_r,
\end{gather*}
that appear in the Onsager reciprocal relations \cite{OnsagerRelations}. The relation between the thermal fluxes $\bm{F}_\rho,\bm{F}_e,\bm{F}_r$ of the thermodynamic variables $\rho,E_r,E_r$ and the thermodynamic forces $X_\rho, X_{E_E}, X_{E_r}$ are given by the Onsager reciprocal relations
\begin{align*}
\bm{F}_\rho &= L^{\rho\rho}\,\nabla_\bx X_\rho + L^{\rho E_e}\,\nabla_\bx X_{E_e}+ L^{\rho E_r}\,\nabla_\bx X_{E_r},\\
\bm{F}_{e}& = L^{E_e\rho}\,\nabla_\bx X_\rho + L^{E_eE_e}\,\nabla_\bx X_{E_e}+ L^{E_e E_r}\,\nabla_\bx X_{E_r},\\
\bm{F}_{r}& = L^{E_r\rho}\,\nabla_\bx X_\rho + L^{E_rE_e}\,\nabla_\bx X_{E_e}+ L^{E_r E_r}\,\nabla_\bx X_{E_r},
\end{align*}
where the $3\times 3$ \textbf{Onsager matrix} $L$ is symmetric positive semi-definite. Coquinot and Morrison \cite{CoMo_2020} observed that when functional derivatives of the entropy functional relate to thermodynamic fluxes in the above manner, a valid dissipative bracket is determined completely by the Onsager matrix according to $(\m{F},\m{G})_{\textup{flux}} = \sum_{A,B \in\{\rho, E_e, E_r\}}\int \nabla_\bx (\delta \m{F}/\delta A) L^{AB} \nabla_\bx (\delta \m{G}/\delta B)\, d\bx$, where here the sum indices run over the thermodynamic variables $A,B \in \{\rho,E_e,E_r\}$. For the fluxes in the 2T model, the Onsager matrix is
\begin{gather*}
L = \begin{pmatrix} L^{\rho\rho} & L^{\rho E_e} & L^{\rho E_r} \\ L^{E_e \rho} & L^{E_e E_e} & L^{E_e E_r}\\ L^{E_r \rho} & L^{E_rE_e} & L^{E_rE_r} \end{pmatrix} = \begin{pmatrix} 0 & 0 & 0 \\ 0 & T_e^2\,K_e(T_e) & 0\\ 0& 0 & 4aT_r^5\mathcal{D}(T_e)\end{pmatrix}.
\end{gather*}
The dissipative bracket for the thermal flux terms of the 2T system is therefore
\begin{align}
(\m{F},\m{G})_{\textup{flux}} =& \int_Q \nabla_\bx\left(\frac{\delta \m{F}}{\delta E_e}\right)\cdot \nabla_\bx\left(\frac{\delta \m{G}}{\delta E_e}\right)\,T_e^2\,K_e(T_e)\,d\bx \label{eq:2T_dissipative_bracket}\\ 
& \qquad + \int_Q \nabla_\bx\left(\frac{\delta \m{F}}{\delta E_r}\right)\cdot \nabla_\bx\left(\frac{\delta \m{G}}{\delta E_r}\right)\, 4aT_r^5\mathcal{D}(T_e) \,d\bx. \nonumber
\end{align}

Now, we derive properties of this bracket. First, we note that the bracket is mass and momentum conservative, since \Cref{eq:2T_dissipative_bracket} does not involve functional derivatives with respect to $\rho$ or $\bold{P}$. Now we show that it is indeed an energy-conservative dissipative bracket. 

We will assume throughout that the temperatures defined through their equations of state are positive everywhere. For a discussion of thermodynamic equations of state, see \cite{thermo_eos}. For a discussion of temperature positivity for radiation transport, see \cite{maxprinciple}.

\begin{prop}\label{prop:2T-flux-bracket}
The 2T dissipative flux bracket \eqref{eq:2T_dissipative_bracket} has the following properties. 
\begin{enumerate}[label=(\roman*)]
\item For any functionals $\m{F}$ and $\m{G}$ on $\mathfrak{M}$, $(\m{F},\m{G})_{\textup{flux}} = (\m{G},\m{F})_{\textup{flux}}$.
\item For any functional $\m{G}$ on $\mathfrak{M}$, $(\m{G},\m{G})_{\textup{flux}}\geq 0$.
\item $(\mathcal{S}_{\mathfrak{M}},\mathcal{S}_{\mathfrak{M}})_{\textup{flux}} = 0$ if and only if $\nabla T_e = \nabla T_r = 0$.
\item For any functional $\m{G}$ on $\mathfrak{M}$, $(\m{G},\mathcal{H}_{\mathfrak{M}})_{\textup{flux}} = 0$, where $\mathcal{H}_{\mathfrak{M}}$ is the Hamiltonian functional on $\mathfrak{M}$, \Cref{eq:ham-functional-moments-matter-rad}.
\end{enumerate}
\begin{proof}
The first property is manifestly clear. The fourth property follows from the fact that the dissipation bracket does not involve functional derivatives with respect to $\mathbf{P}$ or $\rho$ and that $\delta\mathcal{H}_{\mathfrak{M}}/\delta E_i =\delta\mathcal{H}_{\mathfrak{M}}/\delta E_e=\delta\mathcal{H}/\delta E_r  = 1$. Thus, we only need to check $(ii)$ and $(iii)$.

The bracket $(\m{G},\m{G})_{\textup{flux}}$ is explicitly given by
\begin{align*}
(\m{G},\m{G})_{\textup{flux}} =& \int_Q \nabla_\bx\left(\frac{\delta \m{G}}{\delta E_e}\right)\cdot \nabla_\bx\left(\frac{\delta \m{G}}{\delta E_e}\right)\,T_e^2\,K_e(T_e)\, d\bx \\
& \qquad + \int_Q \nabla_\bx\left(\frac{\delta \m{G}}{\delta E_r}\right)\cdot \nabla_\bx\left(\frac{\delta \m{G}}{\delta E_r}\right)\, 4aT_r^5\mathcal{D}(T_e) d\bx.
\end{align*}
Since $T_r,K_i,K_e,\mathcal{D}$ and the constant $a$ are assumed positive, each of the integrands in this formula are non-negative, which establishes the second property. Now, setting $\m{G} = \mathcal{S}_{\mathfrak{M}}$ in the above equation and using the formulas for functional derivatives of the entropy functional \eqref{eq:entropy-functional-derivatives} yields
\begin{align*}
(\mathcal{S}_{\mathfrak{M}},\mathcal{S}_{\mathfrak{M}})_{\textup{flux}} & = \int_Q |\nabla_\bx T_e|^2\,\frac{K_e(T_e)}{T_e^2}\, d\bx + \int_Q |\nabla_\bx T_r|^2\,4aT_r\,\mathcal{D}(T_e)\, d\bx.
\end{align*}
Since $T_e, T_r, a, K_e(T_e), \m{D}(T_e)$ are all positive, the bracket $(\mathcal{S}_{\mathfrak{M}},\mathcal{S}_{\mathfrak{M}})_{\textup{flux}}$ vanishes if and only if the temperature gradients vanish, as claimed.
\qed \end{proof}
\end{prop}

It is straightforward to verify that the dissipative bracket $(\cdot,\cdot)_{\textup{flux}}$ reproduces the flux terms in the 2T system \eqref{2T a}-\eqref{2T d}, using an analogous argument to \Cref{ex:fluid-moments}. Consider the following linear functional of the moment variables $\mathbf{P}, \rho, E_e, E_r$,
\begin{align*}
    \m{F}_{\mathbf{a}bcd}&[\mathbf{P}, \rho, E_e, E_r ] \\
    &= \int_Q \left( \mathbf{a}(\bx) \cdot \mathbf{P}(t,\bx) + b(\bx) \rho(t,\bx) + c(\bx) E_e(t,\bx) + d(\bx) E_r(t,\bx) \right) d\bx,
\end{align*}
where the components of $\mathbf{a}$, $b$, $c$ and $d$ are fixed but arbitrary elements of $C(Q)$. We have
\begin{align*}
    (\m{F}_{\mathbf{a}bcd}, \m{S}_{\mathfrak{M}})_{\textup{flux}} &=  \int_Q \nabla_{\bx} c \cdot \nabla_{\bx}\left(\frac{1}{T_e}\right)\,T_e^2\,K_e(T_e)\, d\bx \\
        & \qquad +  \int_Q \nabla_{\bx} d \cdot \nabla_{\bx}\left(\frac{1}{T_r}\right)\, 4aT_r^5\mathcal{D}(T_e) \,d\bx \\
    &=  \int_Q c\ \underbrace{\nabla_{\bx}\cdot\bigg( K_e(T_e)\,\nabla_{\bx} T_e\bigg)}_{= - \nabla_\bx \bm{F}_e}\,d\bx \\
        & \qquad + \int_Q d\ \underbrace{\nabla_{\bx}\cdot\bigg(4aT_r^3\,\mathcal{D}(T_e)\,\nabla_{\bx} T_r\bigg)}_{= - \nabla_\bx \bm{F}_r}\,d\bx
\end{align*}
We see that the thermal flux bracket correctly produces the thermal flux terms appearing in the 2T equations \eqref{2T a}-\eqref{2T d}.

\textbf{Dissipative bracket for the thermal interaction terms.}
Finally, we state a dissipative bracket for the thermal interaction terms. Combined with the Lie--Poisson bracket for the advection terms and the dissipative flux bracket for the thermal flux terms, this yields the complete bracket description for the 2T system \eqref{2T a}-\eqref{2T d}.

As before, we take a black body equation of state, $E_r = a\,T_r^4$, and thermal interaction term given by
\begin{gather}
G_{er} = \sigma_P\,ac\,(T_e^4 - T_r^4),\label{interaction_formulas_basic}
\end{gather}
where $a$ is a positive constant and $\sigma_P$ is a positive real-valued function of $\bx, E_e, E_r$. The interaction term arises from a dissipative bracket as follows. As in the previous section, it is convenient to build the symmetric dissipation bracket $(\cdot,\cdot)$ in the energy representation $\mathfrak{M} \ni (\mathbf{P}, \rho, E_e, E_r)$. Proceeding analogously to the discussion of constructing interaction brackets for kinetic systems in \Cref{sec:interaction-brackets}, we consider a bracket of the form
$$ (\m{F},\m{G})_{\textup{therm}} = \int_Q \gamma \Delta(\m{F}) \Delta(\m{G}) d\bx. $$
The integration domain is $Q$, rather than $T^*Q$, since we are considering functionals on the moment space $\mathfrak{M}$. Note that a single integral will suffice, since we observe that the thermal interaction terms in \Cref{2T a}-\Cref{2T d} are local. Furthermore, noting that the thermal interaction terms in \Cref{2T a}-\Cref{2T d} only appear in the evolution equations for $E_e$ and $E_r$ and not in the evolution equations for $\rho$ and $\mathbf{P}$, the differential operator $\Delta$ should only involve functional derivatives with respect to $E_r$ and $E_e$; given the form of the Hamiltonian functional \eqref{eq:ham-functional-moments-matter-rad}, the simplest choice of an energy conservative differential operator satisfying this requirement is
$$ \Delta(\m{F}) := \frac{\delta \m{F}}{\delta E_e} - \frac{\delta \m{F}}{\delta E_r}. $$
It follows immediately that the bracket is also mass and momentum conservative, since $\Delta$ does not involve functional derivatives with respect to $\rho$ and $\mathbf{P}$. Finally, a direct computation shows that the appropriate choice for $\gamma$, in order to reproduce the thermal interaction terms, is $\gamma = T_eT_r G_{er}/(T_e-T_r)$, as we will verify below. This yields the thermal interaction bracket
\begin{align}
(\m{F},\m{G})_{\textup{therm}} = \int_Q \bigg(\frac{\delta \m{F}}{\delta E_e} - \frac{\delta \m{F}}{\delta E_r}\bigg) \bigg(\frac{\delta \m{G}}{\delta E_e} - \frac{\delta \m{G}}{\delta E_r}\bigg) \frac{T_e\,T_r\,G_{er}}{T_e-T_r}\, d\bx.\label{eq:2T_dissipative_coupling_bracket}
\end{align}

\begin{prop}\label{prop:2T-thermal-interaction-bracket}
The dissipative thermal interaction bracket \eqref{eq:2T_dissipative_coupling_bracket} has the following properties. 
\begin{enumerate}[label=(\roman*)]
\item For any functionals $\m{F}$ and $\m{G}$ on $\mathfrak{M}$, $(\m{F},\m{G})_{\textup{therm}} = (\m{G},\m{F})_{\textup{therm}}$.
\item For any functional $\m{G}$ on $\mathfrak{M}$, $(\m{G},\m{G})_{\textup{therm}}\geq 0$.
\item $(\mathcal{S}_{\mathfrak{M}},\mathcal{S}_{\mathfrak{M}})_{\textup{therm}} = 0$ if and only if $T_e = T_r$.
\item For any functional $\m{G}$ on $\mathfrak{M}$, $(\m{G},\mathcal{H}_{\mathfrak{M}})_{\textup{therm}} = 0$, where $\mathcal{H}_{\mathfrak{M}}$ is the Hamiltonian functional on $\mathfrak{M}$, \Cref{eq:ham-functional-moments-matter-rad}.
\end{enumerate}
\begin{proof}
The first property is clear. The fourth property is a direct consequence of the choice of differential operator $\Delta$, $\Delta(\m{H}_{\mathfrak{M}})=\delta\mathcal{H}_{\mathfrak{M}}/\delta E_e - \delta\mathcal{H}_{\mathfrak{M}}/\delta E_r = 1-1 = 0 $. So it is only necessary to demonstrate the second and third properties.

Note that the weighting factor in the integrand of \eqref{eq:2T_dissipative_coupling_bracket} may be rearranged as follows:
\begin{align*}
\frac{T_e\,T_r\,G_{er}}{T_e-T_r} & = \frac{T_e\,T_r\,\sigma_P\,ac\,(T_e^4 - T_r^4)}{T_e - T_r} \\
&= \frac{T_e\,T_r\,\sigma_P\,ac\,(T_e - T_r)(T_e + T_r)(T_e^2 + T_r^2)}{T_e - T_r} \\
&= T_e\,T_r\,\sigma_P\,ac\,(T_e + T_r)(T_e^2 + T_r^2).
\end{align*}
Thus, the weighting factor is positive. Then, $(\m{G},\m{G})_{\textup{therm}}$ is given by
\begin{align*}
(\m{G},\m{G})_{\textup{therm}} = \int_Q \bigg(\frac{\delta G}{\delta E_e} - \frac{\delta G}{\delta E_r}\bigg)^2  \frac{T_e\,T_r\,G_{er}}{T_e-T_r}\, d\bx \geq 0,
\end{align*}
which establishes the second property. In particular, setting $\m{G} = \mathcal{S}_{\mathfrak{M}}$ and using the equations for the functional derivatives of the entropy, \eqref{eq:entropy-functional-derivatives}, we have
\begin{align*}
(\mathcal{S}_{\mathfrak{M}},\mathcal{S}_{\mathfrak{M}})_{\textup{therm}} = \int_Q \bigg(\frac{1}{T_e} - \frac{1}{T_r}\bigg)^2  \frac{T_e\,T_r\,G_{er}}{T_e-T_r}\, d\bx.
\end{align*}
It follows that $(\mathcal{S}_{\mathfrak{M}},\mathcal{S}_{\mathfrak{M}})_{\textup{therm}} = 0$ if and only if either $1/T_e - 1/T_i = 0$ or $G_{er}(T_e,T_r) = 0$, which are both equivalent to $T_e = T_r$. We conclude that $(\mathcal{S},\mathcal{S})= 0$ if and only if the temperatures are equal.
\qed \end{proof}
\end{prop}

We check that the bracket \eqref{eq:2T_dissipative_coupling_bracket} produces the correct thermal interaction terms, via a similar computation to the thermal flux terms. As before, consider the following linear functional of the moment variables $\mathbf{P}, \rho, E_e, E_r$,
\begin{align*}
    \m{F}_{\mathbf{a}bcd}&[\mathbf{P}, \rho, E_e, E_r ] \\
    &= \int_Q \left( \mathbf{a}(\bx) \cdot \mathbf{P}(t,\bx) + b(\bx) \rho(t,\bx) + c(\bx) E_e(t,\bx) + d(\bx) E_r(t,\bx) \right) d\bx.
\end{align*}
We have 
\begin{align*}
   (\m{F}_{\mathbf{a}bcd}, \m{S}_{\mathfrak{M}})_{\textup{therm}} &= \int_Q c\ \underbrace{\bigg(\frac{1}{T_e} - \frac{1}{T_r}\bigg) \frac{T_e\,T_r\,G_{er}}{T_e-T_r}}_{= - G_{er}}\, d\bx \\
   & \qquad + \int_Q d\ \underbrace{\left[-\bigg(\frac{1}{T_e} - \frac{1}{T_r}\bigg) \frac{T_e\,T_r\,G_{er}}{T_e-T_r}\right]}_{= G_{er}} \, d\bx
\end{align*}
We see that the thermal interaction bracket correctly produces the thermal interaction terms appearing in \eqref{2T a}-\eqref{2T d}.

\textbf{The complete bracket description of the 2T system.}
Combining the previous discussion together, we see that the two-temperature moment system \eqref{2T a}-\eqref{2T d} can be expressed as
\begin{equation}\label{eq:2T-double-bracket}
    \frac{\partial}{\partial t} \m{F} = \db{\m{F}}{\m{H}_{\mathfrak{M}}}_{\textup{2T}} + (\m{F},\m{S}_{\mathfrak{M}})_{\textup{2T}},
\end{equation}
where $\m{F}$ is a functional on $\mathfrak{M} \ni (\mathbf{P},D,E_e,E_r)$, $\db{\cdot}{\cdot}_{\textup{2T}}$ is the 2T advection bracket \eqref{eq:2T-advection-bracket} and we have defined the complete 2T dissipation bracket as the sum of the thermal flux and thermal interaction brackets,
$$ (\cdot,\cdot)_{\textup{2T}} :=  (\cdot,\cdot)_{\textup{flux}} +  (\cdot,\cdot)_{\textup{therm}}. $$
For the complete 2T dissipation bracket, analogous properties of $(i)$, $(ii)$ and $(iv)$ in \Cref{prop:2T-flux-bracket} and \Cref{prop:2T-thermal-interaction-bracket} hold.  Properties $(iii)$ of both propositions combine to yield 
$$ \frac{\partial\m{S}}{\partial t} = (\m{S}_{\mathfrak{M}}, \m{S}_{\mathfrak{M}})_{\textup{2T}} = 0 $$
if and only if $\nabla T_r = \nabla T_e = 0$ and $T_i = T_r$. That is, the total entropy does not increase if and only if both the electron and radiation temperatures are constant on $Q$ and equal to each other, i.e., the system is in thermodynamic equilibrium. Note also that the complete dissipation bracket is mass and momentum conservative, as the dissipative flux and dissipative thermal interaction brackets do not involve functional derivatives with respect to $\mathbf{P}$ or $\rho$.
\end{example}

\Cref{sec:diffusion} is an example of a pair bracket formulation on a space of truncated moments, which showed how to use physical principles, particularly thermodynamical principles, to construct the ingredients which defined the pair bracket formulation. However, as previously mentioned, a pair bracket formulation on a space of truncated moments need not be related to any underlying kinetic system. One would expect such a relation, in order to interpret the dynamics on the space of truncated moments as an approximation of the dynamics of a kinetic system. As such, we will now propose the following notion of geometric moment closure.

\subsection{Geometric moment closure}\label{sec:geometric-moment-closures}
Now, we consider the moment closure problem \cite{Lev1996, Grad1949}; for a survey of recent advances on this problem, see \cite{Ted2023}. For additional discussions on geometric aspects of moment systems and moment closures, see \cite{GHT08, GHT082, HT09, Scovel_Weinstein_1994, Burby2023}. Instead of solving for the full phase space distribution, a common approach for approximately solving a kinetic system is to solve the moment equations after truncating at some finite number of moments \cite{Lev1996}, say $m$. This leads to the moment closure problem: to obtain a closed set of equations for $M^0,\dots,M^m$, one must specify $M^{m+1}$ as a function of $M^0,\dots,M^m$; the choice of such a function is known as a moment closure \cite{Lev1996, Grad1949}. See also \cite{GaHa2013, CaiFan2013, CaiFan2014, All2019, Abdel2023, Ted2023, Burby2023} for theoretical developments, examples and applications of kinetic moment systems and their closures. Additionally, to incorporate interactions, one should add a dissipative term to the right hand side, depending on $M^0,\dots,M^m$.

We provide a geometric formulation of the moment closure problem, based on a pair bracket approach. By approaching the moment closure problem from this geometric perspective, any such geometric moment closure is automatically energy conservative and entropy dissipative; furthermore, the hyperbolicity of such systems could be studied via the Hamiltonian dynamics of the system, in the limit when dissipation vanishes. 

As discussed previously, the choice of the Hamiltonian functional, entropy functional, and pair brackets determine the nature of the evolution equations for a system. This is similar to the moment closure problem, where the choices of moment closure and interaction term, both depending on $M^0,\dots,M^m$, determine the nature of the evolution equations for the truncated kinetic moment system. In the previous section, \Cref{sec:interaction-moment-spaces}, we discussed pair bracket formulations on a space of truncated moments $\mathfrak{M}$; in principle, such a pair bracket formulation need not be related in any way to an underlying kinetic system. In order to interpret such a pair bracket formulation on a space of truncated moments as a moment closure of a kinetic system, the dynamics on the space of truncated moments should be naturally related to the dynamics of the kinetic system.
As such, we posit the following definition of a geometric moment closure.

\begin{definition}[Geometric Moment Closure]\label{def:geometric-moment-closure}
    Consider a kinetic system of the form \eqref{eq:dissipative-bracket}, i.e.,
    \begin{equation}\label{eq:kinetic-evo-def}
    \frac{\partial}{\partial t} \mathcal{F}[g] = \db{\mathcal{F}[g]}{\mathcal{H}[g]} + (\mathcal{F}[g],\mathcal{S}[g]),
    \end{equation}
    where $\m{H}$ and $\m{S}$ are the kinetic Hamiltonian and kinetic entropy functionals. Furthermore, consider a pair bracket formulation on a total truncated moment space $\mathfrak{M} \ni \vec{G}$ \eqref{eq:total-truncated-moment-space},
    \begin{equation}\label{eq:moment-evo-def}
    \frac{\partial}{\partial t} \m{G}[\vec{G}] = \{\m{G}[\vec{G}], \m{H}_{\mathfrak{M}} [\vec{G}]\}_{\mathfrak{M}} + (\m{G}[\vec{G}], \m{S}_{\mathfrak{M}} [\vec{G}])_{\mathfrak{M}},
    \end{equation}
    where $\m{H}_{\mathfrak{M}}$ and $\m{S}_{\mathfrak{M}}$ are the moment Hamiltonian and moment entropy functionals.

    We say that the moment system \eqref{eq:moment-evo-def} is a \textbf{geometric moment closure} of the kinetic system \eqref{eq:kinetic-evo-def} if there exists a mapping
     \begin{equation}\label{eq:gamma-map-definition}
         \Gamma_{\mathfrak{M}} : \mathfrak{M} \rightarrow C^*(T^*Q)
     \end{equation}
     such that 
     \begin{subequations}\label{eq:conditions-gmc}
         \begin{align}
             \m{H} \circ \Gamma_{\mathfrak{M}} &= \m{H}_{\mathfrak{M}}, \\
             \m{S} \circ \Gamma_{\mathfrak{M}} &= \m{S}_{\mathfrak{M}}, \\
             \{\m{F}_1, \m{F}_2\} \circ \Gamma_{\mathfrak{M}} &= \{ \m{F}_1 \circ \Gamma_{\mathfrak{M}}, \m{F}_2 \circ \Gamma_{\mathfrak{M}} \}_{\mathfrak{M}}, \\
             (\m{F}_1, \m{F}_2) \circ \Gamma_{\mathfrak{M}} &= (\m{F}_1 \circ \Gamma_{\mathfrak{M}}, \m{F}_2 \circ \Gamma_{\mathfrak{M}} )_{\mathfrak{M}},
         \end{align}
     \end{subequations}
     for all functionals $\m{F}_1$, $\m{F}_2$ on $C^*(T^*Q)$. That is, the moment Hamiltonian and moment entropy functionals are given by the pullback of the kinetic Hamiltonian and kinetic entropy functionals by $\Gamma_{\mathfrak{M}}$ and, furthermore, the brackets are appropriately preserved by $\Gamma_{\mathfrak{M}}$.

     When the dissipative brackets are identically zero, we say that the geometric moment closure is \textbf{pure}. In such a case, the conditions \eqref{eq:conditions-gmc} simplify to 
        \begin{subequations}\label{eq:conditions-gmc-pure}
         \begin{align}
             \m{H} \circ \Gamma_{\mathfrak{M}} &= \m{H}_{\mathfrak{M}}, \\
             \{\m{F}_1, \m{F}_2\} \circ \Gamma_{\mathfrak{M}} &= \{ \m{F}_1 \circ \Gamma_{\mathfrak{M}}, \m{F}_2 \circ \Gamma_{\mathfrak{M}} \}_{\mathfrak{M}}. 
         \end{align}
     \end{subequations}
\end{definition}

\begin{remark}
When considering multiple interacting kinetic theories (e.g., kinetic matter-radiation interaction), there are multiple distributions and thus, one introduces a total truncated moment space for each distribution and we posit an analogous definition of a geometric moment closure for multiple interacting kinetic theories with the obvious modifications.
\end{remark}

Compared to the freedom in choosing a dissipative bracket $({\cdot},{\cdot})$, an entropy functional $\m{S}$ at the level of the kinetic theory as discussed in \Cref{sec:interaction-brackets} and a classical moment closure (i.e., $M^{n+1}$ as a function of $M^0,\dots,M^n$), this notion of geometric moment closure is more constrained than the classical notion of moment closure, in the sense that a classical moment closure does not in principle need to admit a pair bracket bracket formulation which arises from the pullback of the kinetic pair brackets in the above way. In particular, such a geometric moment closure is guaranteed to be energy conservative and entropy dissipative, whereas not all classical moment closures need to be. This requirement of maintaining a pair bracket formulation is a geometric structure-preserving perspective on the moment closure problem. 

\begin{remark}
    While a pair bracket formulation for the notion of geometric moment closure is sufficient for our purposes, one can in principle utilize other geometric formulations to define a notion of geometric moment closure. Some possible alternatives include port-Hamiltonian systems (see \cite{vdS_2014} for an overview of port-Hamiltonian systems; see \cite{rashad_2021a, rashad_2021b} for port-Hamiltonian formulations of ideal fluid flow); and in particular, irreversible port-Hamiltonian systems \cite{ramirez_2012,ramirez_2022} (see \cite{pHradhydro} for an irreversible port-Hamiltonian formulation of diffusion radiation hydrodynamics), as well as, more recently, metriplectic four-bracket formulations \cite{Morrison2024} which generalize the metriplectic formulation.
\end{remark}

In \Cref{sec:conclusion} (Conclusion), we provide some interesting possible research directions regarding geometric moment closures in light of recent developments in data-driven moment closure models \cite{sadr2021,huang2022,Por2023, Dona2023, huang2023} and learning of pair bracket formulations via deep graph neural networks \cite{gruber24}. 

As a final application, we will apply the variable moment closure framework of \cite{Burby2023} to derive geometric moment closures for pure transport with arbitrary Hamiltonian $H \in C(T^*Q)$.

\subsection{Variable moment closures for pure transport with arbitrary Hamiltonian}\label{sec:arbitrary-transport-closure}
Here, we will consider kinetic transport and use the variable moment closure framework of \cite{Burby2023} to construct pure geometric moment closures for the transport equation, with arbitrary Hamiltonian. As an example of this general construction, we present novel closures for pure radiation transport in \Cref{example:pure-rad-closures}.  Although we will sketch the idea of the variable moment closure framework here, we refer the reader to \cite{Burby2023} for a more precise and in-depth explanation. Note that the variable moment closure framework applies to any Hamiltonian functional on the space of phase space distributions. As an example, \citet{Burby2023} particularly considers the Hamiltonian functional for the Vlasov--Poisson system to derive degree one, two and three fluid closures for its corresponding moment system. We will generalize the results of \cite{Burby2023} by deriving moment closures for pure transport with arbitrary Hamiltonian. We will derive the degree one and two closures explicitly and show how it can be done in the higher degree case. As a concrete example, we will write the degree one and degree two variable moment closures for pure radiation transport explicitly. Finally, in \Cref{sec:wkb-relation}, we discuss how the variable moment closure framework can be interpreted as an asymptotic expansion.

In \cite{Burby2023}, a Poisson bracket $\db{\cdot}{\cdot}^m_0$ is defined on an auxiliary truncated moment space, defined as the Cartesian product of the space of truncated moments $\mathfrak{M}^m_0$ of degree $m \geq 0$ with a space of auxiliary fields $C(Q)$. Subsequently, \citet{Burby2023} shows that there exists a Poisson mapping 
$$\Gamma_m: \mathfrak{M}^m_0 \times C(Q) \rightarrow C^*(T^*Q)$$ 
from the auxiliary truncated moment space into the space of phase space functionals $C^*(T^*Q)$, equipped with the Lie--Poisson bracket. The map $\Gamma_m$ can be interpreted as a \textit{collectivization} or \textit{Clebsch} map \cite{ClebschCanon, Scovel_Weinstein_1994, Burby2023}. The Lie--Poisson dynamics on the space of phase space functionals with respect to a Hamiltonian functional $\m{H}$ then pull back to dynamics on $\mathfrak{M}^m_0 \times C(Q)$ with respect to the \textit{collective Hamiltonian} of degree $m$,
$$ \m{H}^m := \Gamma_m^* \m{H} = \m{H} \circ \Gamma_m, $$
i.e., $\m{H}^m$ is the pullback of $\m{H}$ by $\Gamma_m$, where $\Gamma_m$ is a \textit{collectivization} or \textit{Clebsch} map \cite{ClebschCanon, Scovel_Weinstein_1994, Burby2023}. Thus, for a functional $\m{F}$ on $\mathfrak{M}^m_0 \times C(Q)$, its dynamics are given by
$$ \frac{\partial}{\partial t} \m{F} = \db{\m{F}}{\m{H}^m}^m_0. $$
This defines a moment closure for an arbitrary truncation degree, through the use of an auxiliary field \cite{Burby2023}.

Recall the general setup for a kinetic theory discussed in \Cref{sec:kinetic-theory}. Let $H \in C(T^*Q)$ be an arbitrary Hamiltonian and let $g$ denote the time-dependent phase space distribution, i.e., $g(t) \in C^*(T^*Q)$. The dynamics of functionals of $g$ are given by \eqref{eq:lie-poisson-evolution} where the Hamiltonian functional is given by
\begin{equation}\label{eq:hamiltonian-functional-variable-sec}
    \m{H}[g] = \int_{T^*Q} H(\bx,\bp) g(t,\bx,\bp) d\bp d\bx.
\end{equation}
We will derive geometric moment closures for the corresponding transport equation on $T^*Q$ given by 
$$ \frac{\partial}{\partial t}g + \nabla_\bx \cdot (g\nabla_\bp H) - \nabla_\bp\cdot(g \nabla_\bx H) = 0.$$
As we will see, this moment closure procedure replaces this transport equation on $T^*Q$ with a system of $m+1$ transport equations on $Q$ coupled with an auxiliary scalar field evolving under the Hamilton--Jacobi equation, where $m$ is the truncation degree.

\textbf{Degree one closure ($m=0$).} Consider the auxiliary space of truncated moments of degree zero, $\mathfrak{M}^0_0 \times C(Q) \ni (M^0, \phi)$. The Poisson bracket $\db{\cdot}{\cdot}^0_0$ between functionals on $\mathfrak{M}^0_0 \times C(Q)$ and the Poisson map $\Gamma_0: \mathfrak{M}^0_0 \times C(Q) \rightarrow C^*(T^*Q)$ are given by, respectively,
\begin{subequations}\label{eq:bracket-map-zero}
\begin{align}
    \db{\m{F}}{\m{G}}^0_0 &= \int_Q \left( \frac{\delta \m{F}}{\delta \phi}\frac{\delta \m{G}}{\delta M^0} - \frac{\delta \m{G}}{\delta \phi} \frac{\delta \m{F}}{\delta M^0} \right) d\bx, \\
    \Gamma_0(M^0, \phi)(\bx,\bp) &= M^0(\bx) \delta(\mathbf{p} + \nabla_\bx \phi(\bx)).
\end{align}
\end{subequations} 
\begin{remark}
    That $\Gamma_0$ \eqref{eq:bracket-map-zero}, and similarly $\Gamma_1$ \eqref{eq:bracket-map-one} discussed in the $m=1$ case below, are Poisson maps is proven in Theorem 3 of \cite{Burby2023} using algebraic methods; particularly, using a modification of the techniques developed in \cite{Scovel_Weinstein_1994}. For intuition, we will show this by a direct calculation for the special case of linear functionals. Namely, let 
    $$\m{F}_{i}[g] := \int_{T^*Q} \varphi_i(\bx,\bp) g(\bx,\bp) d\bp d\bx,$$
    denote a linear functional where $\varphi_i$ is a fixed but arbitrary element of $C(T^*Q)$, $i=1,2$. We will verify that
    $$ \{ \m{F}_1, \m{F}_2 \} \circ \Gamma_0 = \{ \m{F}_1 \circ \Gamma_0, \m{F}_2 \circ \Gamma_0 \}^0_0, $$
    where the bracket on the left hand side is again the Lie--Poisson bracket \eqref{eq:lie-poisson-bracket}. Note that
    \begin{align*}
    \m{F}_{i} \circ \Gamma_0(M^0, \phi) &= \int_{T^*Q} \varphi_i(\bx,\bp) M^0(\bx) \delta(\bp + \nabla_\bx \phi(\bx)) d\bp d\bx \\
    &= \int_{Q} \varphi_i(\bx,-\nabla_\bx \phi(\bx)) M^0(\bx) d\bx. 
    \end{align*}
    Then, a direct calculation yields
    \begin{align*}
        \{\m{F}_1 &\circ \Gamma_0, \m{F}_2 \circ \Gamma_0\}^0_0 = \int_Q \left( \frac{\delta (\m{F}_1 \circ \Gamma_0)}{\delta \phi}\frac{\delta (\m{F}_2 \circ \Gamma_0)}{\delta M^0} - \frac{\delta (\m{F}_2 \circ \Gamma_0)}{\delta \phi} \frac{\delta (\m{F}_1 \circ \Gamma_0)}{\delta M^0} \right) d\bx \\
        &= \int_Q \Big[ \nabla_\bx \cdot \left(M^0 \nabla_\bp \varphi_1\Big|_{(\bx,-\nabla_\bx\phi)} \right) \varphi_2\Big|_{(\bx,-\nabla_\bx\phi)} \\
        & \qquad \qquad \qquad - \nabla_\bx \cdot \left(M^0 \nabla_\bp \varphi_2\Big|_{(\bx,-\nabla_\bx\phi)} \right) \varphi_1\Big|_{(\bx,-\nabla_\bx\phi)} \Big] d\bx \\
        &= \int_Q M^0 \Big( \nabla_\bx \varphi_1 \cdot \nabla_\bp \varphi_2 - \nabla_\bx \varphi_2 \cdot \nabla_\bp \varphi_1 \Big)\Big|_{(\bx,-\nabla_\bx\phi)} d\bx \\
        &= \int_Q M^0 [\varphi_1,\varphi_2]\Big|_{(\bx,-\nabla_\bx\phi)} d\bx = \int_Q M^0 \left[\frac{\delta \m{F}_1}{\delta g},\frac{\delta \m{F}_2}{\delta g}\right]\Big|_{(\bx,-\nabla_\bx\phi)} d\bx \\
        &= \int_{T^*Q} M^0 \delta(\bp + \nabla_\bx\phi) \left[\frac{\delta \m{F}_1}{\delta g},\frac{\delta \m{F}_2}{\delta g}\right] d\bp d\bx = \{\m{F}_1,\m{F}_2\} \circ \Gamma_0,
    \end{align*}
    where, in the third equality, we used integration by parts (noting that the boundary terms vanish since $M^0 \in \mathfrak{M}^0_0 = C^*(Q)$). A similar calculation holds for the case $m=1$ discussed below.
\end{remark}

Now, pulling back the above Hamiltonian functional \eqref{eq:hamiltonian-functional-variable-sec} by $\Gamma_0$ defines a Hamiltonian functional on $\mathfrak{M}^0_0 \times C(Q)$. Explicitly, we have
\begin{align}
    \m{H}^0[M^0,\phi] &= \m{H}[ \Gamma_0(M^0, \Phi) ] = \int_{T^*Q} H(\bx,\bp) M^0(\bx) \delta(\bp + \nabla_\bx \phi(\bx)) d\bp d\bx \\
    &= \int_Q H(\bx,-\nabla_\bx\phi (\bx)) M^0(\bx) d\bx. \nonumber
\end{align}
The evolution of a functional $\m{F}[M^0,\phi]$ is given by 
$$ \frac{\partial}{\partial t} \m{F} = \db{\m{F}}{\m{H}^0}^0_0 = \int_Q \left( \frac{\delta \m{F}}{\delta \phi}\frac{\delta \m{H}^0}{\delta M^0} - \frac{\delta \m{H}^0}{\delta \phi} \frac{\delta \m{F}}{\delta M^0} \right) d\bx. $$
We thus compute the functional derivatives of $\m{H}^0$ with respect to $M^0$ and $\phi$.
\begin{align*}
    \frac{\delta \m{H}^0}{\delta M^0} &= H(\bx, -\nabla_\bx\phi), \\
    \frac{\delta \m{H}^0}{\delta \phi} &= \nabla_\bx \cdot \left( M^0 \nabla_\bp H \Big|_{\bp = -\nabla_\bx\phi} \right).
\end{align*}
This yields the evolution equations for $M^0(t,\bx)$ and $\phi(t,\bx)$,
\begin{subequations}\label{eq:evolution-degree-one}
\begin{align}
    \frac{\partial}{\partial t} M^0 &= - \nabla_\bx \cdot \left( M^0 \nabla_\bp H \Big|_{\bp = -\nabla_\bx\phi} \right), \label{eq:M0-evolution-degree-one} \\
    \frac{\partial}{\partial t} \phi &= H(\bx, -\nabla_\bx \phi). \label{eq:phi-evolution-degree-one}
\end{align}
\end{subequations}
The system consists of a conservation law \eqref{eq:M0-evolution-degree-one} coupled to the Hamilton--Jacobi equation \eqref{eq:phi-evolution-degree-one} (for a discussion of the Hamilton--Jacobi equations, see, e.g., \cite{EvansPDE}). This is a pure geometric moment closure for transport $\partial \m{F}/\partial t = \db{\m{F}}{\m{H}}$, replacing the Lie--Poisson bracket with the truncated bracket $\db{\cdot}{\cdot}^0_0$ as well as pulling back the original Hamiltonian functional $\m{H}$ to $\m{H}^0$. The system \eqref{eq:evolution-degree-one} can also be written in the traditional form of a moment closure (see, e.g., Section 4 of \cite{Lev1996}),
\begin{align}
    \frac{\partial}{\partial t} M^0 &= - \nabla_\bx \cdot \mathbf{M}^1, \label{eq:M0-evolution-degree-one-closure} 
\end{align}
where the closure for the degree one moment is given by defining $\mathbf{M}^1$ to be the quantity appearing inside the divergence in \eqref{eq:M0-evolution-degree-one}, together with the Hamilton--Jacobi equation for the auxiliary field, i.e.,
\begin{equation}\label{eq:degree-one-closure}
    \mathbf{M}^1 := M^0 \nabla_\bp H \Big|_{\bp = -\nabla_\bx\phi}, \quad    \frac{\partial}{\partial t} \phi = H(\bx, -\nabla_\bx \phi).
\end{equation}
This defines a degree one closure for a kinetic system with any particle Hamiltonian $H \in C(T^*Q)$. We will discuss later the interpretation of these variable moment closures in \Cref{sec:wkb-relation}.

Before we move on to higher degree closures, let us note that the definition of $\mathbf{M}^1$ is not satisfactory in the following sense: we defined $\mathbf{M}^1$ to be the quantity appearing inside the divergence \eqref{eq:M0-evolution-degree-one}; however, this definition is ambiguous, since one can shift $\mathbf{M}^1$ by any element of the kernel of the divergence operator and this shifted $\mathbf{M}^1$ would still satisfy \eqref{eq:M0-evolution-degree-one-closure}. We would like to avoid such an ambiguity, since $\mathbf{M}^1$ represents a closure approximation of a physical quantity; namely, the first kinetic moment (see \Cref{sec:moment-equations}), which we denote here as
$$ \mathbf{M}^1[g](t,\bx) := \int_{T^*_\bx Q} (\nabla_\bp H(\bx,\bp)) g(t,\bx,\bp)\, d\bp, $$
to emphasize the dependence of the first kinetic moment on $g$. This ambiguity has a straightforward resolution: the mapping $\Gamma_0$ provides a map from $(M^0,\phi)$ to a phase space distribution; subsequently, we can compute the first kinetic moment with respect to this phase space distribution. That is, we have
\begin{align*}
    \mathbf{M}^1[\Gamma_0(M^0,\Phi)](t,\bx) &= \int_{T^*_\bx Q} (\nabla_\bp H(\bx,\bp)) \Gamma_0(M^0, \phi)\, d\bp \\
        &= \int_{T^*_\bx Q} (\nabla_\bp H(\bx,\bp)) M^0(t,\bx) \delta(\bp + \nabla_\bx\phi(t,\bx)) d\bp  \\
        &= M^0(t,\bx) \nabla_\bp H(\bx,\bp)\Big|_{\bp = -\nabla_\bx \phi(t,\bx)}.
\end{align*}
We see that this is precisely the choice that we made for $\mathbf{M}^1$ in \Cref{eq:degree-one-closure}. This resolves the ambiguity; namely, we define the degree one closure to be
$$ \mathbf{M}^1 = \mathbf{M}^1[\Gamma_0(M^0,\Phi)]. $$

\textbf{Degree two closure ($m=1$).} Consider the auxiliary space of truncated moments of degree one,
$\mathfrak{M}^1_0 \times C(Q) \ni (M^0, \mathbf{P}_0, \phi)$. The Poisson bracket between functionals on $\mathfrak{M}^1_0 \times C(Q)$ and the Poisson map $\Gamma_1: \mathfrak{M}^1_0 \times C(Q) \rightarrow C^*(T^*Q)$ are given by, respectively,
\begin{subequations}\label{eq:bracket-map-one}
\begin{align}
    &\db{\m{F}}{\m{G}}^1_0 = \db{\m{F}}{\m{G}}^0_0 - \int_Q \mathbf{P}_0 \cdot \left( \frac{\delta \m{F}}{\delta \mathbf{P}_0} \cdot \nabla \frac{\delta \m{G}}{\delta \mathbf{P}_0} - \frac{\delta \m{G}}{\delta \mathbf{P}_0} \cdot \nabla \frac{\delta \m{F}}{\delta \mathbf{P}_0} \right) d\bx, \label{eq:bracket-map-one-a} \\
    &\Gamma_1(M^0, \mathbf{P}_0, \phi)(\bx,\bp) = M^0(\bx) \delta(\bp + \nabla_\bx \phi (\bx)) - \mathbf{P}_0(\bx) \cdot \nabla_\bp \delta(\bp + \nabla_\bx \phi (\bx)), \label{eq:bracket-map-one-b}
\end{align}
\end{subequations}
where again, $M^0 \in \mathfrak{M}_0$ denotes the zeroth kinetic moment and $\phi \in C(Q)$ denotes the auxiliary field. Here, $\mathbf{P}_0 \in \mathfrak{M}_1$ can be thought of as a perturbation away from the center of the phase space distribution, where the center of the phase space distribution is specified through the auxiliary field, $\bp = - \nabla_\bx \phi(\bx)$, arising from the Dirac delta distribution appearing above. The first kinetic moment $\bold{M}^1 \in \mathfrak{M}_1$ will then be a function of $M^0, \mathbf{P}_0$ and $\phi$, as we will see below.

Pulling back the Hamiltonian functional \eqref{eq:hamiltonian-functional-variable-sec} by $\Gamma_1$ defines a Hamiltonian functional on this space.
\begin{align}
    \m{H}^1[M^0, \mathbf{P}_0, \phi] &= \m{H}[\Gamma_1(M^0, \mathbf{P}_0, \phi)] \\
        &= \int_{T^*Q} H(\bx,\bp)M^0(\bx)\delta(\bp + \nabla_\bx \phi(\bx)) d\bp d\bx \nonumber \\
            & \qquad - \int_{T^*Q} H(\bx,\bp) \mathbf{P}_0 \cdot  \nabla_\bp \delta(\bp + \nabla_\bx\phi(\bx)) d\bp d\bx \nonumber \\
        &= \int_Q H(\bx, - \nabla_\bx\phi(\bx)) M^0(\bx) d\bx + \int_Q \mathbf{P}_0 \cdot \nabla_\bp H \Big|_{\bp = -\nabla_\bx \phi} d\bx.
\end{align}
The evolution of a functional $\m{F}[M^0, \mathbf{P}_0, \phi]$ is given by 
$$ \frac{\partial}{\partial t} \m{F} = \db{\m{F}}{\m{H}^1}^1_0, $$
where the bracket $\db{\cdot}{\cdot}^1_0$ is given by \eqref{eq:bracket-map-one-b}. We compute the functional derivatives of $\m{H}^1$ with respect to $M^0$, $\mathbf{P}_0$, and $\phi$.
\begin{align*}
    \frac{\delta \m{H}^1}{\delta M^0} &= H(\bx, -\nabla_\bx\phi), \\
    \frac{\delta \m{H}^1}{\delta \mathbf{P}_0} &= \nabla_\bp H \Big|_{\bp = - \nabla_\bx \phi}, \\
    \frac{\delta \m{H}^1}{\delta \phi} &= \nabla_\bx \cdot \left( M^0 \nabla_\bp H \Big|_{\bp = -\nabla_\bx\phi} \right) + \nabla_\bx \cdot \left( \mathbf{P}_0 : \nabla_\bp \nabla_\bp H \Big|_{\bp = -\nabla_\bx\phi} \right),
\end{align*}
where $:$ denotes tensor contraction over all indices. This yields the evolution equations for $M^0(t,\bx)$, $\mathbf{P}_0(t,\bx)$, and $\phi(t,\bx)$,
\begin{subequations}\label{eq:evolution-degree-two}
\begin{align}
    \frac{\partial}{\partial t} M^0 &= - \nabla_\bx \cdot \left( M^0 \nabla_\bp H \Big|_{\bp = -\nabla_\bx\phi} + \mathbf{P}_0 : \nabla_\bp \nabla_\bp H \Big|_{\bp = -\nabla_\bx\phi} \right), \label{eq:M0-evolution-degree-two} \\
    \frac{\partial}{\partial t} \mathbf{P}_0 &= - \mathbf{P}_0 \cdot \nabla_\bx \left(\nabla_\bp H \Big|_{p = -\nabla_\bx\phi} \right) - \mathbf{P}_0 \nabla_\bx \cdot  \left(\nabla_\bp H \Big|_{p = -\nabla_\bx\phi} \right) \label{eq:P0-evolution-degree-two} \\
    & \qquad -  \left(\nabla_\bp H \Big|_{p = -\nabla_\bx\phi} \right) \cdot \nabla_\bx \mathbf{P}_0,    \nonumber \\
    \frac{\partial}{\partial t} \phi &= H(\bx, -\nabla_\bx \phi). \label{eq:phi-evolution-degree-two}
\end{align}
\end{subequations}
Equation \eqref{eq:evolution-degree-two} provides a geometric moment closure for pure transport of degree two. Let us write the system in the traditional form of a moment closure,
\begin{subequations}\label{eq:evolution-degree-two-closure}
\begin{align}
        \frac{\partial}{\partial t} M^0 &= - \nabla_\bx \cdot \mathbf{M}^1, \label{eq:M0-evolution-degree-two-closure}  \\
        \frac{\partial}{\partial t} \mathbf{M}^1 &= - \nabla_\bx \cdot \mathbb{M}^2. \label{eq:M1-evolution-degree-two-closure} 
\end{align}
\end{subequations}
Here, $\mathbf{M}^1$ is the moment variable corresponding to the first kinetic moment, which is defined analogously to the degree one case discussed previously. Namely, we define $\mathbf{M}^1$ to be the first kinetic moment of the phase space distribution given by the image of $(M^0,\mathbf{P}_0,\phi)$ under $\Gamma_1$, i.e.,
\begin{align}
    \mathbf{M}^1(t,\bx) &= \mathbf{M}^1[\Gamma_1(M^0, \mathbf{P}_0, \phi)](t,\bx) =  \int_{T^*_\bx Q} (\nabla_\bp H(\bx,\bp)) \Gamma_1(M^0, \mathbf{P}_0, \phi)\, d\bp \label{eq:M1-degree-two-explicit} \\
    &= \int_{T^*_\bx Q} (\nabla_\bp H(\bx,\bp)) M^0(t,\bx) \delta(\bp + \nabla_\bx \phi (t,\bx)) d\bp \nonumber \\
    & \quad - \int_{T^*_\bx Q} (\nabla_\bp H(\bx,\bp)) \mathbf{P}_0(t,\bx) \cdot \nabla_\bp \delta(\bp + \nabla_\bx \phi (t,\bx))\, d\bp \nonumber \\
    &= M^0(t,\bx) \nabla_\bp H(\bx,\bp)\Big|_{\bp = -\nabla_\bx\phi} + \mathbf{P}_0(t,\bx) : \nabla_\bp \nabla_\bp H(\bx,\bp)\Big|_{\bp = -\nabla_\bx\phi}, \nonumber
\end{align}
As expected, this coincides with the quantity appearing in the divergence of equation \eqref{eq:M0-evolution-degree-two}, which confirms \eqref{eq:M0-evolution-degree-two-closure}. It is also interesting to note that the above equation for $\mathbf{M}^1$ is different than the equation for $\mathbf{M}^1$ in the case of the degree one closure; we will return to this point in \Cref{sec:wkb-relation}.

Now, $\mathbb{M}^2$ represents the closure for the second kinetic moment. By computing explicitly the expression $\partial \mathbf{M}^1 /\partial t$ and expressing it in divergence form, one can obtain the expression for the degree two closure
\begin{align}\label{eq:degree-two-closure-explicit}
    \mathbb{M}^2 &= -M^0 \left( \nabla_\bp H \Big|_{p = -\nabla_\bx\phi} \otimes \nabla_\bp H \Big|_{p = -\nabla_\bx\phi} \right) \\
    &\qquad + \mathbf{M}^1 \otimes \nabla_\bp H \Big|_{p = -\nabla_\bx\phi} + \nabla_\bp H \Big|_{p = -\nabla_\bx\phi} \otimes \mathbf{M}^1. \nonumber
\end{align}
It is straightforward to verify that the above choice of $\mathbf{M}^1$ and $\mathbb{M}^2$ satisfy \eqref{eq:M1-evolution-degree-two-closure}. \Cref{eq:degree-two-closure-explicit}, together with the Hamilton--Jacobi equation for $\phi$ \eqref{eq:phi-evolution-degree-two}, close the system \eqref{eq:M0-evolution-degree-two-closure}-\eqref{eq:M1-evolution-degree-two-closure}.

Alternatively, we will derive this closure via an analogous construction to the first kinetic moment. Namely, we take $\mathbb{M}^2$ to be the second kinetic moment of the phase space distribution given by the image of $(M^0,\mathbf{P}_0,\phi)$ under $\Gamma_1$. Let us denote the second kinetic moment of a phase space distribution $g$ as
$$ \mathbb{M}^2[g](t,\bx) := \int_{T^*_\bx Q} (\nabla_\bp H(\bx,\bp) \otimes \nabla_\bp H(\bx,\bp)) g(t,\bx,\bp)\, d\bp. $$
Then, we define the degree two closure to be 
\begin{align*}
    \mathbb{M}^2(t,\bx) := \mathbb{M}^2[\Gamma_1(M^0, \mathbf{P}_0, \phi)](t,\bx)
\end{align*}
Explicitly, this is (omitting the arguments $(t,\bx,\bp)$ for brevity in the calculation)
\begin{align*}
    \mathbb{M}^2(t,\bx) &= \mathbb{M}^2[\Gamma_1(M^0, \mathbf{P}_0, \phi)](t,\bx) = \int_{T^*_\bx Q} (\nabla_\bp H \otimes \nabla_\bp H) \Gamma_1(M^0, \mathbf{P}_0, \phi)\, d\bp \\
    &=  \int_{T^*_\bx Q} (\nabla_\bp H \otimes \nabla_\bp H) ( M^0 \delta(\bp + \nabla_\bx \phi) - \mathbf{P}_0 \cdot \nabla_\bp \delta(\bp + \nabla_\bx \phi) )\, d\bp \\
    &= M^0 \left( \nabla_\bp H \Big|_{p = -\nabla_\bx\phi} \otimes \nabla_\bp H \Big|_{p = -\nabla_\bx\phi} \right) \\
    & \qquad + \left(\mathbf{P}_0 : \nabla_\bp \nabla_\bp H \Big|_{p = -\nabla_\bx\phi}\right) \otimes \nabla_\bp  H \Big|_{p = -\nabla_\bx\phi} \\
    & \qquad + \nabla_\bp  H \Big|_{p = -\nabla_\bx\phi} \otimes \left(\mathbf{P}_0 : \nabla_\bp \nabla_\bp H \Big|_{p = -\nabla_\bx\phi}\right) \\
    &= -M^0 \left( \nabla_\bp H \Big|_{p = -\nabla_\bx\phi} \otimes \nabla_\bp H \Big|_{p = -\nabla_\bx\phi} \right) \\
    &\qquad + \mathbf{M}^1 \otimes \nabla_\bp H \Big|_{p = -\nabla_\bx\phi} + \nabla_\bp H \Big|_{p  -\nabla_\bx\phi} \otimes \mathbf{M}^1,
\end{align*}
where, in the last equality, we substituted the definition of $\mathbf{M}^1$ into the last two terms, $\mathbf{P}_0 : \nabla_\bp \nabla_\bp H \Big|_{p = -\nabla_\bx\phi} = \mathbf{M}^1 - M^0 \nabla_\bp H \Big|_{p = -\nabla_\bx\phi}.$ This confirms the expression for the degree two closure \eqref{eq:degree-two-closure-explicit}.

\textbf{Degree $m+1$ closure.} Finally, from the previous two cases, the procedure to obtain the degree $m+1$ closure is clear, which we briefly outline here. Consider the auxiliary space of truncated moments of degree $m$, $\mathfrak{M}^m_0 \times C(Q) \ni (M^0, P^1_0, \dots, P^m_0, \phi)$. Here, $P^k_0 \in \mathfrak{M}_k$ ($1 \leq k \leq m$) is a symmetric degree $k$ tensor representing a perturbation away from center of the $k^{th}$ kinetic moment of the phase space distribution (for more details, see \cite{Burby2023}), analogous to the interpretation of $\mathbf{P}_0 = P^1_0$ in the case $m=1$. The mapping $\Gamma_m: \mathfrak{M}^m_0 \times C(Q) \rightarrow C^*(T^*Q)$ is provided in \cite{Burby2023}, explicitly,
\begin{align}\label{eq:gamma-m-general}
    \Gamma_m(M^0, &P^1_0, \dots, P^m_0, \phi)(\bx,\bp) \\
    &= M^0(\bx) \delta(\bp + \nabla_\bx\phi) - P^1_0 \cdot \nabla_\bp \delta(\bp + \nabla_\bx\phi) \nonumber \\
     & \qquad + \frac{1}{2} P^2_0 : \nabla_\bp \nabla_\bp \delta(\bp + \nabla_\bx\phi) \nonumber\\
     & \qquad + \dots + \frac{(-1)^m}{m!} P^m_0 : \underbrace{\nabla_\bp \cdots \nabla_\bp}_{m \textup{ times}} \delta(\bp + \nabla_\bx\phi). \nonumber
\end{align}
The Hamiltonian functional on $\mathfrak{M}^m_0 \times C(Q)$ is again defined by pulling back the Hamiltonian functional \eqref{eq:hamiltonian-functional-variable-sec} by $\Gamma_m$, i.e.,
$$ \m{H}^m [M^0, P^1_0, \dots, P^m_0, \phi] := \m{H}[\Gamma_m(M^0, P^1_0, \dots, P^m_0, \phi)]. $$
Using the bracket $\db{\cdot}{\cdot}^m_0$ on this auxiliary space of truncated moments, defined in \cite{Burby2023}, one can compute the evolution equations for $(M^0, P^1_0, \dots, P^m_0, \phi)$, analogously to the $m=0$ and $m=1$ case we considered previously. In a similar manner, one can instead derive evolution equations for the moments $(M^0,\dots,M^m)$ where the $k^{th}$ moment is defined as the $k^{th}$ kinetic moment of the phase space distribution given by the image of $(M^0, P^1_0, \dots, P^m_0, \phi)$ under $\Gamma_m$, i.e.,
$$ M^k := M^k[\Gamma_m(M^0, P^1_0, \dots, P^m_0, \phi)],\ k=1,\dots,m, $$
where, recalling from recalling from \Cref{sec:moment-equations}, the $k^{th}$ kinetic moment of a phase space distribution $g$ is defined as
$$ M^k[g](t,\bx) = \int_{T^*_\bx Q} \underbrace{ \nabla_\bp H \otimes \cdots \otimes \nabla_\bp H}_{k \textup{ times}} g(t,\bx,\bp) d\bp. $$
Note that this construction crucially depends on using the correct moment kernel $\nabla_\bp H \otimes \cdots \otimes \nabla_\bp H$ to define the $k^{th}$ kinetic moment. This results in a system of conservation laws of the form
\begin{subequations}\label{eq:m-moment-system}
\begin{align}
    \frac{\partial}{\partial t} M^0 &= -\nabla_\bx \cdot M^1, \\
    \frac{\partial}{\partial t} M^1 &= -\nabla_\bx \cdot M^2, \\
                                    &\vdots \\
    \frac{\partial}{\partial t} M^m &= -\nabla_\bx \cdot M^{m+1},
\end{align}
\end{subequations}
where the degree $m+1$ closure for this system is given by
$$ M^{m+1} := M^{m+1}[\Gamma_m(M^0, P^1_0, \dots, P^m_0, \phi)], $$
together with the Hamilton--Jacobi equation for $\phi$.

The cost of the moment closure is that, for a given truncation degree $m$, one cannot represent arbitrary phase space distributions and is restricted to phase space distributions in the image of $\Gamma_m$, see \Cref{eq:gamma-m-general}. In particular, the momentum dependence of the phase space distribution is spatially local, due to the factors of $\delta(\bp + \nabla_\bx\phi(\bx))$ appearing in each term of \eqref{eq:gamma-m-general}. 

\begin{remark}[Initial Conditions]
    For the moment system of truncation degree m \eqref{eq:m-moment-system}, coupled with the Hamilton--Jacobi equation for $\phi$, one must specify initial conditions for $M^0, \dots, M^m, \phi$ in order to evolve the system. One possibility for determining the initial conditions is as follows.

    Suppose the phase space distribution at time $t=0$ is known, i.e., $g|_{t=0}$ (as justification for using the moment system here instead of evolving the initial condition under the full kinetic transport equation, note that for numerical methods, the moment system is often preferred over the full kinetic equation due to the decreased dimensionality of the problem arising from removing the momentum dependence \cite{momentreview}). Then, the initial conditions for $M^0|_{t=0},\dots,M^m|_{t=m}$ can be obtained by computing the corresponding kinetic moment of $g|_{t=0}$. Since the negative gradient of $\phi$ represents the spatially local momentum corresponding to $g$, we can initialize $\phi$ as follows. Let
    $$ \mathbf{P}(t,x) := \int_{T_\bx Q} \bp g(t,\bx,\bp) d\bp. $$
    Since we can compute this quantity at the initial time, we can take the initial condition for $\phi$ to be the solution of
    $$ \nabla \phi(0,\bx) = -\mathbf{P}(0,\bx), $$
    which, by the Poincar\'{e} lemma, assuming $Q$ is a star-shaped domain and the distribution $g$ is chosen such that $\mathbf{P}(0,\bx)$ is curl-free, admits a unique solution up to an additive constant (see, e.g., \cite{vectorcalc}). The choice of the additive constant does not affect the dynamics, since all of the evolution equations depend on $\phi$ only through its derivatives.
\end{remark}

We conclude this section with two concrete examples. Namely, we first recall the fluid moment closures previously derived in \cite{Burby2023}. Subsequently, we present novel degree one and degree two closures for pure radiation transport, using the general closures derived in this section.

\begin{example}[Fluid Variable Moment Closures]\label{example:fluid-closures}
 In \cite{Burby2023}, the fluid variable moment closures are derived (for degrees one, two and three) for the electron kinetic distribution $f$ with particle Hamiltonian
$$ H_e = \frac{\|\bp\|^2}{2m} + V. $$
For brevity, we set the potential to zero, $V \equiv 0$, as this does not affect the form of the moment closure \cite{Burby2023}.

\textbf{Degree one fluid closure.} For the matter distribution, the collective Hamiltonian of degree zero corresponding to $H_e = \|\bp\|^2/2m$ is
$$ \m{H}^0_e[M^0_e, \phi] = \int_Q \frac{1}{2m} M^0_e \|\nabla_\bx \phi\|^2 d\bx. $$
\citet{Burby2023} derives the equations of motion:
\begin{align}
    \frac{\partial}{\partial t} M^0_e &= - \nabla_\bx \cdot (m^{-1} M^0_e \nabla_\bx\phi), \nonumber\\
    \frac{\partial}{\partial t} \phi &= \frac{1}{2m} \|\nabla_\bx\phi\|^2. \nonumber
\end{align}
As noted in \cite{Burby2023}, this degree one closure is equivalent to Euler's equation for a cold, irrotational fluid. This can be expressed in the traditional form of a moment closure 
\begin{align*}
    \frac{\partial}{\partial t} M^0_e &= \nabla_\bx \cdot \bold{M}^1_e, \\
    \frac{\partial}{\partial t} \phi &= \frac{1}{2m}\|\nabla_\bx\phi\|^2,
\end{align*}
where the closure for $\bold{M}^1_e$ is given by
\begin{equation}\label{eq:degree-one-closure-matter} 
\bold{M}^1_e = - \frac{1}{m} M^0_e \nabla_\bx\phi.
\end{equation}
It is straightforward to verify that this agrees with the general form for the degree one closure that we derived, \eqref{eq:degree-one-closure}, for this choice of Hamiltonian.

\textbf{Degree two fluid closure.} For the matter distribution, the collective Hamiltonian of degree one corresponding to $H_e = \|\bp\|^2/2m$ is
$$ \m{H}^1_e[M^0_e, \mathbf{P}_0, \phi] = \int_Q \frac{1}{2m M^0_e} \| \mathbf{P}_0 - M^0_e \nabla \phi\|^2 d\bx - \int_Q \frac{1}{2m M^0_e} \| \mathbf{P}_0\|^2 d\bx; $$
\citet{Burby2023} derives the equations of motion:
\begin{align*}
    \frac{\partial}{\partial t} M^0_e &= - \nabla_\bx \cdot (m^{-2} \mathbf{P}_0 - m^{-1}M^0_e \nabla_\bx \phi), \\
    \frac{\partial}{\partial t} \mathbf{P}_0 &= m^{-1} \mathbf{P}_0 : \nabla_\bx\nabla_\bx\phi + m^{-1} \nabla_\bx \phi \cdot \nabla_\bx\mathbf{P}_0 + m^{-1} (\nabla_\bx\cdot \nabla_\bx\phi) \mathbf{P}_0, \\
    \frac{\partial}{\partial t} \phi &= \frac{1}{2m} \|\nabla_\bx\phi\|^2.
\end{align*}
The system can then be expressed in the traditional form of a moment closure,
\begin{align*}
    \frac{\partial}{\partial t} M^0_e &= - \nabla_\bx\cdot \bold{M}^1_e, \\
    \frac{\partial}{\partial t} \bold{M}^1_e &= - \nabla_\bx\cdot \mathbb{M}^2_e, \\
    \frac{\partial}{\partial t} \phi &= \frac{1}{2m} \|\nabla_\bx\phi\|^2,
\end{align*}
where $\mathbf{M}^1_e = m^{-2} \mathbf{P}_0 - m^{-1}M^0_e \nabla_\bx \phi$. It is straightforward to verify that this equals \eqref{eq:M1-degree-two-explicit} for this choice of Hamiltonian. Finally, the degree two closure for $\mathbb{M}^2_e$ is given by
\begin{equation}\label{eq:degree-two-closure-matter} 
\mathbb{M}^2_e = - \frac{1}{m} \bold{M}^1_e \otimes \nabla_\bx\phi -  \frac{1}{m}\bold{M}^1_e \otimes \nabla_\bx \phi - \frac{1}{m^2} M^0_e \nabla_\bx\phi \otimes \nabla_\bx\phi. 
\end{equation}
Again, it is straightforward to verify that this agrees with the general form for the degree two closure derived above \eqref{eq:degree-two-closure-explicit}, for this choice of Hamiltonian. 
\end{example}

\begin{example}[Variable Moment Closures for Pure Radiation Transport]\label{example:pure-rad-closures}
Using the theory developed in this section, it is straightforward to derive moment closures for the pure radiation transport equation
$$ \frac{\partial}{\partial t} \Psi + c\bold{\Omega} \cdot \nabla_\bx \Psi = 0. $$

\textbf{Degree one radiation closure.} Recall the Hamiltonian functional for radiation transport,
$$ \mathcal{H}_r[\Psi] := \int_{T^*Q} \Psi H_r(\bx,\bp) d\bx\, d\bp, $$
where $H_r(\bx,\bp) = c\|\bp\|$. Pulling back $\mathcal{H}_r$ along $\Gamma_0$ \eqref{eq:bracket-map-zero} yields the collective Hamiltonian of degree zero,
$$ \mathcal{H}^0_r[M^0_r,\phi] =  \int_{Q} M^0_r H_r(\bx,-\nabla_\bx \phi) d\bx = \int_{Q} c M^0_r \|\nabla_\bx \phi\| d\bx. $$
The functional derivatives of the collective Hamiltonian are
\[
\begin{aligned}
    \frac{\delta \mathcal{H}^0_r}{\delta M^0_r} = c \|\nabla_\bx\phi\|, && \frac{\delta \mathcal{H}^0_r}{\delta \phi} = - \nabla_\bx \cdot \left( cM^0_r \frac{\nabla_\bx \phi}{\|\nabla_\bx \phi\|} \right).
\end{aligned}
\]
Thus, Hamilton's equations on the auxiliary space of truncated moments of degree zero are
\begin{subequations}\label{eq:zero-moment-geometric-optics}
\begin{align}
    \frac{\partial M^0_r}{\partial t} &= \db{M^0_r}{\m{H}^0_r}^0_0 = \nabla_\bx \cdot \left( cM^0_r \frac{\nabla_\bx \phi}{\|\nabla_\bx \phi\|} \right), \label{eq:zero-moment-geometric-optics-a} \\
    \frac{\partial \phi}{\partial t} &= \db{\phi}{\m{H}^0_r}^0_0 = c\|\nabla_\bx \phi\|. \label{eq:zero-moment-geometric-optics-b}
\end{align}
\end{subequations}
The system consists of a conservation law \eqref{eq:zero-moment-geometric-optics-a} coupled with an Eikonal equation \eqref{eq:zero-moment-geometric-optics-b}. The system \eqref{eq:zero-moment-geometric-optics} can also be written in the traditional form of a moment closure,
$$ \frac{\partial M^0_r}{\partial t} = -\nabla_\bx \cdot \bold{M}^1_r, $$
where the closure for the degree one moment $\bold{M}^1_r$ follows from a direct calculation using \eqref{eq:degree-one-closure},
\begin{equation}\label{eq:degree-one-closure-radiation}
\bold{M}^1_r = -cM^0_r \frac{\nabla_\bx \phi}{\|\nabla_\bx \phi\|}, \quad \frac{\partial \phi}{\partial t} = \db{\phi}{\m{H}^0_r}^0_0 = c\|\nabla_\bx \phi\|.  
\end{equation}

\textbf{Degree two radiation closure.} Consider the space of truncated moments of degree one, $\mathfrak{M}^1_0$. Pulling back $\mathcal{H}_r$ along $\Gamma_1$ \eqref{eq:bracket-map-one} yields the collective Hamiltonian of degree one,
$$ \mathcal{H}^1_r[M^0_r,\mathbf{P}_0, \phi] = c\int_Q M^0_r \|\nabla_\bx\phi\| d\bx - c\int_Q \mathbf{P}_0  \cdot\frac{\nabla_\bx \phi}{\|\nabla_\bx \phi\|} d\bx. $$
The functional derivatives of the collective Hamiltonian are
\begin{align*}
    \frac{\delta \m{H}^1_r}{\delta M^0_r} &= c \|\nabla_\bx\phi\|, \\
    \frac{\delta \m{H}^1_r}{\delta \phi} &= -c\nabla_\bx \cdot \left( M^0_r \frac{\nabla_\bx\phi}{\|\nabla_\bx\phi\|} - \frac{\mathbf{P}_0}{\|\nabla_\bx\phi\|} + \mathbf{P}_0 \cdot \nabla_\bx \phi \frac{\nabla_\bx\phi}{\|\nabla_\bx\phi\|^3} \right), \\
    \frac{\delta \m{H}^1_r}{\delta \mathbf{P}_0} &= -c \frac{\nabla_\bx\phi}{\|\nabla_\bx\phi\|}.
\end{align*}
Hamilton's equations for the truncated moment system of degree one are thus,
\begin{align*}
    \frac{\partial M^0}{\partial t} &= \db{M^0_r}{\m{H}^1_r}^1_0 = c\nabla_\bx \cdot \left( M^0_r \frac{\nabla_\bx\phi}{\|\nabla_\bx\phi\|} - \frac{\mathbf{P}_0}{\|\nabla_\bx\phi\|} + \mathbf{P}_0 \cdot \nabla_\bx \phi \frac{\nabla_\bx\phi}{\|\nabla_\bx\phi\|^3} \right), \\
    \frac{\partial \mathbf{P}_0}{\partial t} &= \db{\mathbf{P}_0}{\m{H}^1_r}^1_0 =c \mathbf{P}_0 : \nabla \left( \frac{\nabla\phi}{\|\nabla\phi\|} \right) + c \frac{\nabla\phi}{\|\nabla\phi\|} \cdot \nabla \mathbf{P}_0 + c \mathbf{P}_0 \nabla\cdot  \left( \frac{\nabla\phi}{\|\nabla\phi\|} \right), \\
    \frac{\partial \phi}{\partial t} &= \db{\phi}{\m{H}^1_r}^1_0 = c \|\nabla_\bx\phi\|.
\end{align*}
The above system involves evolution equations for $M^0$ and $\mathbf{P}_0$ coupled to the auxiliary field $\phi$ which again satisfies the Eikonal equation. This can be expressed as a traditional moment closure, 
    \begin{align*}
    \frac{\partial M^0_r}{\partial t} &= - \nabla_\bx \cdot \bold{M}^1_r,  \\
    \frac{\partial \bold{M}^1_r}{\partial t} &= - \nabla_\bx \cdot \mathbb{M}^2_r, \\
    \frac{\partial \phi}{\partial t} &= c \|\nabla_\bx\phi\|, 
    \end{align*}
where $\mathbf{M}^1_r$ can be directly computed from \eqref{eq:M1-degree-two-explicit},
\begin{equation}\label{eq:moment-one-definition}
\bold{M}^1_r := -c \left( M^0_r \frac{\nabla_\bx\phi}{\|\nabla_\bx\phi\|} - \frac{\mathbf{P}_0}{\|\nabla_\bx\phi\|} + \mathbf{P}_0 \cdot \nabla_\bx \phi \frac{\nabla_\bx\phi}{\|\nabla_\bx\phi\|^3} \right). 
\end{equation}
and the degree two closure can be computed from \eqref{eq:degree-two-closure-explicit},
    \begin{equation}\label{eq:closure-for-degree-1}
        \mathbb{M}^2_r = - c^2 M^0_r \frac{\nabla_\bx\phi}{\|\nabla_\bx\phi\|} \otimes \frac{\nabla_\bx\phi}{\|\nabla_\bx\phi\|} -c \bold{M}^1_r \otimes \frac{\nabla_\bx\phi}{\|\nabla_\bx\phi\|} -c \frac{\nabla_\bx\phi}{\|\nabla_\bx\phi\|} \otimes \bold{M}^1_r.
    \end{equation}
\end{example}

\subsubsection{Variable moment closures as an asymptotic expansion}\label{sec:wkb-relation}
To conclude, we give an interpretation of the variable moment closure framework as an asymptotic expansion. This interpretation is similar in nature to the Wentzel–Kramers–Brillouin (WKB) approximation. Briefly, the WKB approximation computes approximate solutions to wave-like evolution equations by expressing the solution as $\psi(t,\bx) = A(t,\bx) \exp(i S(t,\bx)/\epsilon)$ and expanding the amplitude $A$ as a power series in $\epsilon$ 
\cite{wkbschrodtext}. Substituting the expansion into the evolution equation allows one to solve the resulting equations asymptotically, at successive orders in powers of $\epsilon$, to obtain an approximate solution. Higher order terms in the asymptotic expansion then produce corrections to the amplitude. Interestingly, for, e.g., the Schr\"{o}dinger equation \cite{wkbschrodtext}, the leading order equations for the amplitude $A$ and phase $S$ in the WKB approximation satisfy the same transport equation and Hamilton--Jacobi equations that arose from the leading order ($m=0$) equations of the variable moment closure framework, \eqref{eq:M0-evolution-degree-one-closure} and \eqref{eq:degree-one-closure}. This can be understood as the fact that the semi-classical fluid limit of the Schr\"{o}dinger equation \cite{wkbschrodtext} is the same as the classical limit of the variable moment closure, i.e., truncation degree zero. For more details on the WKB approximation, see for example \cite{wkbschrodtext, wkbraytext, wkbEM, wkbPlasma}.

We now show that the variable moment closure framework has a similar interpretation, where successive orders in the asymptotic expansion produce corrections to the kinetic moments. Recall for the degree one closure (truncation degree $m=0$), we found the degree one moment to be \eqref{eq:degree-one-closure},
$$\mathbf{M}^1 = M^0 \nabla_\bp H \Big|_{\bp = -\nabla_\bx\phi}.$$
On the other hand, for the degree two moment closure (truncation degree $m=1$), we instead had \eqref{eq:M1-degree-two-explicit},
$$ \mathbf{M}^1 = M^0 \nabla_\bp H\Big|_{\bp = -\nabla_\bx\phi} + \mathbf{P}_0 \cdot \nabla_\bp \nabla_\bp H\Big|_{\bp = -\nabla_\bx\phi}. $$
We thus see that increasing the truncation order from $m=0$ to $m=1$ produced a correction to the first moment. To see this more generally, we consider the moment generating function of $g \in C^*(T^*Q)$, defined as
\begin{equation}\label{eq:moment-generating-function}
    \mathbb{M}[g; \bs](\bx) :=  \int_{T^*_\bx Q} \exp(\bs \cdot \nabla_\bp H(\bx,\bp)) g(\bx,\bp) d\bp,
\end{equation}
where the magnitude of the vector $\bs \in \mathbb{R}^n$ is interpreted as an expansion parameter; we denote the components of $\bs$ as $\{s^i\}_{i=1}^n$. The moment generating function $\mathbb{M}$ generates the kinetic moments of $g$ as an asymptotic expansion in $\bs$ as follows. For brevity in the following calculation, let $\bz$ denote $\nabla_\bp H(\bx,\bp)$ as in \Cref{sec:moment-equations}, let $\{z_i\}_{i=1}^n$ denote its components, and omit the arguments $(\bx,\bp)$. Compute
\begin{align*}
    \mathbb{M}[g; \bs] &= \int_{T^*_\bx Q} \left( 1 + \bs\cdot \bz + \frac{1}{2}(\bs \cdot \bz)^2 + \mathcal{O}(\|\bs\|^3)\right) g d\bp \\
    &= \int_{T^*_\bx Q} g d\bp + s^i\int_{T^*_\bx Q} z_i g d\bp + \frac{1}{2} s^i s^j \int_{T^*_\bx Q} z_i z_j g d\bp + \mathcal{O}(\|\bs\|^3) \\
    &= \int_{T^*_\bx Q} g d\bp + \bs \cdot \int_{T^*_\bx Q} \bz g d\bp + \frac{1}{2} (\bs \otimes \bs) : \int_{T^*_\bx Q} \bz \otimes \bz g d\bp + \mathcal{O}(\|\bs\|^3) \\
    &= M^0[g] + \bs \cdot \mathbf{M}^1[g] + \frac{1}{2} (\bs \otimes \bs) : \mathbb{M}^2[g] + \mathcal{O}(\|\bs\|^3).
\end{align*}
That is, the order $\|\bs\|^k$ term in the above expansion corresponds to the $k^{th}$ kinetic moment of $g$ ($k=0,1,2,\dots$). This is a generating function for the moments in the sense that the $(i_1, \dots, i_k)^{th}$ component of the $k^{th}$ kinetic moment can be computed by
$$ \mathbb{M}^k_{i_1\dots i_k}[g] = k! \frac{\partial}{\partial s^{i_1}} \cdots \frac{\partial}{\partial s^{i_k}} \mathbb{M}[g; \bs] \Big|_{\bs = 0}. $$
To see that increasing the truncation order from $m = p$ to $m=p+1$ produces an order $\|\bs\|^{p+1}$ correction to the moment generating function, compute, using the definition of $\Gamma_m$ \eqref{eq:gamma-m-general}, the difference between the moment generating functions for the distribution in the image of $\Gamma^{p+1}$ and $\Gamma^p$, i.e.,
\begin{align*}
    \mathbb{M}&[\Gamma^{p+1}(M^0, P^1_0, \dots, P^p_0, P^{p+1}_0,\phi); \bs](\bx) - \mathbb{M}[\Gamma^{p}(M^0, P^1_0, \dots, P^p_0,\phi); \bs](\bx) \\
    &= \int_{T^*_\bx Q} \exp(\bs \cdot \bz ) \Big( \Gamma^{p+1}(M^0, P^1_0, \dots, P^p_0, P^{p+1}_0,\phi) -\Gamma^{p}(M^0, P^1_0, \dots, P^p_0,\phi) \Big) d\bp \\
    &= \frac{(-1)^{p+1}}{(p+1)!} \int_{T^*_\bx Q} \exp(\bs \cdot \bz ) P^{p+1}_0: \underbrace{\nabla_\bp \cdots \nabla_\bp}_{p+1 \textup{ times}} \delta(\bp + \nabla_\bx\phi) d\bp \\
    &= \frac{1}{(p+1)!} \int_{T^*_\bx Q} \Big(\underbrace{\nabla_\bp \cdots \nabla_\bp}_{p+1 \textup{ times}} \exp(\bs\cdot\bz)\Big) :  P^{p+1}_0 \delta(\bp + \nabla_\bx\phi) d\bp,
\end{align*}
which is indeed order $\|\bs\|^{p+1}$, since each derivative of $\exp(\bs\cdot\bz)$ with respect to $\bp$ produces an additional factor of $\bs$.

\section{Conclusion}\label{sec:conclusion}
In this work, we considered kinetic systems from a geometric perspective, taking the kinetic description of matter and radiation as our two fundamental examples. Taking a geometric perspective on kinetic systems particularly allowed us to provide a characterization of the moment kernels associated to a kinetic system in terms of the corresponding particle Hamiltonian. This naturally led us to a geometric interpretation of the moment closure problem \cite{Lev1996, Grad1949}. We contextualized the theory presented with two examples, one utilizing principles of thermodynamics to derive a pair bracket formulation of diffusion radiation hydrodynamics and another using geometric arguments to derive pure geometric moment closures for transport with arbitrary Hamiltonian. In particular, this led to novel degree one and degree two closures for pure radiation transport. 

For future research, it would be interesting, in light of recent developments and interest in data-driven moment closure models \cite{sadr2021,huang2022,Por2023, Dona2023, huang2023}, to incorporate the work of \cite{gruber24}, on learning of pair bracket formulations via deep graph neural networks, in order to learn geometric moment closures for kinetic systems. Particularly, since the graph neural network architectures presented in \cite{gruber24} are guaranteed to preserve energy and generate entropy with increasing depth, synthesis of these ideas could lead to an approach for structure-preserving learning of geometric moment closures.

% Finally, we plan to numerically investigate the geometric moment closures for radiation transport presented here, supplemented with a dissipative bracket for interactions, and compare these to existing moment closures for radiation transport.

\section*{Acknowledgements}
The authors would like to thank the reviewers for their helpful comments and suggestions. BKT was supported by the Marc Kac Postdoctoral Fellowship at the Center for Nonlinear Studies at Los Alamos National Laboratory. BSS was supported by the Laboratory Directed Research and Development program of Los Alamos National Laboratory under project number 20220174ER. JWB was supported by the U.S. Department of Energy (DOE), the Office of Science and the Office of Advanced Scientific Computing Research (ASCR). Specifically, JWB acknowledges funding support from ASCR for DOE-FOA-2493 “Data-intensive scientific machine learning and analysis”. Los Alamos National Laboratory Report LA-UR-24-29144.

\bibliographystyle{plainnat}
\bibliography{georh}

\begin{thebibliography}{64}
\providecommand{\natexlab}[1]{#1}
\providecommand{\url}[1]{\texttt{#1}}
\expandafter\ifx\csname urlstyle\endcsname\relax
  \providecommand{\doi}[1]{doi: #1}\else
  \providecommand{\doi}{doi: \begingroup \urlstyle{rm}\Url}\fi

\bibitem[Abdelmalik et~al.(2023)Abdelmalik, Cai, and Pichard]{Abdel2023}
M.R.A. Abdelmalik, Z.~Cai, and T.~Pichard.
\newblock Moment methods for the radiative transfer equation based on
  $\varphi$-divergences.
\newblock \emph{Computer Methods in Applied Mechanics and Engineering},
  417:\penalty0 116454, 2023.
\newblock ISSN 0045-7825.
\newblock \doi{10.1016/j.cma.2023.116454}.

\bibitem[Abraham and Marsden(1987)]{RaMa1987}
Ralph Abraham and Jerrold~E. Marsden.
\newblock \emph{Foundations of Mechanics}.
\newblock Addison-Wesley Publishing Company, Inc., second edition, 1987.

\bibitem[Abraham et~al.(1988)Abraham, Marsden, and Ratiu]{manifoldstensor}
Ralph Abraham, Jerrold~E. Marsden, and Tudor Ratiu.
\newblock \emph{Manifolds, Tensor Analysis, and Applications}.
\newblock Applied Mathematical Sciences. Springer New York, NY, second edition,
  1988.
\newblock \doi{10.1007/978-1-4612-1029-0}.

\bibitem[Alldredge et~al.(2019)Alldredge, Frank, and Hauck]{All2019}
Graham~W. Alldredge, Martin Frank, and Cory~D. Hauck.
\newblock A regularized entropy-based moment method for kinetic equations.
\newblock \emph{SIAM Journal on Applied Mathematics}, 79\penalty0 (5):\penalty0
  1627--1653, 2019.
\newblock \doi{10.1137/18M1181201}.

\bibitem[Andreev et~al.(1983)Andreev, Kozmanov, and Rachilov]{maxprinciple}
E.S. Andreev, M.Yu. Kozmanov, and E.B. Rachilov.
\newblock The maximum principle for a system of equations of energy and
  non-stationary radiation transfer.
\newblock \emph{USSR Computational Mathematics and Mathematical Physics},
  23\penalty0 (1):\penalty0 104--109, 1983.
\newblock ISSN 0041-5553.
\newblock \doi{10.1016/S0041-5553(83)80018-4}.

\bibitem[Berk and Dominguez(1983)]{wkbPlasma}
H.~L. Berk and R.~R. Dominguez.
\newblock Local {WKB} dispersion relation for the {V}lasov–{M}axwell
  equations.
\newblock \emph{The Physics of Fluids}, 26\penalty0 (7):\penalty0 1825--1829,
  1983.
\newblock \doi{10.1063/1.864358}.

\bibitem[Bessa et~al.(2013)Bessa, Rocha, and Torres]{hyperbolic2}
Mário Bessa, Jorge Rocha, and Maria~Joana Torres.
\newblock Hyperbolicity and stability for hamiltonian flows.
\newblock \emph{Journal of Differential Equations}, 254\penalty0 (1):\penalty0
  309--322, 2013.
\newblock ISSN 0022-0396.
\newblock \doi{10.1016/j.jde.2012.08.010}.

\bibitem[Bosboom et~al.(2023)Bosboom, Kraus, and Schlottbom]{bosboom2023}
V.~Bosboom, M.~Kraus, and M.~Schlottbom.
\newblock A metriplectic formulation of polarized radiative transfer.
\newblock \emph{Journal of Physics A: Mathematical and Theoretical},
  56\penalty0 (34):\penalty0 345206, 2023.
\newblock \doi{10.1088/1751-8121/aceae2}.

\bibitem[Burby(2023)]{Burby2023}
J.~W. Burby.
\newblock Variable-moment fluid closures with {H}amiltonian structure.
\newblock \emph{Sci Rep}, 13\penalty0 (18286), 2023.
\newblock \doi{10.1038/s41598-023-45416-5}.

\bibitem[Cai et~al.(2014)Cai, Fan, and Li]{CaiFan2014}
Zhenning Cai, Yuwei Fan, and Ruo Li.
\newblock Globally hyperbolic regularization of {G}rad's moment system.
\newblock \emph{Communications on Pure and Applied Mathematics}, 67\penalty0
  (3):\penalty0 464--518, 2014.
\newblock \doi{https://doi.org/10.1002/cpa.21472}.

\bibitem[Cai et~al.(2015)Cai, Fan, and Li]{CaiFan2013}
Zhenning Cai, Yuwei Fan, and Ruo Li.
\newblock A framework on moment model reduction for kinetic equation.
\newblock \emph{SIAM Journal on Applied Mathematics}, 75\penalty0 (5):\penalty0
  2001--2023, 2015.
\newblock \doi{10.1137/14100110X}.

\bibitem[Carles(2020)]{wkbschrodtext}
Rémi Carles.
\newblock \emph{Semi-Classical Analysis for Nonlinear {S}chr\"{o}dinger
  Equations}.
\newblock WORLD SCIENTIFIC, 2nd edition, 2020.
\newblock \doi{10.1142/12030}.

\bibitem[Castor(2004)]{castor_2004}
John~I. Castor.
\newblock \emph{Radiation Hydrodynamics}.
\newblock Cambridge University Press, 2004.
\newblock \doi{10.1017/CBO9780511536182}.

\bibitem[{Chandrasekhar}(1960)]{Chan1960}
Subrahmanyan {Chandrasekhar}.
\newblock \emph{{Radiative transfer}}.
\newblock Dover, NY, 1960.

\bibitem[Coquinot and Morrison(2020)]{CoMo_2020}
B.~Coquinot and P.~J. Morrison.
\newblock A general metriplectic framework with application to dissipative
  extended magnetohydrodynamics.
\newblock \emph{J. Plasma Phys.}, 86:\penalty0 835860302, 2020.
\newblock \doi{https://doi.org/10.1017/S0022377820000392}.

\bibitem[Donaghy and Germaschewski(2023)]{Dona2023}
John Donaghy and Kai Germaschewski.
\newblock In search of a data-driven symbolic multi-fluid ten-moment model
  closure.
\newblock \emph{Journal of Plasma Physics}, 89\penalty0 (1), 2023.
\newblock \doi{10.1017/S0022377823000119}.

\bibitem[Dubrovin(2009)]{hyperbolic3}
Boris Dubrovin.
\newblock Hamiltonian perturbations of hyperbolic pdes: from classification
  results to the properties of solutions.
\newblock In Vladas Sidoravi{\v{c}}ius, editor, \emph{New Trends in
  Mathematical Physics}, pages 231--276, Dordrecht, 2009. Springer Netherlands.
\newblock ISBN 978-90-481-2810-5.

\bibitem[Dubrovin et~al.(2015)Dubrovin, Grava, Klein, and Moro]{hyperbolic4}
Boris Dubrovin, Tamara Grava, Christian Klein, and Antonio Moro.
\newblock On critical behaviour in systems of hamiltonian partial differential
  equations.
\newblock \emph{J Nonlinear Sci}, 25:\penalty0 631--707, 2015.
\newblock \doi{10.1007/s00332-015-9236-y}.

\bibitem[Dukek et~al.(1997)Dukek, Karlin, and Nonnenmacher]{Dukek1997}
G.~Dukek, Iliya~V. Karlin, and T.F. Nonnenmacher.
\newblock Dissipative brackets as a tool for kinetic modeling.
\newblock \emph{Physica A: Statistical Mechanics and its Applications},
  239\penalty0 (4):\penalty0 493--508, 1997.
\newblock \doi{10.1016/S0378-4371(97)00015-0}.

\bibitem[Eberhard~Engel(2011)]{Dreizler2011}
Reiner M.~Dreizler Eberhard~Engel.
\newblock \emph{Density Functional Theory: An Advanced Course}.
\newblock Theoretical and Mathematical Physics. Springer Berlin, Heidelberg,
  2011.
\newblock \doi{10.1007/978-3-642-14090-7}.

\bibitem[Evans(1998)]{EvansPDE}
Lawrence~C. Evans.
\newblock \emph{Partial Differential Equations}.
\newblock Graduate Studies in Mathematics. American Mathematical Society,
  second edition, 1998.
\newblock \doi{10.1090/gsm/019}.

\bibitem[Garrett and Hauck(2013)]{GaHa2013}
C.~Kristopher Garrett and Cory~D. Hauck.
\newblock A comparison of moment closures for linear kinetic transport
  equations: The line source benchmark.
\newblock \emph{Transport Theory and Statistical Physics}, 42\penalty0
  (6-7):\penalty0 203--235, 2013.
\newblock \doi{10.1080/00411450.2014.910226}.

\bibitem[Gay-Balmaz and Tronci(2012)]{BT_2012}
F.~Gay-Balmaz and C.~Tronci.
\newblock {Vlasov moment flows and geodesics on the Jacobi group}.
\newblock \emph{J. Math. Phys.}, 53:\penalty0 123502, 2012.

\bibitem[G\'{e}rard et~al.(1997)G\'{e}rard, Markowich, Mauser, and
  Poupaud]{wigner}
Patrick G\'{e}rard, Peter~A. Markowich, Norbert~J. Mauser, and Fr\'{e}d\'{e}ric
  Poupaud.
\newblock Homogenization limits and {W}igner transforms.
\newblock \emph{Communications on Pure and Applied Mathematics}, 50\penalty0
  (4):\penalty0 323--379, 1997.
\newblock \doi{10.1002/(SICI)1097-0312(199704)50:4<323::AID-CPA4>3.0.CO;2-C}.

\bibitem[Gibbons et~al.(2008{\natexlab{a}})Gibbons, Holm, and
  Tronci]{Tronci_geom_2008}
J.~Gibbons, D.~D. Holm, and C.~Tronci.
\newblock {Geometry of Vlasov kinetic moments: A bosonic Fock space for the
  symmetric Schouten bracket}.
\newblock \emph{Phys. Lett. A}, 372:\penalty0 4184--4196, 2008{\natexlab{a}}.
\newblock \doi{https://doi.org/10.1016/j.physleta.2008.03.034}.

\bibitem[Gibbons et~al.(2008{\natexlab{b}})Gibbons, Holm, and Tronci]{GHT08}
John Gibbons, Darryl~D. Holm, and Cesare Tronci.
\newblock Geometry of {V}lasov kinetic moments: A bosonic {F}ock space for the
  symmetric {S}chouten bracket.
\newblock \emph{Physics Letters A}, 372\penalty0 (23):\penalty0 4184--4196,
  2008{\natexlab{b}}.
\newblock ISSN 0375-9601.
\newblock \doi{10.1016/j.physleta.2008.03.034}.

\bibitem[Gibbons et~al.(2008{\natexlab{c}})Gibbons, Holm, and Tronci]{GHT082}
John Gibbons, Darryl~D. Holm, and Cesare Tronci.
\newblock Vlasov moments, integrable systems and singular solutions.
\newblock \emph{Physics Letters A}, 372\penalty0 (7):\penalty0 1024--1033,
  2008{\natexlab{c}}.
\newblock ISSN 0375-9601.
\newblock \doi{10.1016/j.physleta.2007.08.054}.

\bibitem[Grad(1949)]{Grad1949}
H.~Grad.
\newblock On the kinetic theory of rarefied gases.
\newblock \emph{Commun. Pure and Appl. Math.}, 2\penalty0 (4):\penalty0
  331--407, 1949.
\newblock \doi{10.1002/cpa.3160020403}.

\bibitem[Gruber et~al.(2024)Gruber, Lee, and Trask]{gruber24}
Anthony Gruber, Kookjin Lee, and Nathaniel Trask.
\newblock Reversible and irreversible bracket-based dynamics for deep graph
  neural networks.
\newblock In \emph{Proceedings of the 37th International Conference on Neural
  Information Processing Systems}, NIPS '23, Red Hook, NY, USA, 2024. Curran
  Associates Inc.

\bibitem[Holm and Tronci(2009{\natexlab{a}})]{Holm_Tronci_2009}
D.~D. Holm and C.~Tronci.
\newblock {Geodesic Vlasov equations and their integrable moment closures}.
\newblock \emph{J. Geom. Mech.}, 1:\penalty0 181--208, 2009{\natexlab{a}}.
\newblock \doi{10.3934/jgm.2009.1.181}.

\bibitem[Holm and Tronci(2009{\natexlab{b}})]{HT09}
Darryl~D. Holm and Cesare Tronci.
\newblock Geodesic {V}lasov equations and their integrable moment closures.
\newblock \emph{Journal of Geometric Mechanics}, 1\penalty0 (2):\penalty0
  181--208, 2009{\natexlab{b}}.
\newblock ISSN 1941-4889.
\newblock \doi{10.3934/jgm.2009.1.181}.

\bibitem[Huang et~al.(2022)Huang, Cheng, Christlieb, and Roberts]{huang2022}
Juntao Huang, Yingda Cheng, Andrew~J. Christlieb, and Luke~F. Roberts.
\newblock Machine learning moment closure models for the radiative transfer
  equation {I}: Directly learning a gradient based closure.
\newblock \emph{Journal of Computational Physics}, 453:\penalty0 110941, 2022.
\newblock ISSN 0021-9991.
\newblock \doi{10.1016/j.jcp.2022.110941}.

\bibitem[Huang et~al.(2023)Huang, Cheng, Christlieb, Roberts, and
  Yong]{huang2023}
Juntao Huang, Yingda Cheng, Andrew~J. Christlieb, Luke~F. Roberts, and Wen-An
  Yong.
\newblock Machine learning moment closure models for the radiative transfer
  equation {II}: Enforcing global hyperbolicity in gradient-based closures.
\newblock \emph{Multiscale Modeling \& Simulation}, 21\penalty0 (2):\penalty0
  489--512, 2023.
\newblock \doi{10.1137/21M1423956}.

\bibitem[Hubbard and Hubbard(2015)]{vectorcalc}
John~H. Hubbard and Barbara~Burke Hubbard.
\newblock \emph{Vector Calculus, Linear Algebra, and Differential Forms: A
  Unified Approach}.
\newblock Matrix Editions, 5 edition, 2015.

\bibitem[Jayawardana et~al.(2022)Jayawardana, Morrison, and
  Ohsawa]{ClebschCanon}
Buddhika Jayawardana, Philip~J. Morrison, and Tomoki Ohsawa.
\newblock Clebsch canonization of {L}ie–{P}oisson systems.
\newblock \emph{Journal of Geometric Mechanics}, 14\penalty0 (4):\penalty0
  635--658, 2022.
\newblock ISSN 1941-4889.
\newblock \doi{10.3934/jgm.2022017}.

\bibitem[Kaufman(1984)]{Kaufman1984}
Allan~N. Kaufman.
\newblock Dissipative hamiltonian systems: A unifying principle.
\newblock \emph{Physics Letters A}, 100\penalty0 (8):\penalty0 419--422, 1984.
\newblock \doi{10.1016/0375-9601(84)90634-0}.

\bibitem[Kremer(2010)]{Kre2010}
Gilberto~Medeiros Kremer.
\newblock \emph{An Introduction to the Boltzmann Equation and Transport
  Processes in Gases}.
\newblock Interaction of Mechanics and Mathematics. Springer Berlin,
  Heidelberg, 2010.
\newblock \doi{10.1007/978-3-642-11696-4}.

\bibitem[Kuehn(2016)]{momentreview}
Christian Kuehn.
\newblock Moment closure---a brief review.
\newblock In Eckehard Sch{\"o}ll, Sabine H.~L. Klapp, and Philipp H{\"o}vel,
  editors, \emph{Control of Self-Organizing Nonlinear Systems}, pages 253--271.
  Springer International Publishing, Cham, 2016.
\newblock \doi{10.1007/978-3-319-28028-8_13}.

\bibitem[Lessig and Castro(2015)]{phasespacelifts}
Christian Lessig and Alex~L. Castro.
\newblock The geometry of radiative transfer.
\newblock In Dong~Eui Chang, Darryl~D. Holm, George Patrick, and Tudor Ratiu,
  editors, \emph{Geometry, Mechanics, and Dynamics: The Legacy of {J}erry
  {M}arsden}. Springer New York, NY, 2015.
\newblock \doi{10.1007/978-1-4939-2441-7}.

\bibitem[Levermore(1996)]{Lev1996}
C.D. Levermore.
\newblock Moment closure hierarchies for kinetic theories.
\newblock \emph{J Stat Phys}, 83:\penalty0 1021--1065, 1996.
\newblock \doi{10.1007/BF02179552}.

\bibitem[Li et~al.(2023)Li, Dong, and Wang]{Li_2023}
Z.~Li, B.~Dong, and Y.~Wang.
\newblock {Learning invariance preserving moment closure model for
  Boltzmann–BGK equation}.
\newblock \emph{Comm. Math. Stat.}, 11:\penalty0 59--101, 2023.
\newblock \doi{https://doi.org/10.1007/s40304-022-00331-5}.

\bibitem[Marsden and Ratiu(1999)]{MaRa1999}
J.~E. Marsden and T.~S. Ratiu.
\newblock \emph{Introduction to Mechanics and Symmetry}.
\newblock Springer New York, NY, second edition, 1999.
\newblock \doi{10.1007/978-0-387-21792-5}.

\bibitem[Mihalas and Mihalas(1984)]{mihalas_1984}
Dimitri Mihalas and Barbara~Weibel Mihalas.
\newblock \emph{Foundations of Radiation Hydrodynamics}.
\newblock Oxford University Press, New York, 1984.
\newblock ISBN 0-19-503437-6.

\bibitem[Morrison(1986)]{Morrison_met_1986}
P.~J. Morrison.
\newblock {A paradigm for joined Hamiltonian and dissipative systems}.
\newblock \emph{Physica D}, 18:\penalty0 410--419, 1986.

\bibitem[Morrison(1984)]{Morrison1984}
Philip~J. Morrison.
\newblock Bracket formulation for irreversible classical fields.
\newblock \emph{Physics Letters A}, 100\penalty0 (8):\penalty0 423--427, 1984.
\newblock ISSN 0375-9601.
\newblock \doi{10.1016/0375-9601(84)90635-2}.

\bibitem[Morrison and Updike(2024)]{Morrison2024}
Philip~J. Morrison and Michael~H. Updike.
\newblock Inclusive curvaturelike framework for describing dissipation:
  Metriplectic 4-bracket dynamics.
\newblock \emph{Phys. Rev. E}, 109:\penalty0 045202, Apr 2024.
\newblock \doi{10.1103/PhysRevE.109.045202}.

\bibitem[Oancea et~al.(2020)Oancea, Joudioux, Dodin, Ruiz, Paganini, and
  Andersson]{wkbEM}
Marius~A. Oancea, J\'er\'emie Joudioux, I.~Y. Dodin, D.~E. Ruiz, Claudio~F.
  Paganini, and Lars Andersson.
\newblock Gravitational spin {H}all effect of light.
\newblock \emph{Phys. Rev. D}, 102:\penalty0 024075, Jul 2020.
\newblock \doi{10.1103/PhysRevD.102.024075}.

\bibitem[Olver and Nutku(1988)]{hyperbolic1}
Peter~J. Olver and Yavuz Nutku.
\newblock Hamiltonian structures for systems of hyperbolic conservation laws.
\newblock \emph{Journal of Mathematical Physics}, 29\penalty0 (7):\penalty0
  1610--1619, 1988.
\newblock \doi{10.1063/1.527909}.

\bibitem[Onsager(1931)]{OnsagerRelations}
Lars Onsager.
\newblock Reciprocal relations in irreversible processes. {I}.
\newblock \emph{Phys. Rev.}, 37:\penalty0 405--426, 1931.
\newblock \doi{10.1103/PhysRev.37.405}.

\bibitem[Pavelka et~al.(2016)Pavelka, Klika, Esen, and Grmela]{pavelka2016}
Michal Pavelka, Václav Klika, Oğul Esen, and Miroslav Grmela.
\newblock A hierarchy of {P}oisson brackets in non-equilibrium thermodynamics.
\newblock \emph{Physica D: Nonlinear Phenomena}, 335:\penalty0 54--69, 2016.
\newblock ISSN 0167-2789.
\newblock \doi{10.1016/j.physd.2016.06.011}.

\bibitem[{Pichard, Teddy}(2023)]{Ted2023}
{Pichard, Teddy}.
\newblock Some recent advances on the method of moments in kinetic theory.
\newblock \emph{ESAIM: ProcS}, 75:\penalty0 86--95, 2023.
\newblock \doi{10.1051/proc/202375086}.

\bibitem[Porteous et~al.(2023)Porteous, Laiu, and Hauck]{Por2023}
William~A. Porteous, Ming Tse~P. Laiu, and Cory~D. Hauck.
\newblock Data-driven, structure-preserving approximations to entropy-based
  moment closures for kinetic equations.
\newblock \emph{Commun. Math. Sci}, 21\penalty0 (4):\penalty0 885--913, 2023.
\newblock \doi{10.4310/CMS.2023.v21.n4.a1}.

\bibitem[Ramirez et~al.(2012)Ramirez, Maschke, and Sbarbaro]{ramirez_2012}
Héctor Ramirez, Bernhard Maschke, and Daniel Sbarbaro.
\newblock Irreversible port {H}amiltonian systems.
\newblock \emph{IFAC Proceedings Volumes}, 45\penalty0 (19):\penalty0 13--18,
  2012.
\newblock \doi{https://doi.org/10.3182/20120829-3-IT-4022.00014}.
\newblock 4th IFAC Workshop on Lagrangian and Hamiltonian Methods for Non
  Linear Control.

\bibitem[Ramirez and Le~Gorrec(2022)]{ramirez_2022}
Hector Ramirez and Yann Le~Gorrec.
\newblock An overview on irreversible port-hamiltonian systems.
\newblock \emph{Entropy}, 24\penalty0 (10), 2022.
\newblock \doi{10.3390/e24101478}.

\bibitem[Rashad et~al.(2021{\natexlab{a}})Rashad, Califano, Schuller, and
  Stramigioli]{rashad_2021a}
Ramy Rashad, Federico Califano, Frederic~P. Schuller, and Stefano Stramigioli.
\newblock Port-{H}amiltonian modeling of ideal fluid flow: Part {I}.
  foundations and kinetic energy.
\newblock \emph{Journal of Geometry and Physics}, 164:\penalty0 104201,
  2021{\natexlab{a}}.
\newblock ISSN 0393-0440.
\newblock \doi{https://doi.org/10.1016/j.geomphys.2021.104201}.

\bibitem[Rashad et~al.(2021{\natexlab{b}})Rashad, Califano, Schuller, and
  Stramigioli]{rashad_2021b}
Ramy Rashad, Federico Califano, Frederic~P. Schuller, and Stefano Stramigioli.
\newblock Port-{H}amiltonian modeling of ideal fluid flow: Part {II}.
  compressible and incompressible flow.
\newblock \emph{Journal of Geometry and Physics}, 164:\penalty0 104199,
  2021{\natexlab{b}}.
\newblock ISSN 0393-0440.
\newblock \doi{https://doi.org/10.1016/j.geomphys.2021.104199}.

\bibitem[Sadr et~al.(2021)Sadr, Wang, and Gorji]{sadr2021}
Mohsen Sadr, Qian Wang, and M.~Hossein Gorji.
\newblock Coupling kinetic and continuum using data-driven maximum entropy
  distribution.
\newblock \emph{Journal of Computational Physics}, 444:\penalty0 110542, 2021.
\newblock ISSN 0021-9991.
\newblock \doi{10.1016/j.jcp.2021.110542}.

\bibitem[Scovel and Weinstein(1994)]{Scovel_Weinstein_1994}
C.~Scovel and A.~Weinstein.
\newblock {Finite-dimensional Lie-Poisson approximations to Vlasov-Poisson
  equations}.
\newblock \emph{Comm. Pure Appl. Math.}, 47:\penalty0 683--709, 1994.
\newblock \doi{https://doi.org/10.1002/cpa.3160470505}.

\bibitem[Sengers et~al.(2000)Sengers, Kayser, Peters, and Jr.]{thermo_eos}
J.V. Sengers, R.F. Kayser, C.J. Peters, and H.J.~White Jr., editors.
\newblock \emph{Equations of State for Fluids and Fluid Mixtures}, volume~5 of
  \emph{Experimental Thermodynamic}.
\newblock Elsevier, 2000.

\bibitem[Southworth et~al.(2024)Southworth, Park, Tokareva, and
  Charest]{imex-radhydro}
Ben~S. Southworth, HyeongKae Park, Svetlana Tokareva, and Marc Charest.
\newblock Implicit-explicit {R}unge-{K}utta for radiation hydrodynamics {I}:
  Gray diffusion.
\newblock \emph{Journal of Computational Physics}, 518:\penalty0 113339, 2024.
\newblock ISSN 0021-9991.
\newblock \doi{10.1016/j.jcp.2024.113339}.

\bibitem[Tassi(2014)]{Tassi_2014}
E.~Tassi.
\newblock {Hamiltonian closures for two-moment fluid models derived from
  drift-kinetic equations}.
\newblock \emph{J. Phys. A}, 47:\penalty0 195501, 2014.
\newblock \doi{10.1088/1751-8113/47/19/195501}.

\bibitem[Tracy et~al.(2014)Tracy, Brizard, Richardson, and Kaufman]{wkbraytext}
E.~R. Tracy, A.~J. Brizard, A.~S. Richardson, and A.~N. Kaufman.
\newblock \emph{Ray Tracing and Beyond: Phase Space Methods in Plasma Wave
  Theory}.
\newblock Cambridge University Press, 2014.

\bibitem[Tran et~al.(2024)Tran, Burby, and Southworth]{pHradhydro}
Brian~K. Tran, Joshua~W. Burby, and Ben~S. Southworth.
\newblock Geometric formulation of three-temperature radiation hydrodynamcs.
\newblock In \emph{16th World Congress in Computational Mechanics (WCCM2024)},
  2024.
\newblock \doi{10.23967/wccm.2024.027}.

\bibitem[van~der Schaft and Jeltsema(2014)]{vdS_2014}
Arjan van~der Schaft and Dimitri Jeltsema.
\newblock Port-{H}amiltonian systems theory: An introductory overview.
\newblock \emph{Found. Trends Syst. Control}, 1\penalty0 (2–3):\penalty0
  173–378, 2014.
\newblock ISSN 2325-6818.
\newblock \doi{10.1561/2600000002}.

\end{thebibliography}

\end{document}